\documentclass[twoside]{report}
\usepackage{latexsym,amssymb, stmaryrd,epsf,lj-igpl}
\Title{Title}

\usepackage[dvips]{graphicx}

\newenvironment{ack}{\begin{flushleft}{\bf Acknowledgements.} \ \ }{\end{flushleft}}

\newcommand{\Kn}[2]
{{\bf K}_{ #1 }( #2 )}

\newcommand{\Star}[1]
{\stackrel{\ast}{ #1 }}

\newcommand{\Rii}[3]
{
\begin{tabular}[b]{c c}
        #1 & #2 \\ \hline
        \multicolumn{2}{c}{ #3 }
\end{tabular}
\raisebox{0.8em}{\bf r2}
}

\newcommand{\Sv}
{{\bf S5}}

\newcommand{\Nec}
{\Box}

\newcommand{\NEC}
{\Box}

\newcommand{\Pos}
{\diamond}

\newcommand{\POS}
{\Diamond}

\newcommand{\IMP}
{\supset}

\newcommand{\AND}
{\wedge}

\newcommand{\OR}
{\vee}

\newcommand{\EQU}
{\equiv}

\newcommand{\Lmn}
{{\cal L}_{m,n}}

\newcommand{\dna} 
{\&^{-1}}

\newcommand{\NDWITH} 
{\blacktriangle}

\makeatletter
\def\linearimp{\mathop{- \hspace{-.025in} \circ}}

\def\linearequiv{\mathop{\circ \hspace{-.055in} - \hspace{-.055in} \circ}}

\def\mathdef{\mathop{\rm def} \nolimits}

\def\cconst{\mathop{\rm const} \nolimits}

\def\depth{\mathop{\rm depth} \nolimits}

\def\size{\mathop{\rm size} \nolimits}

\makeatother

\newcommand{\LINEARIMP}   
{\linearimp}
\newcommand{\LINEAREQUIV} 
{\linearequiv}

\newcommand{\NOT}
{\sim}

\newcommand{\NEG}
{\neg}

\newcommand{\LIMP}
{\linearimp}

\newcommand{\ALL}
{\forall}

\newcommand{\TENS}
{\otimes}

\newcommand{\PAR}
{\bindnasrepma}

\newcommand{\WITH}
{\binampersand}

\newcommand{\PLUS}
{\oplus}

\Title{P-time Completeness of Light Linear Logic and its Nondeterministic Extension}
\ShortAuthor{Satoshi Matsuoka}
\LongAuthor{
\author{Satoshi Matsuoka}
\address{
National Institute of Advanced Industrial Science and Technology,\\
1-1-1 Umezono\\
Tsukuba, Ibaraki\\
305-8561 Japan\\
{\tt matsuoka@ni.aist.go.jp}
}
}
\Received{07/10/2003}

\begin{document}
\begin{paper}
\begin{abstract}
In CSL'99 Roversi pointed out that the Turing machine encoding of
Girard's seminal paper "Light Linear Logic" has a flaw.
Moreover he presented a working version of the encoding 
in Light Affine Logic, but not in Light Linear Logic.
In this paper we present a working version of the encoding
in Light Linear Logic. 
The idea of the encoding is based on a remark of Girard's tutorial paper
on Linear Logic.
The encoding is also an example 
which shows usefulness of additive connectives.\\
Moreover we also consider a nondeterministic extension of Light Linear Logic. 
We show that the extended system is {\bf NP}-complete in the same meaning as 
{\bf P}-completeness of Light Linear Logic.
\end{abstract}



\def\mathdef{\mathop{\rm def} \nolimits}
\def\lazy{\mathop{\rm lazy} \nolimits}

\section{Introduction}
In \cite{Rov99}, Roversi pointed out that the Turing machine encoding of
Girard's seminal paper \cite{Gir98} has a flaw. 
The flaw is due to how to encode configurations of Turing machines:
Girard chooses ${\bf list^p} \TENS {\bf list^p} \TENS {\bf bool^q}$ 
as the type of the configurations, 
where the first argument ${\bf list^p}$ represents the left parts of tapes,
the second argument ${\bf list^p}$ the right parts, and the third argument ${\bf bool^q}$ states.
But it is impossible to communicate data between the first and the second in this type:
the communication is needed in transitions of configurations.
Roversi changed the type of configurations in order to make the communication possible
and showed that an encoding of Turing machines based on the type works in {\it Light Affine Logic}, 
which is Intuitionistic Light Linear Logic with unconstrained weakening and without additives. 
But he did not sufficiently discuss whether his encoding works in {\it Light Linear Logic}.\\
In this paper, we show an encoding of Turing machines in {\it Light Linear Logic}. 
This completes {\bf P}-time completeness of Light Linear Logic 
with Girard's Theorem \cite{Gir98} that states computations on proof nets with fixed depth in Light Linear Logic 
belong to class {\bf P}.
The idea of the encoding is based on a remark of Girard's tutorial paper on Linear Logic \cite{Gir95}:
\begin{quotation}
 {\it Affine linear logic} is the system of linear logic enriched (?) with weakening.
 There is no much use for this system since the affine implication between
 $A$ and $B$ can be faithfully mimicked by $1 \WITH A \LIMP B$.
\end{quotation}
Roversi's encoding exploits weakening to discard some information after applications of iterations. 
Our encoding uses  $A \WITH 1$ as type of data that may be discarded. 
On the other hand Light Linear Logic retains principle $!A \TENS !B \LIMP !(A \WITH B)$. 
Because of this principle, we can obtain a proof of $!1 \TENS !A \LIMP !B$ or $!1 \TENS !A \LIMP  \S B$
from a proof of $1 \WITH A \LIMP B$ in Light Linear Logic. 
The obtained proof behaves like a function from $!A$ to $!B$ or $\S B$, not that of $!(1 \WITH A)$:
in other words, outside boxes we can hide additive connectives which are inside boxes.
That is a reason why the encoding works in Light Linear Logic.\\
On the other hand we also try to simplify lazy cut elimination procedure of Light Linear Logic 
in \cite{Gir98}.
The attempt is based on the notion of chains of $\PLUS$-links.
The presentation of Girard's Light Linear Logic~\cite{Gir98} by sequent calculus 
has the {\it comma} delimiter, which implicitly denotes the $\PLUS$-connective.
The comma delimiter also appears in Girard's proof nets for Light Linear Logic.
The introduction of two expressions for the same object complicates the presentation
of Light Linear Logic. We try to exclude the comma delimiter from our proof nets. \\
Next, we consider a nondeterministic extension of the Light Linear Logic system. 
Our approach is to introduce a self-dual additive connective. 
The approach is also discussed in a recently appeared paper~\cite{Mau03}.
But the approach was known to us seven years ago~\cite{Mat96}.
Moreover, our approach is different from that of \cite{Mau03}, 
because we directly use the self-dual additive connective, not SUM rule in \cite{Mau03} and
we use a polymorphic encoding of nondeterminism. 
In particular, our approach does not bother us about commutative reduction 
between nondeterministic rule and other rules unlike \cite{Mau03}. 
\section{The System}
In this section, we define a simplified version of the system of Light Linear Logic (for short LLL)~\cite{Gir98}.
First we present the formulas in the LLL system. 
These formulas ($F$) are inductively constructed from literals ($T$) and logical connectives:
\begin{quotation}
   $T = \alpha \, | \, \beta \, | \, \gamma \, | \,  \ldots \, | \, \alpha^\bot \, | \, \beta^\bot \, | \,  \gamma \, | \,  \ldots$\\
   $F = T \, | \,  1 \, | \, \bot \, | \, F \TENS F \, | \,  F \PAR F \, | \, F \WITH F \, | \, F \PLUS F \, | \, !F \, | \, ?F 
  \, | \, \$F \, | \, \forall \alpha. F \, | \, \exists \alpha. F.$
\end{quotation}
We say unary connective $\$$ is {\it neutral}. Girard \cite{Gir98} used the symbol $\S$ for the connective.
But we use $\$$ since this symbol is an ascii character.

Negations of formulas are defined as follows:
\begin{itemize}
  \item $(\alpha)^\bot \equiv_{\mathdef} \alpha^\bot, (\alpha^\bot)^\bot \equiv_{\mathdef} \alpha$
  \item $1^\bot \equiv_{\mathdef} \bot, \bot^\bot \equiv_{\mathdef} 1$
  \item $(A \TENS B)^\bot \equiv_{\mathdef} A^\bot \PAR B^\bot, (A \PAR B)^\bot \equiv_{\mathdef} A^\bot \TENS B^\bot$
  \item $(A \WITH B)^\bot \equiv_{\mathdef} A^\bot \PLUS B^\bot, (A \PLUS B)^\bot \equiv_{\mathdef} A^\bot \WITH B^\bot$
  \item $(\forall \alpha.A)^\bot \equiv_{\mathdef} \exists \alpha. A^\bot , (\exists \alpha.A)^\bot \equiv_{\mathdef} \forall \alpha. A^\bot$
  \item $(!A)^\bot \equiv_{\mathdef} ?A^\bot ,(?A)^\bot \equiv_{\mathdef} !A^\bot$
  \item $(\$ A)^\bot \equiv_{\mathdef} \$A^\bot$
\end{itemize}

We also define linear implication $\LIMP$ in terms of negation and $\PAR$-connective:
\[ A \LIMP B \equiv_{\mathdef} A^\bot \PAR B \]

In this paper we do not present sequent calculus for Light Linear Logic. 
Instead of that, we present a subclass of Girard's proof nets for Light Linear Logic, {\it simple proof nets}
(precisely, simple proof nets can be mapped into a subclass of Girard's proof nets).
Although there is a proof net that is not simple in the sense of \cite{Gir96}, 
simple proof nets are sufficient for our purpose, encoding of Turing machines, 
because nonsimple proof nets never occur in our encoding.
Moreover it is possible to translate proof nets in the sense of \cite{Gir96} into simple proof nets
although simple proof nets are generally more redundant than nonsimple proof nets.\\
A simple proof net consists of formulas and links.
Figure~\ref{fig:links} shows the {\it links} in LLL: 
$F_\PLUS(A_1,\ldots,A_p)$ represents  a formula that is generated from formulas $A_1,\ldots,A_p$ 
by using $\PLUS$-connective 
and is called {\it general $\PLUS$-formula}.
$S_\$(A_1,\ldots,A_p)$ represents a list of several general $\PLUS$-formulas
that are generated from $A_1,\ldots,A_p$.

\begin{figure}[htbp]
\begin{center}
\includegraphics[scale=.5]{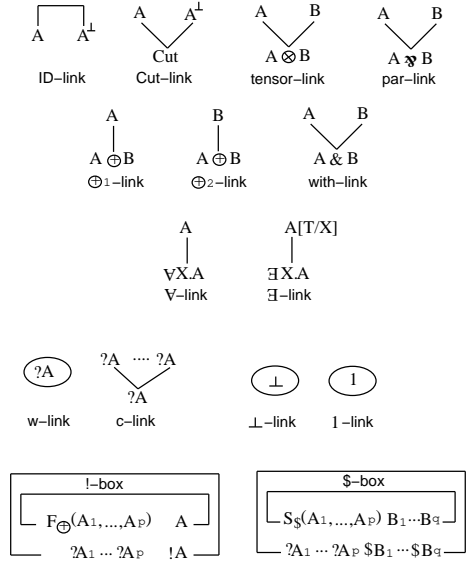}
\caption[the links in the LLL system]{the links in the LLL system}  
\label{fig:links}
\end{center}
\end{figure}

Figure~\ref{simple} shows simple proof nets are defined inductively.
The formulas and links in simple proof nets have weights.
These weights are generated from eigenweights that are associated with $\WITH$-links occurring in 
simple proof nets by using boolean product operator '$\cdot$'. 
If a formula or a link has the weight $1$, then we omit the weight.

\begin{figure}[htbp]
\begin{center}
\includegraphics[scale=.5]{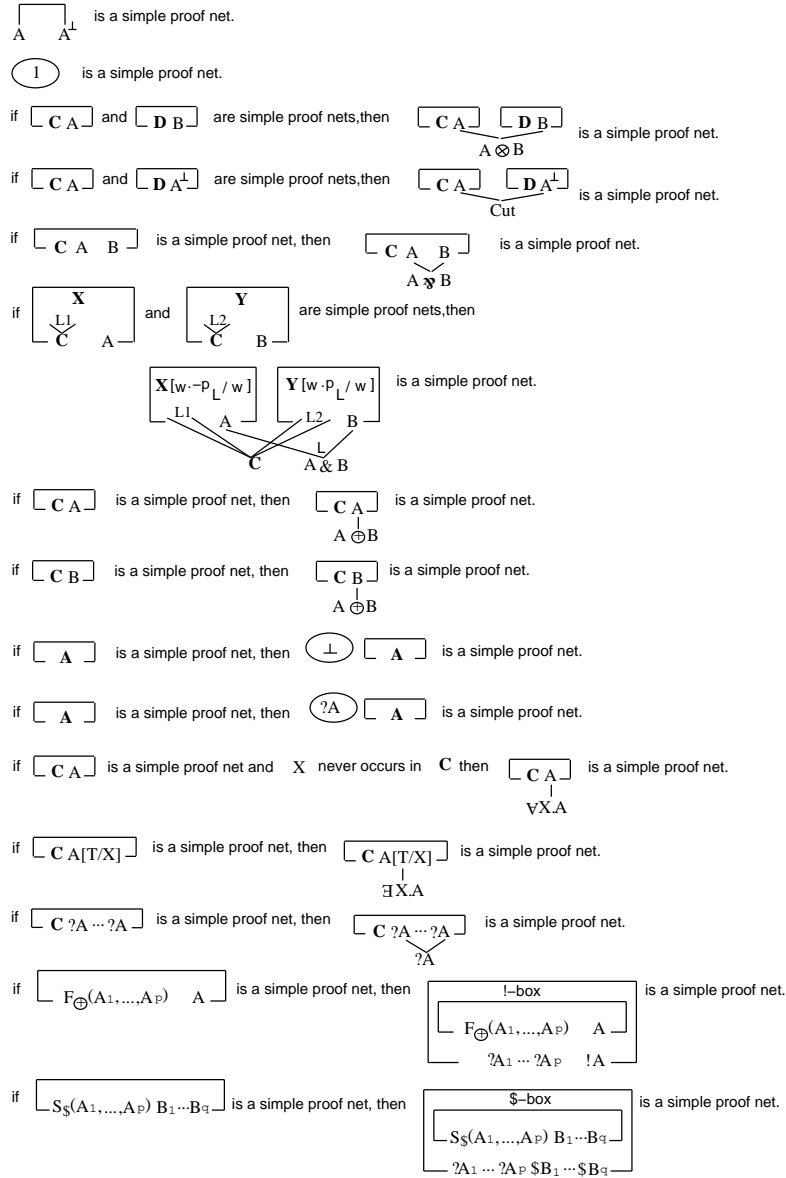}
\caption[the definition of simple proof nets]{the definition of simple proof nets}  
\label{simple}
\end{center}
\end{figure}

Moreover we must take care of the case of $\WITH$-links.
For example from two simple proof nets of Figure~\ref{glueing-pre-ex1} 
we can construct a simple proof net with the conclusions $?A^\bot, ?B^\bot, !A \WITH \$ A$ 
of Figure~\ref{glueing-after-ex1}. As shown in the figure, 
the {\it context}-formulas must be shared. 

\begin{figure}[htbp]
\begin{center}
\includegraphics[scale=.5]{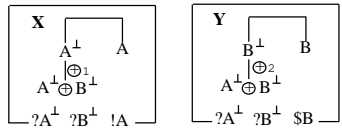}
\caption[two simple proof nets]
{two simple proof nets}  
\label{glueing-pre-ex1}
\end{center}
\end{figure}

\begin{figure}[htbp]
\begin{center}
\includegraphics[scale=.5]{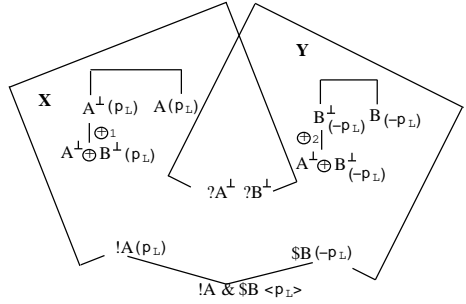}
\caption[An example of constructions of simple proof nets with $\WITH$-links]
{An example of constructions of simple proof nets with $\WITH$-links}  
\label{glueing-after-ex1}
\end{center}
\end{figure}

Moreover sharing of context-formulas may be complex. 
For example, from two simple proof nets of Figure~\ref{glueing-ex2} 
we can construct a simple proof net with 
\[ !A \WITH \$ D, ?B^\bot, ?C^\bot, ?D^\bot, \$C \TENS !A. \]
But it is difficult to write down this on a plane in a concise way. 
So we omit this.

\begin{figure}[htbp]
\begin{center}
\includegraphics[scale=.5]{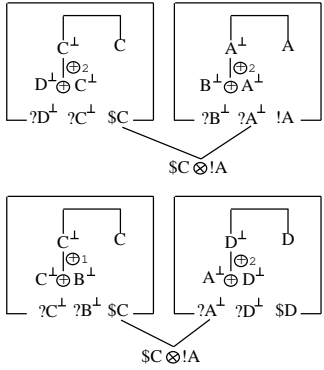}
\caption[more two simple proof nets]
{more two simple proof nets}  
\label{glueing-ex2}
\end{center}
\end{figure}

Figure~\ref{nonsimplepnex} shows an example that is a proof net in the sense of \cite{Gir96}.
The proof net satisfies the correctness condition of \cite{Gir96}.
But it is not simple.
However we can easily construct a simple proof net that has the same conclusions as the proof net.
For example a simple proof net corresponding to that of Figure~\ref{nonsimplepnex} 
is that of Figure~\ref{simplepnex}.
But such a simple proof net is not uniquely determined.
For instance, Figure~\ref{another-simplepnex} shows another simple proof net 
corresponding to that of Figure~\ref{nonsimplepnex}.
Besides, in the introduction rules of $!$-box and $\$$-box 
when we replace $\PLUS$-occurrences of generalized $\PLUS$-formulas by comma delimiters, 
we can easily find that any modified simple proof net in this manner is a proof net of Girard by induction on derivations 
of simple proof nets. 

\begin{figure}[htbp]
\begin{center}
\includegraphics[scale=.5]{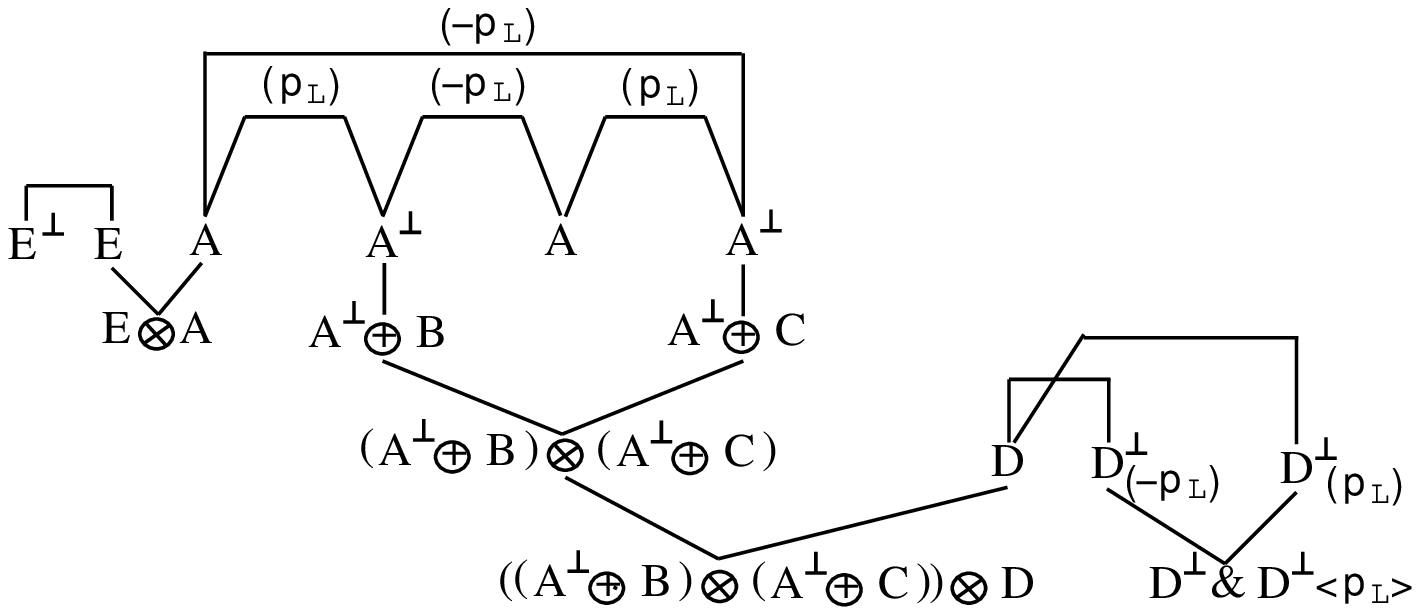}
\caption[An example of non-simple proof nets]
{An example of non-simple proof nets}  
\label{nonsimplepnex}
\end{center}
\end{figure}

\begin{figure}[htbp]
\begin{center}
\includegraphics[scale=.5]{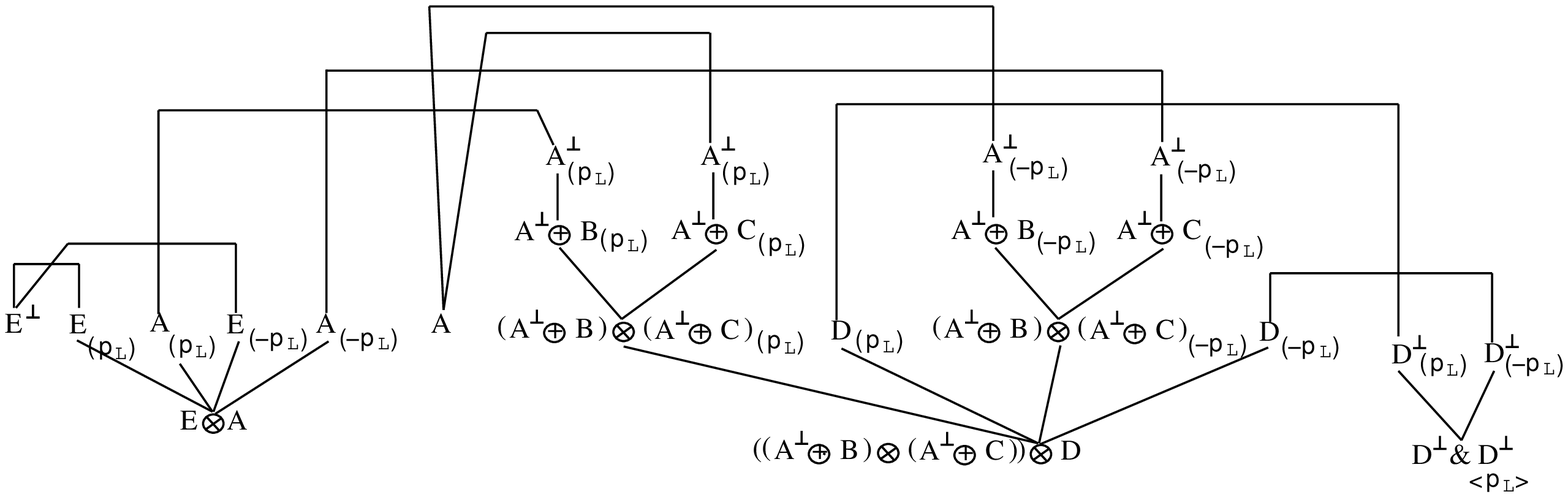}
\caption[A simple proof net corresponding to the above non-simple net]
{A simple proof net corresponding to the above non-simple net}  
\label{simplepnex}
\end{center}
\end{figure}

\begin{figure}[htbp]
\begin{center}
\includegraphics[scale=.5]{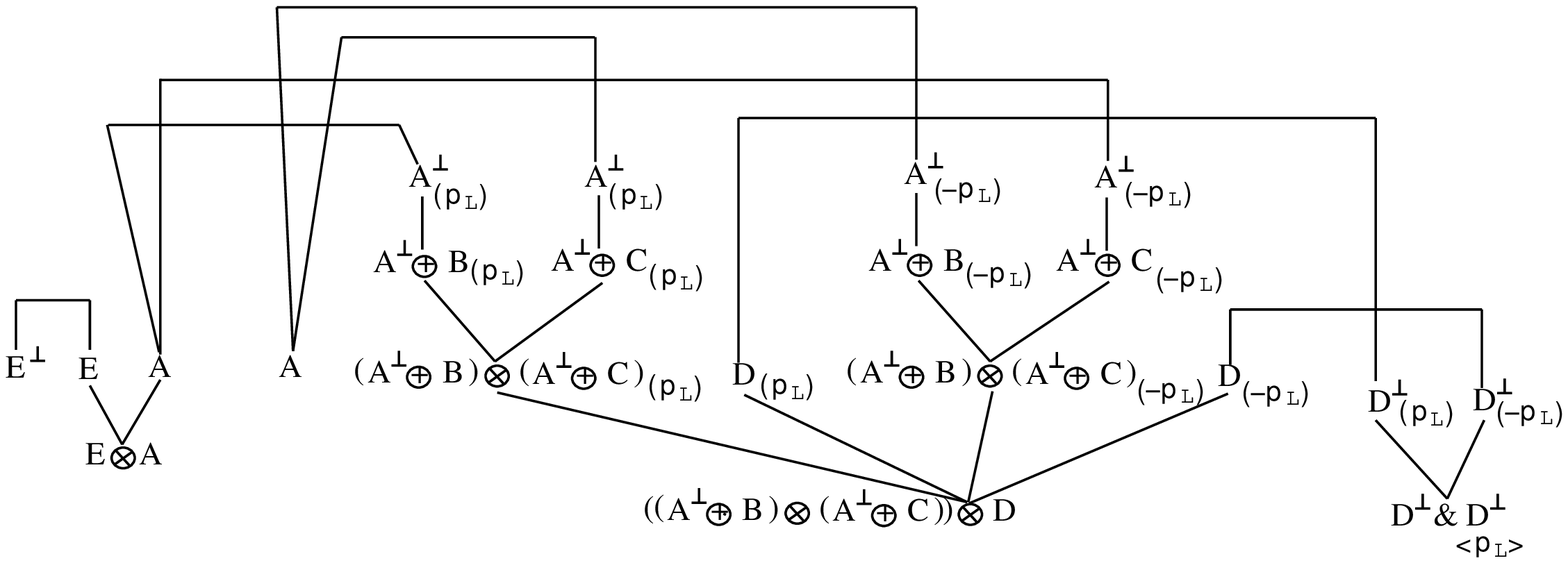}
\caption[Another simple proof net corresponding to the above non-simple net]
{Another simple proof net corresponding to the above non-simple net}  
\label{another-simplepnex}
\end{center}
\end{figure}

Figure~\ref{figCutElim} shows the {\it rewrite rules} in the LLL system 
except for contraction, neutral, $!$-$!$, and $!$-$\$$ rewrite rules.
Fusion and c-w rewrite rules first appeared in \cite{DK97}. 
The other rewrite rules in Figure~\ref{figCutElim} are standard in Linear Logic.
Figure~\ref{neutral} shows neutral rewrite rule.
Figure~\ref{contraction} shows contraction rewrite rule, 
where 
${\bf \$ Y}$, ${\bf \$ Y'}$, ${\bf !Z}$ and ${\bf !Z'}$ represent sequences of proof nets and 
'{\bf w}' a sequence of weakening links. 
The length of ${\bf \$ Y}$ must be  the same as that of ${\bf \$ Y'}$ and
the length of ${\bf !Z}$ the same as that of ${\bf !Z'}$.
Let ${\bf \$ Y}= Y_1,\ldots,Y_m$, ${\bf \$ Y'}= Y'_1,\ldots,Y'_m$, 
${\bf \$ Z}= Z_1,\ldots,Z_n$, ${\bf \$ Z'}= Z'_1,\ldots,Z'_n$.
Each $Y_i \, (1 \le i \le m)$ and $Z_\ell \, (1 \le \ell \le n)$ must have the following conditions:
\begin{enumerate}
\item The conditions on $Z_\ell$.\\
Each $Z_\ell$ must have the form of the upper proof net of 
Figure~\ref{new-contraction-contractum-whynot-nonfake}
or that of Figure~\ref{new-contraction-contractum-whynot-fake}. 
In both proof nets, the first $j$ arguments of $G_\PLUS(A_1,\ldots,A_n)$ are all $A^\bot$ occurrences and 
all the links from $A_i$ to $G_\PLUS(A_1,\ldots,A_n)$ are $\PLUS$-links that have weight $1$
(therefore all the formulas from $A_i$ to $G_\PLUS(A_1,\ldots,A_n)$ are not conclusions of two links. 
We call such a sequence of $\PLUS$-links {\it $\PLUS$-chain}). 
In the former case $A_i$ is equal to $A^\bot$ (in this case $i \le j$)
and in the latter case $A_i$ not (in this case $i > j$).
We call the former $\PLUS$-chain {\it non-fake}
and the latter {\it fake}.
\item The conditions on $Y_i$.\\
Each $Y_i$ must have the form of the upper proof net of 
Figure~\ref{new-contraction-contractum-newtral-nonfake-90}
or that of Figure~\ref{new-contraction-contractum-newtral-fake-90}. 
In the former case there are some non-fake chains, 
but in the latter case all the $\PLUS$-chains are fake.
\end{enumerate}
In other words each proof net in ${\bf \$ Y}$ and ${\bf !Z}$ must have 
at least one $\PLUS$-chain. 
Moreover each $Y'_i \, (1 \le i \le m)$ and $Z'_\ell \, (1 \le \ell \le n)$  must have 
the following forms according to $Y_i$ and $Z_\ell$:
\begin{enumerate}
\item The case where the $PLUS$-chain of $Z_\ell$ is non-fake:\\
Then $Z'_\ell$ must be the lower proof net of Figure~\ref{new-contraction-contractum-whynot-nonfake}.
\item The case where the $\PLUS$-chain of $Z_\ell$ is fake:\\
Then $Z'_\ell$ must be the lower proof net of Figure~\ref{new-contraction-contractum-whynot-fake}.
\item The case where some $\PLUS$-chains of $Y_i$ are non-fake:\\
Then $Y'_i$ must be the lower proof net of Figure~\ref{new-contraction-contractum-newtral-nonfake-90}. 
The notation $\hat{?B_{r_j}}$ of the right side means that 
the weakening link with conclusion ${?B_{r_j}}$ is missing in the proof net.
\item The case where all the $\PLUS$-chains of $Y_i$ are fake:\\
Then $Y'_i$ must be the lower proof net of Figure~\ref{new-contraction-contractum-newtral-fake-90}.
\end{enumerate}

Note that neither the left hand side nor the right hand side of 
Figure~\ref{contraction} is a simple proof net. 
If we find a pattern of the left hand side of 
Figure~\ref{contraction} in a simple proof net, 
we can apply the contraction rule to the simple proof net and
replace the pattern by an appropriate instantiation of 
the right hand side of Figure~\ref{contraction}.

\begin{figure}[htbp]
\begin{center}
\includegraphics[scale=.5]{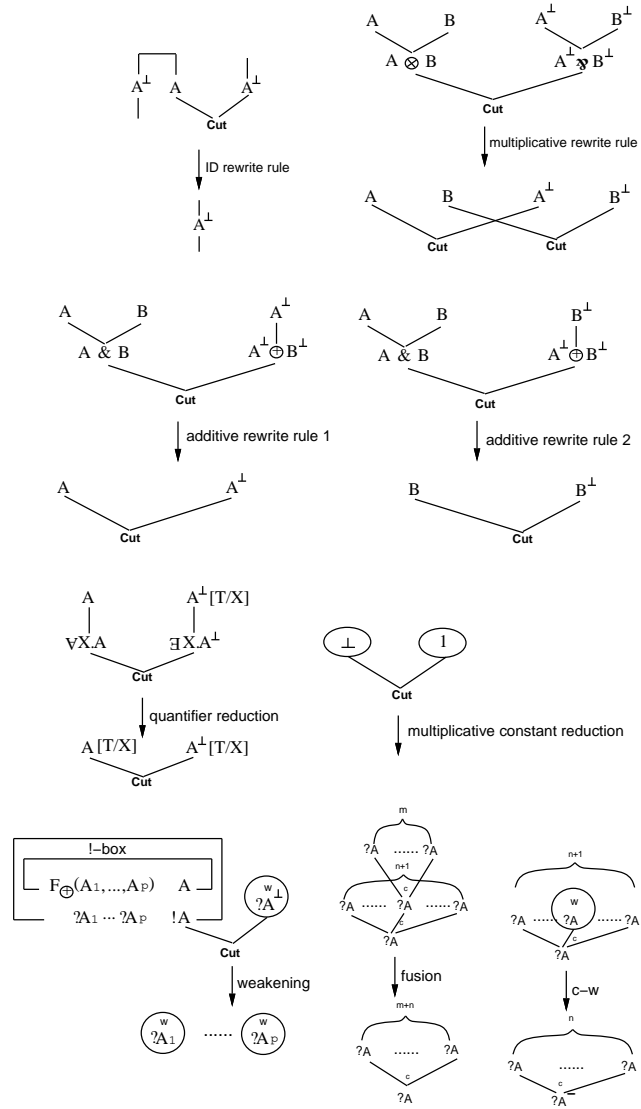}
\caption[the rewrite rules in the LLL system]{the rewrite rules in the LLL system}  
\label{figCutElim}
\end{center}
\end{figure}

\begin{figure}[htbp]
\begin{center}
\includegraphics[scale=0.5]{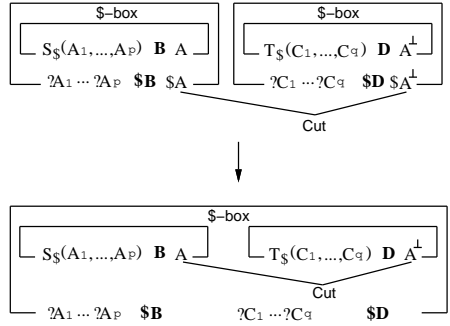}
\caption[neutral rewrite rule]{neutral rewrite rule}
\label{neutral}
\end{center}
\end{figure}

\begin{figure}[htbp]
\begin{center}
\includegraphics[scale=0.5]{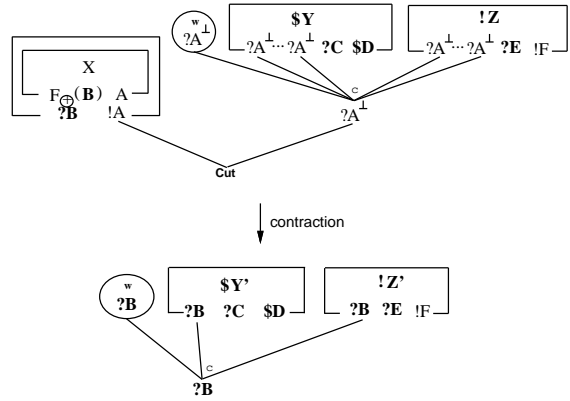}
\caption[contraction rewrite rule]{contraction rewrite rule}
\label{contraction}
\end{center}
\end{figure}
\begin{figure}[htbp]
\begin{center}
\includegraphics[scale=.5]{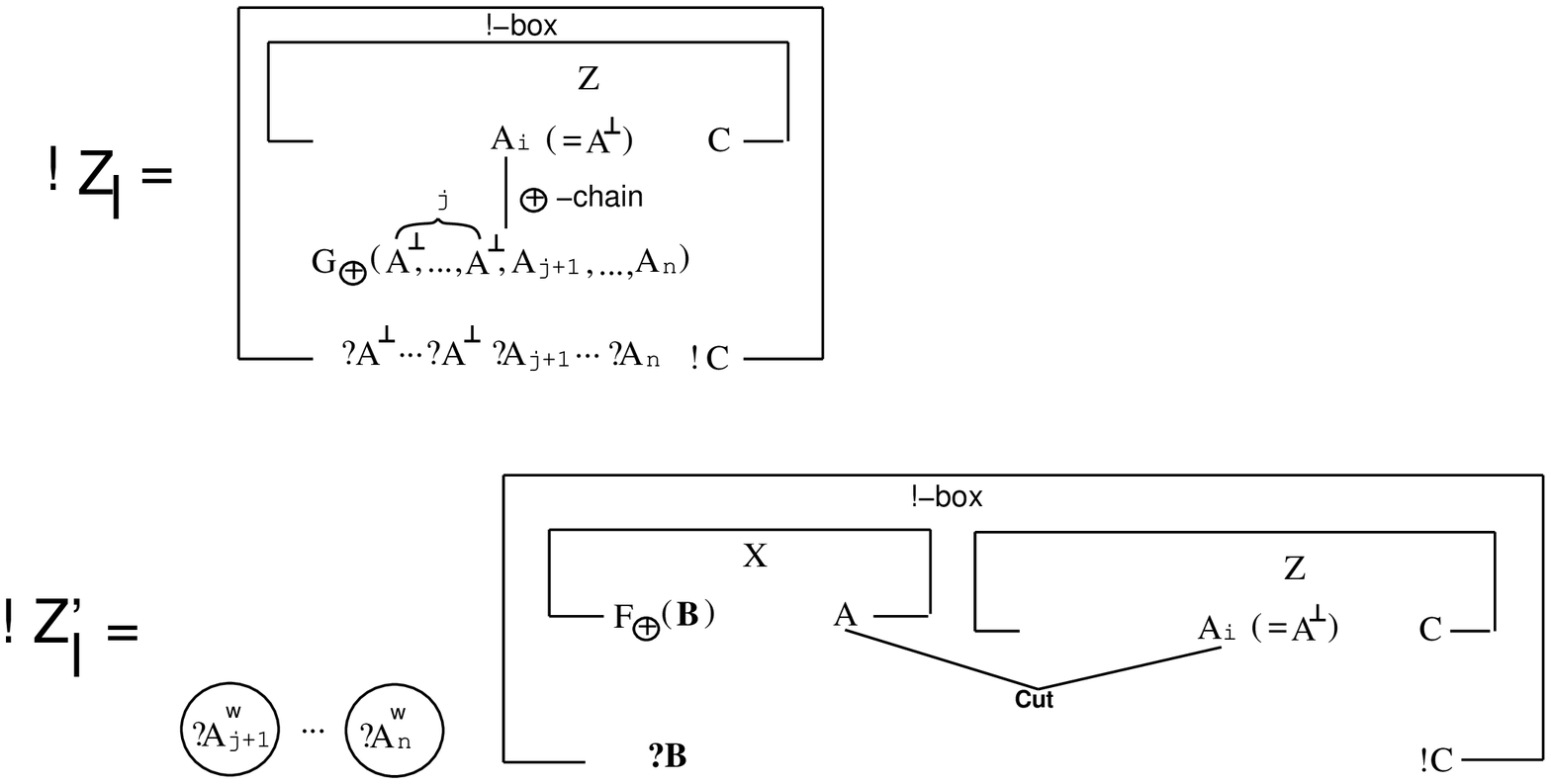}
\caption[The case where $Z_\ell$ has the non-fake $\PLUS$-chain]{The case where $Z_\ell$ has the non-fake $\PLUS$-chain}
\label{new-contraction-contractum-whynot-nonfake}
\end{center}
\end{figure}

\begin{figure}[htbp]
\begin{center}
\includegraphics[scale=.5]{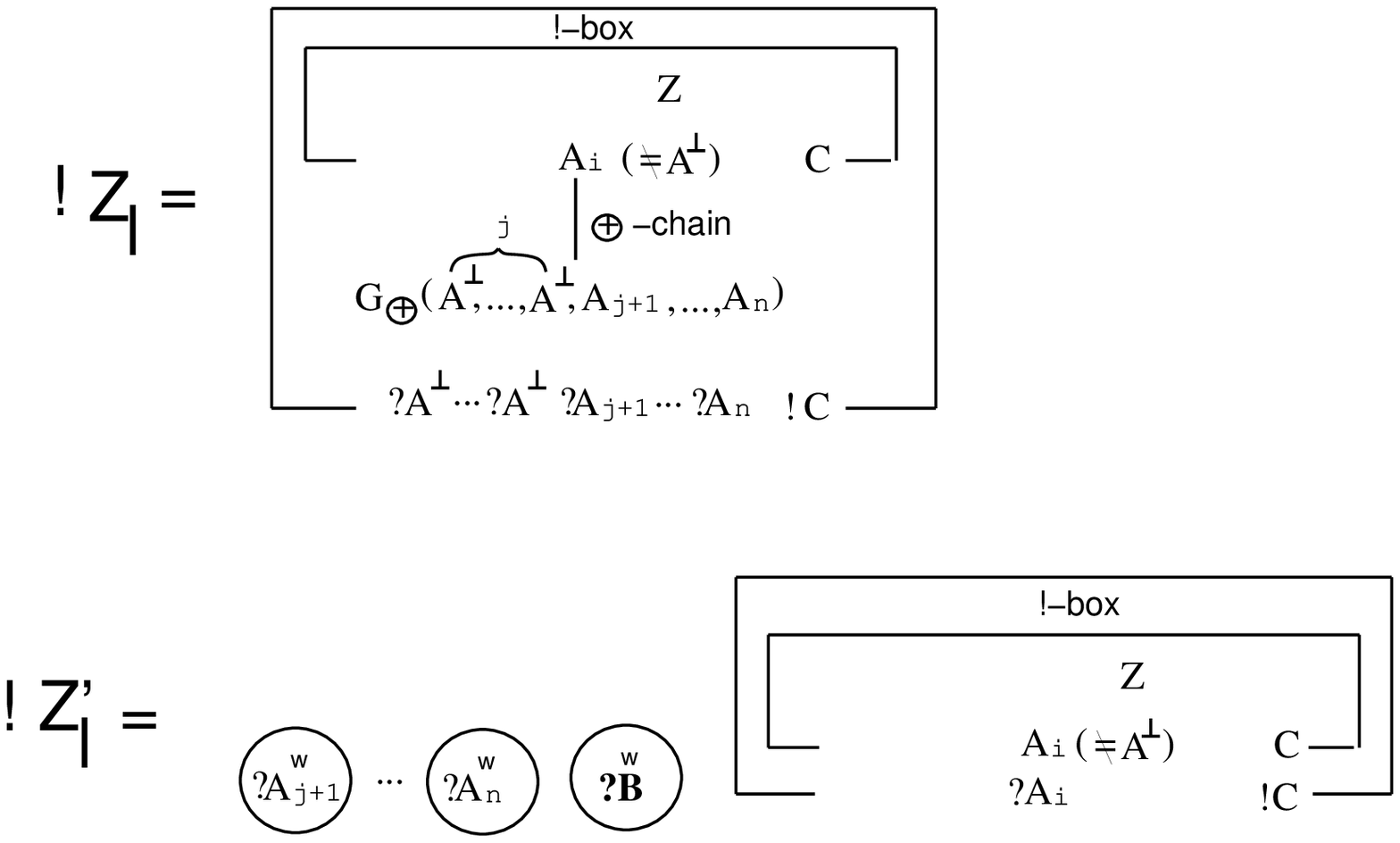}
\caption[The case where $Z_\ell$ has the fake $\PLUS$-chain]{The case where $Z_\ell$ has the fake $\PLUS$-chain}
\label{new-contraction-contractum-whynot-fake}
\end{center}
\end{figure}

\begin{figure}[htbp]
\begin{center}
\includegraphics[scale=.5]{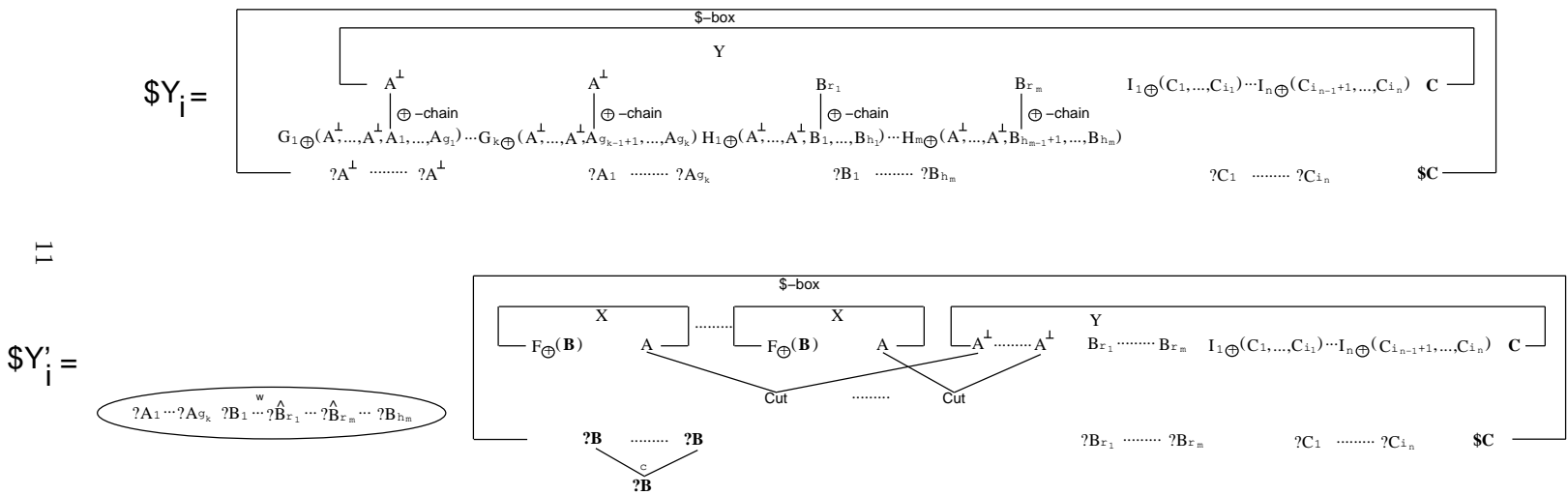}
\caption[The case where $Y_i$ has non-fake $\PLUS$-chains]{The case where $Y_i$ has non-fake $\PLUS$-chains}
\label{new-contraction-contractum-newtral-nonfake-90}
\end{center}
\end{figure}

\begin{figure}[htbp]
\begin{center}
\includegraphics[scale=.5]{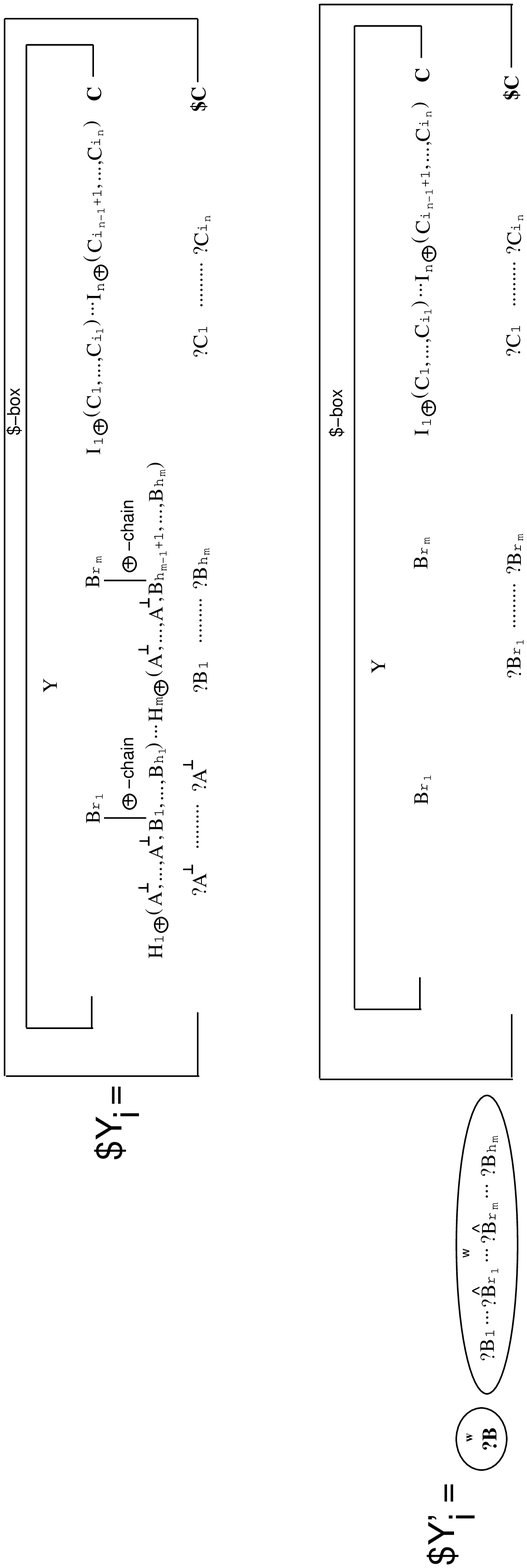}
\caption[The case where $Y_i$ does not have any non-fake $\PLUS$-chains]
{The case where $Y_i$ does not have any non-fake $\PLUS$-chains}
\label{new-contraction-contractum-newtral-fake-90}
\end{center}
\end{figure}




Let us recall lazy cut elimination in \cite{Gir96}.
\begin{definition}
Let $L$ be a Cut-link in an additive proof net. 
When two premises of $L$ are $A$ and $A^\bot$, 
$L$ is {\it ready} if 
\begin{enumerate}
\item $L$ has the weight $1$;
\item Both $A$ and $A^\bot$ are the conclusion of exactly one link.
\end{enumerate}
\end{definition}
For example, in Figure~\ref{readynonreadyex}, 
the right cut is ready, but the left not. 
After the right cut is rewritten, the left become ready.

\begin{figure}[htbp]
\begin{center}
\includegraphics[scale=.5]{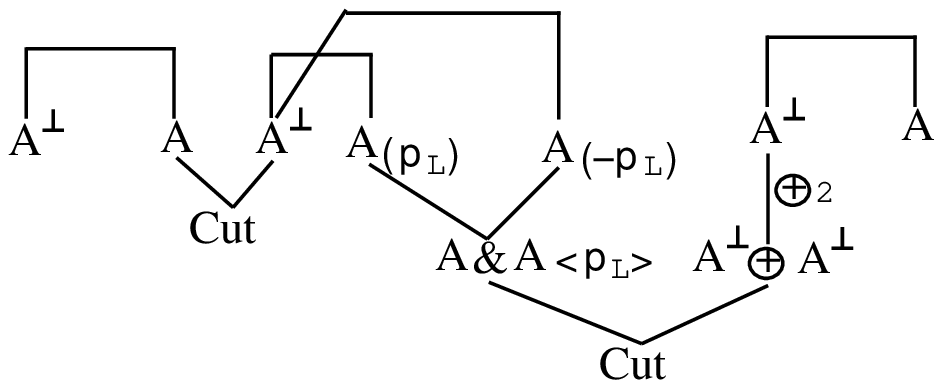}
\caption[An example of ready cuts and non-ready cuts]
{An example of ready cuts and non-ready cuts}  
\label{readynonreadyex}
\end{center}
\end{figure}

{\it Lazy cut elimination} is a reduction procedure in which 
only ready cuts are redexes (of course, in the contraction rewrite rule the above mentioned
conditions must be satisfied). 
The definition also applies to our rewrite rules. So we use the definition.
By $\to_{\lazy}$ we denote one step reduction of lazy cut elimination.

\begin{theorem}
Let $\Theta_1$ be a simple proof net. 
If $\Theta_1 \to_{\lazy} \Theta_2$, then $\Theta_2$ is also a simple proof net.
\end{theorem}

\begin{proof}
Induction on the construction of simple proof net $\Theta_1$ and an easy argument on permutations of links. 
\end{proof}

Next, we relate lazy cut elimination of simple proof nets with 
that of Girard's proof nets. 
\begin{proposition}
One step of lazy cut elimination of simple proof nets can be 
simulated by several steps of that of Girard's proof nets.
\end{proposition}
We do not present the proof because 
in order to prove this we must rephrase the full details of Girard's proof nets.
We just show the difference between them.
The left cut of Figure~\ref{nonsimpleredex-ex1} is a redex of 
Girard's lazy cut elimination, but not of that of simple proof nets.
In Girard's lazy cut elimination, Figure~\ref{nonsimpleredex-ex1}
can be reduced to Figure~\ref{nonsimple-contractum-ex1}.
That does not happen to simple proof nets. 
Instead of that, in lazy cut elimination of simple proof nets 
the right cut of Figure~\ref{nonsimpleredex-ex1} is ready and
Figure~\ref{nonsimpleredex-ex1} can be reduced to Figure~\ref{simpleredex-ex1}.
Then the residual left cut of Figure~\ref{simpleredex-ex1} become ready. 
In lazy cut elimination of simple proof nets, Figure~\ref{simpleredex-ex1}
is reduced to Figure~\ref{girard-lazy-red-ex2} by {\it one-step}. 
But in Girard's lazy cut elimination this reduction takes {\it two-steps}. 
For example we need an intermediate proof net like Figure~\ref{girard-lazy-red-ex1}.

It is obvious that there is a proof net that is reduced to a cut-free form
in Girard's lazy cut elimination, but not in lazy cut elimination of simple proof nets.
Hence, in this sense, our lazy cut elimination is weaker than that of Girard's proof nets.
But, 
when we  execute Theorem~\ref{maintheorem2},
that is, compute polynomial bounded functions on binary integers 
in proof nets, our lazy cut eliminations and Girard's always return the same 
result, since this is due to the following Girard's theorem 
and our binary integer encoding in simple proof nets does not have any  
$\WITH$-occurrences.

\begin{theorem}[\cite{Gir96}]
\label{notwiththeorem}
Let $\Theta$ be a proof-net whose conclusions do not contain the connective
$\WITH$ and $\exists X.$ and without ready cut; then $\Theta$ is cut-free.
\end{theorem}

\begin{figure}[htbp]
\begin{center}
\includegraphics[scale=.5]{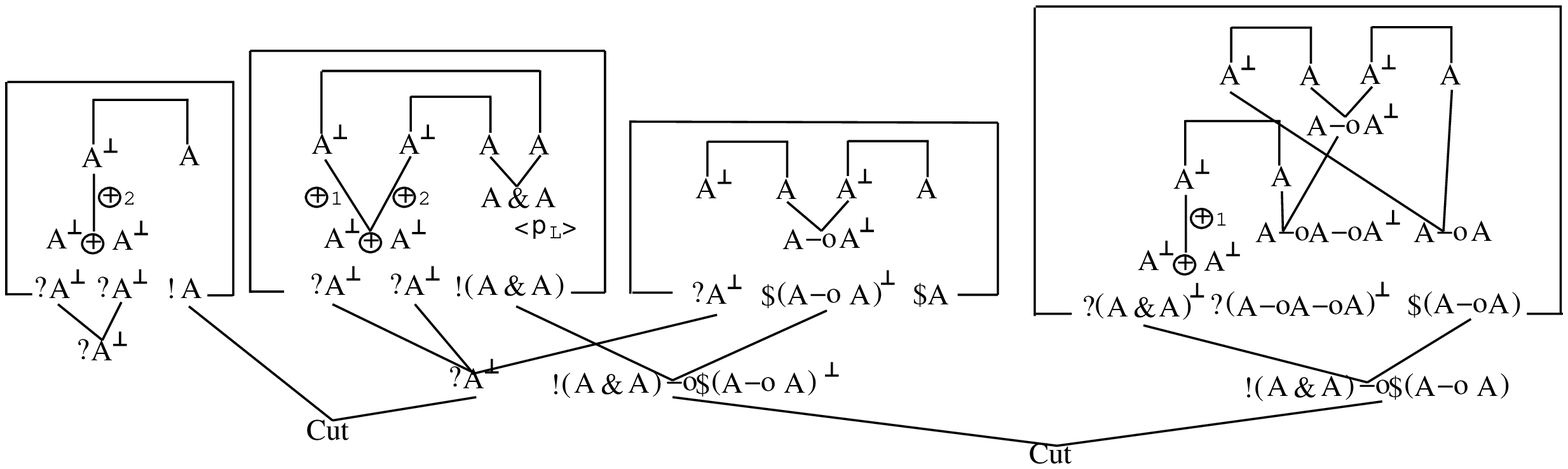}
\caption[an example that is a redex of Girard' proof nets but not that of simple proof nets]
{an example that is a redex of Girard' proof nets but not that of simple proof nets}  
\label{nonsimpleredex-ex1}
\end{center}
\end{figure}

\begin{figure}[htbp]
\begin{center}
\includegraphics[scale=.5]{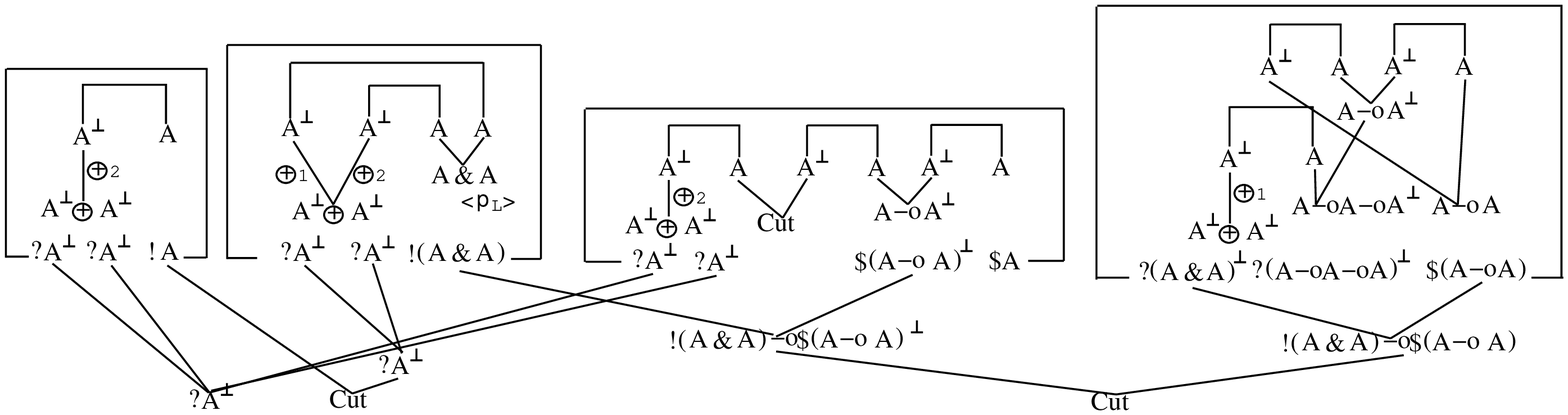}
\caption[a contractum of Girard's proof nets]
{a contractum of Girard's proof nets}  
\label{nonsimple-contractum-ex1}
\end{center}
\end{figure}

\begin{figure}[htbp]
\begin{center}
\includegraphics[scale=.5]{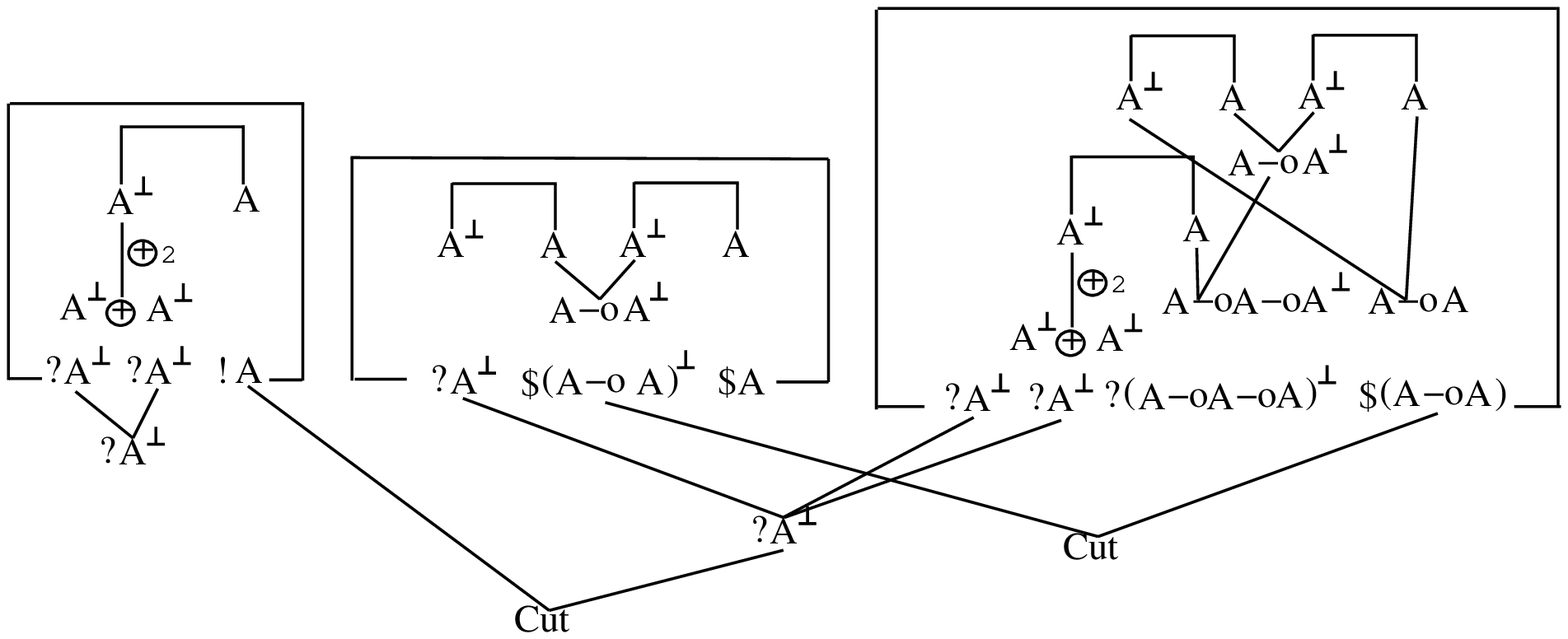}
\caption[an example that is a redex of simple proof nets]
{an example that is a redex of simple proof nets}  
\label{simpleredex-ex1}
\end{center}
\end{figure}

\begin{figure}[htbp]
\begin{center}
\includegraphics[scale=.5]{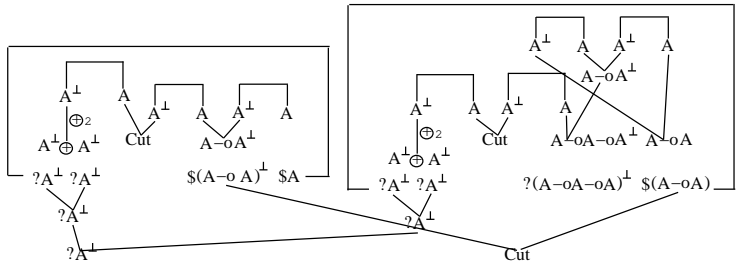}
\caption[a contractum of Figure~\ref{simpleredex-ex1}]
{a contractum of Figure~\ref{simpleredex-ex1}}  
\label{girard-lazy-red-ex2}
\end{center}
\end{figure}

\begin{figure}[htbp]
\begin{center}
\includegraphics[scale=.5]{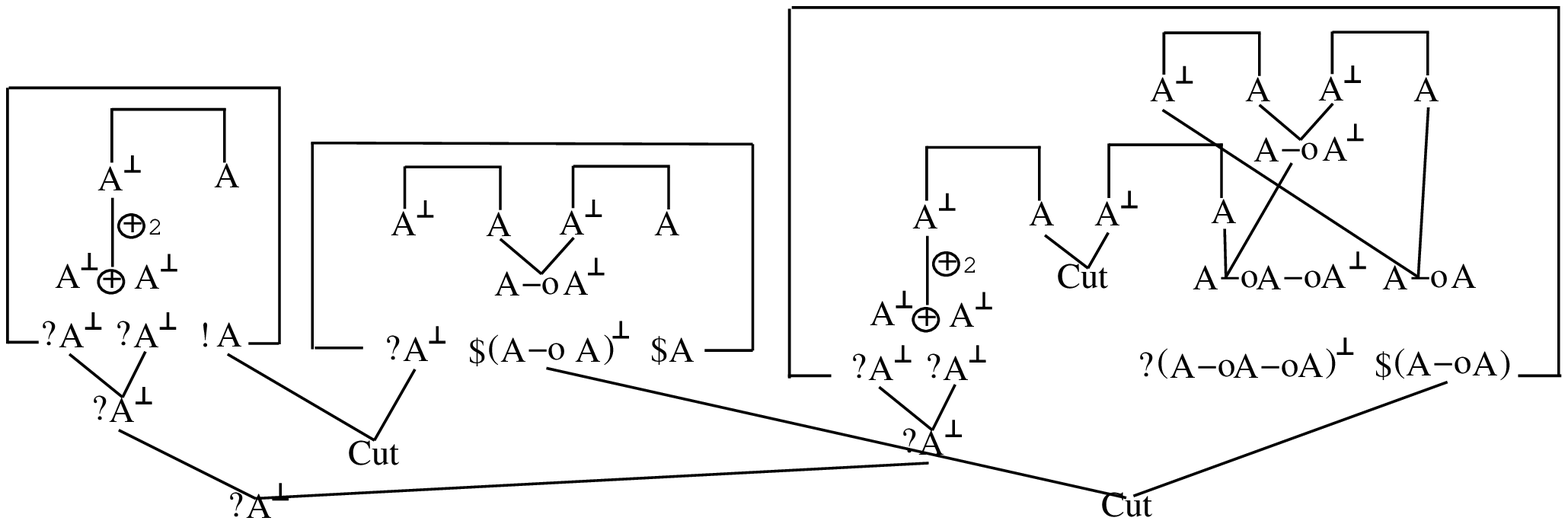}
\caption[an intermediate proof net]
{an intermediate proof net}  
\label{girard-lazy-red-ex1}
\end{center}
\end{figure}

\section{A Turing Machine Encoding}
Let $M$ be a Turing machine and $k$ be the number of the states of $M$.
Without loss of generality, we can assume that only $0$, $1$, and $\ast$ occur in 
the tape of $M$, where $\ast$ is the blank symbol of $M$.

We use
\[{\bf bool^k} \equiv_{\mathdef} \forall X. 
\overbrace{X \WITH (\cdots \WITH (X \WITH X) \cdots ) }^{k} \LIMP X\]
for the type of the states of $M$.
In contrast to ${\bf bool^k}$ in \cite{Gir98}, ${\bf bool^k}$ in this paper does not include 
the neutral connective $\$$.
Figure~\ref{boolex} shows an example of ${\bf bool^k}$ proofs.
After $0$ or $1$ $\PLUS_1$-link, $\PLUS_2$-links follow $k-1$ or $i-1$ times.

\begin{figure}[htbp]
\begin{center}
\includegraphics[scale=0.5]{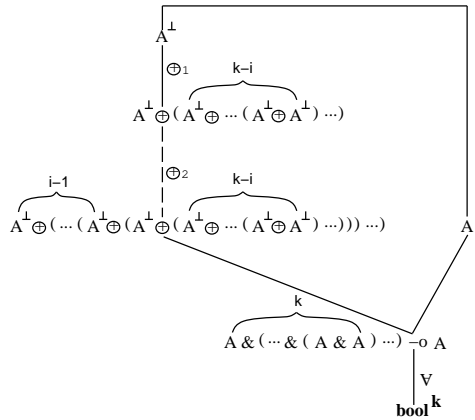}
\caption[an example of ${\bf bool^k}$ proofs]{an example of ${\bf bool^k}$ proofs}
\label{boolex}
\end{center}
\end{figure}


In addition we use 
\[{\bf config} \equiv_{\mathdef} \forall X. 
!(X \LIMP X) \LIMP  !(X \LIMP X) \LIMP !(X \LIMP X) \LIMP \$(X \LIMP X \LIMP (X \TENS X) \TENS {\bf bool^k})\]
for the type of configurations of $M$.
The type represents the current configuration of running $M$, that is, 
the 3-tuple of the left part of the current tape, the right part, and the current state.
Figure~\ref{configex} 
shows an example of ${\bf config}$ proofs.
In the $\lambda$-notation, the example is 
$\lambda f_0. \lambda f_1. \lambda f_\ast. \lambda x. \lambda y. \langle f_0 (f_1 (x)), f_\ast (f_1 (f_0 (y))), b \rangle$, 
where $b$ is a ${\bf bool^k}$-value.
Hence the example denotes configuration $\langle 1 0, \ast 1 0, b \rangle$.

\begin{figure}[htbp]
\begin{center}
\includegraphics[scale=0.5]{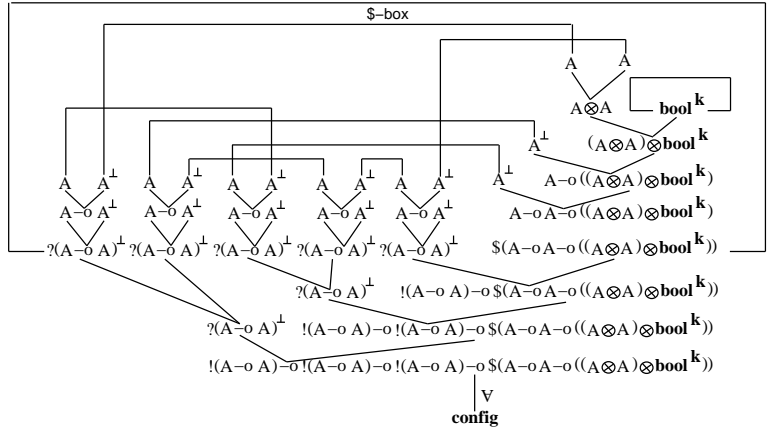}
\caption[an example of ${\bf config}$ proofs]{an example of ${\bf config}$ proofs}
\label{configex}
\end{center}
\end{figure}

For shorthand, we use ${\bf id}_A$ to represent $A \LIMP A$.
Then we write down the transition function of $M$ in Light Linear Logic, which is the main task of the paper.
Figure~\ref{transition} 
shows our encoding of the transition function, 
where ${\bf trpl}_A$ is an abbreviation of
$({\bf bool^4} \WITH 1) \TENS (({\bf id}_A \WITH 1) \TENS A)$.
The formula ${\bf trpl}_A$ is fed to the second-order variable that 
is bound by the $\forall$-link in an input proof of ${\bf config}$.

\begin{figure}[htbp]
\begin{center}
\includegraphics[scale=0.49]{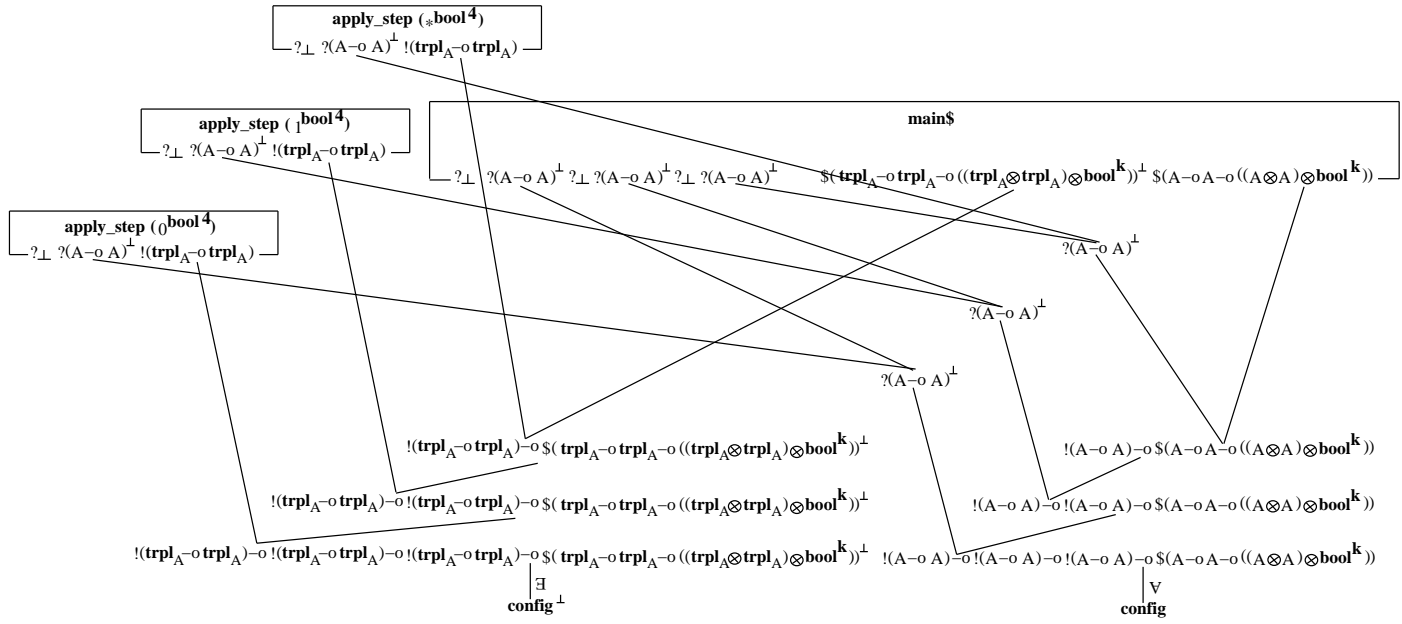}
\caption[{\tt transition} function]{{\tt transition} function}
\label{transition}
\end{center}
\end{figure}

Three proof nets {\tt apply\_step}($0^{\bf bool^4}$), {\tt apply\_step}($1^{\bf bool^4}$), and {\tt apply\_step}($\ast^{\bf bool^4}$)
in Figure~\ref{transition} 
are made up by giving a ${\bf bool^4}$ proof ($0$, $1$, or $\ast$) to 
step of Figure~\ref{step} (see Figure~\ref{applystep}),
where $0^{\bf bool^4}$, $1^{\bf bool^4}$, and $\ast^{\bf bool^4}$ are
different normal proof net of ${\bf bool^4}$. 
$0^{\bf bool^4}$, $1^{\bf bool^4}$, and $\ast^{\bf bool^4}$
represent the symbols $0$, $1$, and $\ast$ on the tape of $M$.
The main purpose of these {\tt apply\_step}($\Theta^{\bf bool^4}$) is 
to decompose the left or right part of the tape of a given configuration into data with type 
${\bf trpl}_A=({\bf bool^4} \WITH 1) \TENS (({\bf id}_A \WITH 1) \TENS A)$,
where both ${\bf bool^4} \WITH 1$ and ${\bf id}_A \WITH 1$ represent 
the top symbol of the left or right part of the tape
and $A$ represents the rest except for the top symbol. 
The principle by which the encoding works is the same as that used in writing down the predecessor function.
There is just one proof net of ${\bf bool^4}$ that are different from these three. 
Let the proof net be ${\tt empty}^{\bf bool^4}$. 
The proof net ${\tt empty}^{\bf bool^4}$ do not have any corresponding symbol on the tape of $M$:
the proof net is used in {\tt apply\_base} of Figure~\ref{baseandapplybase} 
in order to make our encoding easy.

\begin{figure}[htbp]
\begin{center}
\includegraphics[scale=0.5]{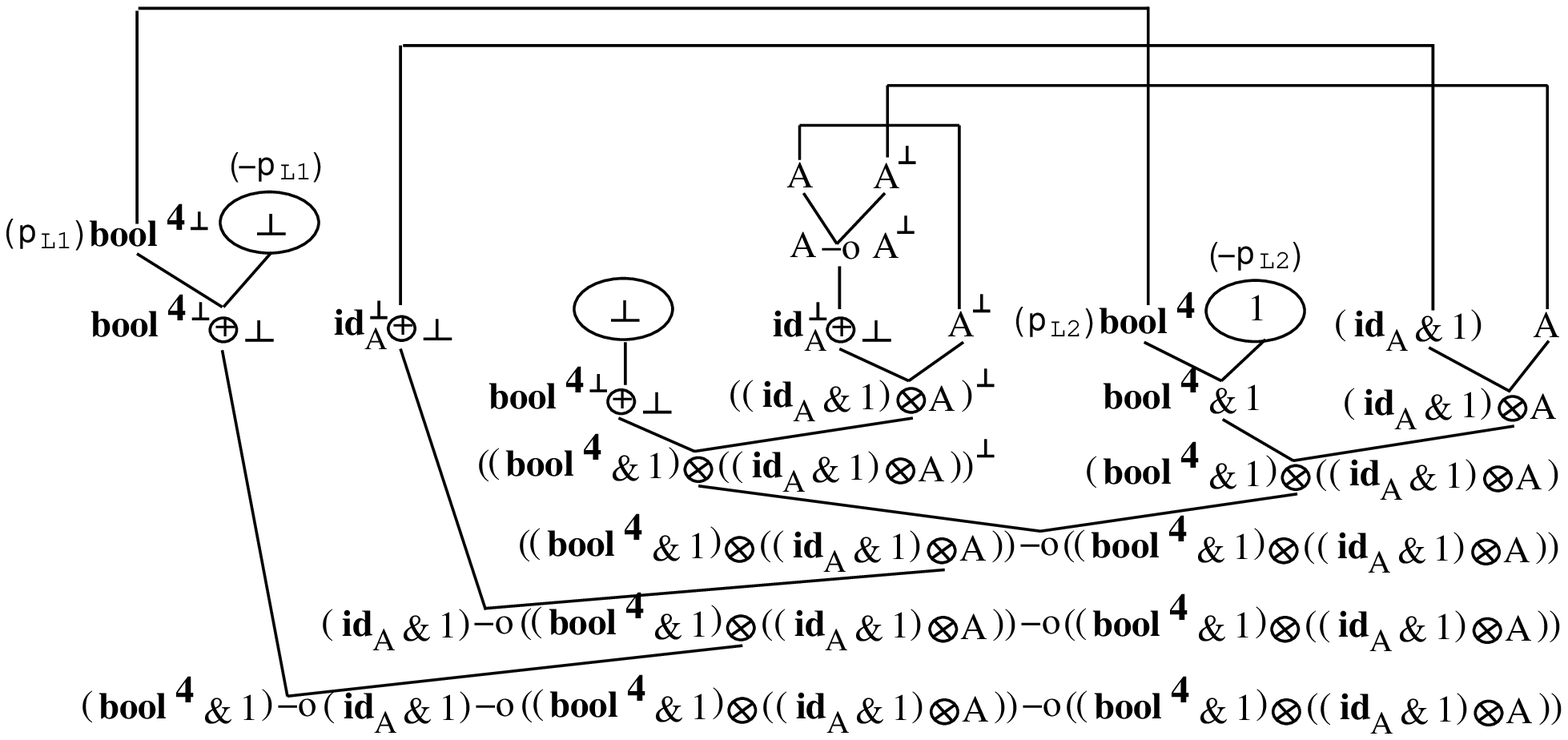}
\caption[{\tt step} function]{{\tt step} function}
\label{step}
\end{center}
\end{figure}

\begin{figure}[htbp]
\begin{center}
\includegraphics[scale=0.5]{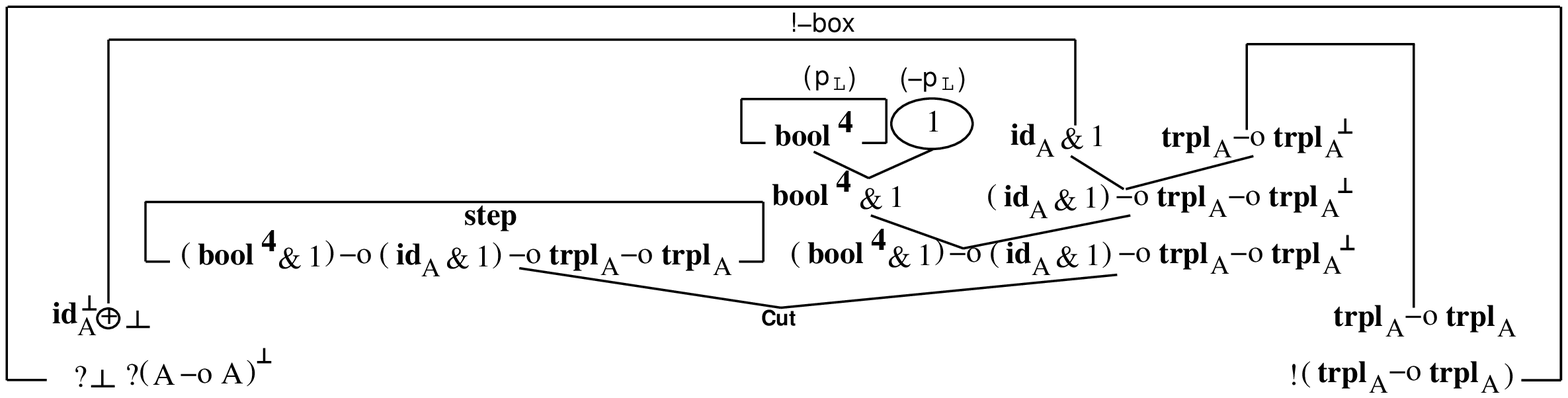}
\caption[{\tt apply\_step}]{{\tt apply\_step}}
\label{applystep}
\end{center}
\end{figure}

The proof net {\tt main}\$ in the $\$$-box 
in Figure~\ref{transition} is shown in Figure~\ref{mainneutral}.

\begin{figure}[htbp]
\begin{center}
\includegraphics[scale=0.5]{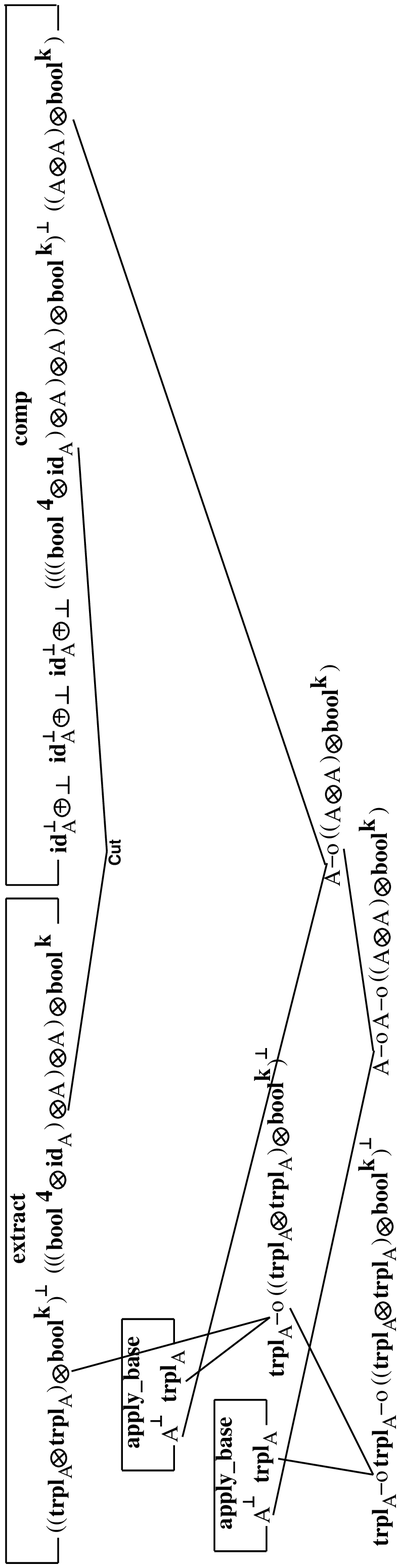}
\caption[{\tt main}\$]{{\tt main}\$}
\label{mainneutral}
\end{center}
\end{figure}

Proof net {\tt apply\_base}
in Figure~\ref{mainneutral} 
are made up by giving a ${\bf bool^4}$ proof {\tt empty} to 
base of Figure~\ref{baseandapplybase},
where 
as we mentioned before, 
${\tt empty}^{\bf bool^4}$ is a normal proof net of ${\bf bool^4}$, which is different from
$0^{\bf bool^4}$, $1^{\bf bool^4}$, and $\ast^{\bf bool^4}$.
The proof net {\tt apply\_base} is used in order to feed an initial value to {\tt apply\_step}($\Theta^{\bf bool^4}$).

\begin{figure}[htbp]
\begin{center}
\includegraphics[scale=0.5]{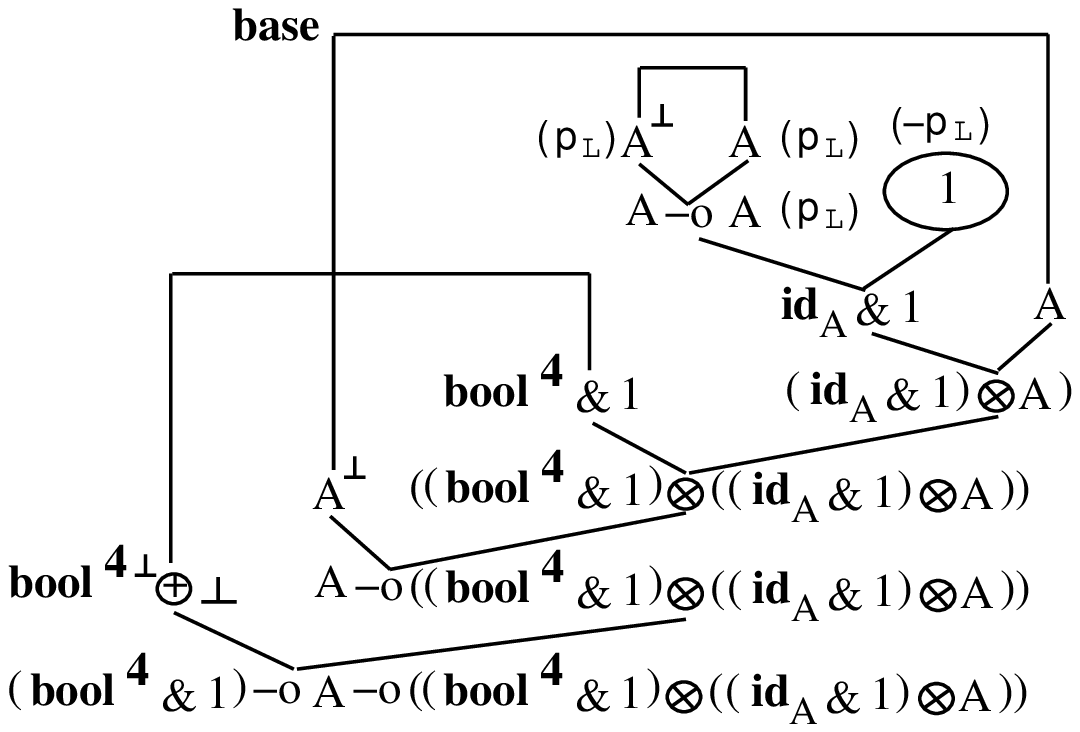}
\quad
\includegraphics[scale=0.5]{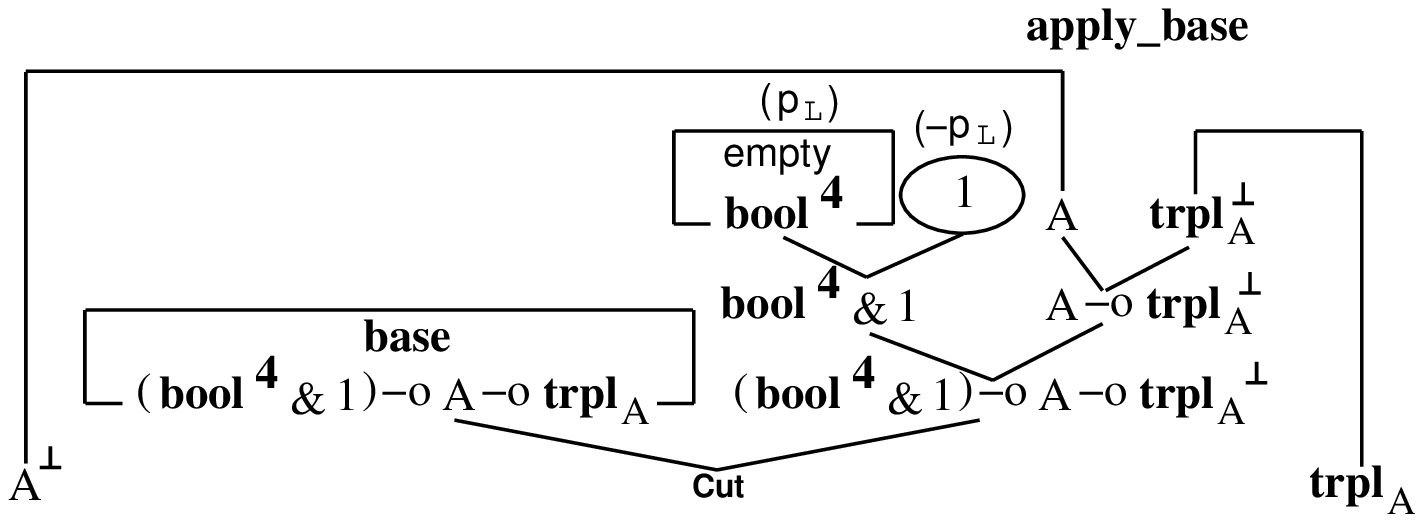}
\caption[{\tt base} function and {\tt apply\_base}]{{\tt base} function and {\tt apply\_base}}
\label{baseandapplybase}
\end{center}
\end{figure}

Proof net {\tt extract} 
in Figure~\ref{mainneutral} 
is shown in Figure~\ref{extract}.
The intention of {\tt extract} was 
to transform an input of the net 
$\langle \langle b_1, \langle f_1, a_1 \rangle \rangle, 
\langle \langle b_2, \langle f_2, a_2 \rangle \rangle,
bk
\rangle
\rangle$ with type 
$({\bf bool^4} \WITH 1) \TENS (({\bf id}_A \WITH 1) \TENS A) \TENS
(({\bf bool^4} \WITH 1) \TENS (({\bf id}_A \WITH 1) \TENS A) \TENS {\bf bool^k})$
into 
$\langle \langle \langle \langle b_2, f_1 \rangle, a_1 \rangle, a_2 \rangle, bk \rangle$
with 
type
$((({\bf bool^4} \TENS {\bf id}_A) \TENS A) \TENS A) \TENS {\bf bool^k}$.
The top symbol  of the left part of the current tape must be left with type ${\bf id}_A$
since this is used in order to be attached to the left or right part of the tape of the next configuration.
The top symbol  of the right part of the current tape, at which the head of $M$ currently points,
must be left with type ${\bf bool^4}$ 
since this is used in order to choose one of select functions (which are defined later).
Note that to do this one must use additive connectives and multiplicative constants.

\begin{figure}[htbp]
\begin{center}
\includegraphics[scale=0.5]{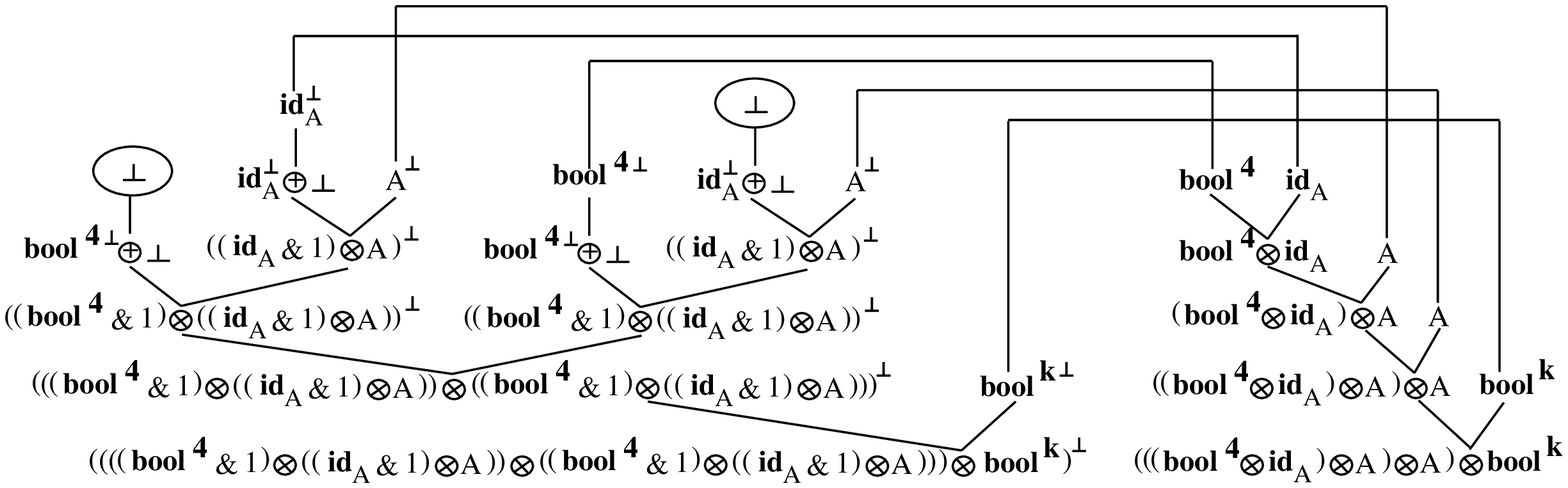}
\caption[{\tt extract} function]{{\tt extract} function}
\label{extract}
\end{center}
\end{figure}

Proof net {\tt comp}
in Figure~\ref{mainneutral} 
is shown in Figure~\ref{comp}, where 
\[{\bf shift} \equiv_{\mathdef} \forall X. 
(X \LIMP X)  \LIMP X \LIMP X \LIMP ((X \TENS X) \TENS {\bf bool^k})\]
and 
\[{\bf row} \equiv_{\mathdef}
{\bf shift} \WITH ({\bf shift} \WITH ({\bf shift} \WITH {\bf shift})). \]
The main purpose of {\tt extract} is 
to transform an input of the net 
$\langle \langle \langle \langle b, f \rangle, a_1 \rangle, a_2 \rangle, bk \rangle$
with 
type
$((({\bf bool^4} \TENS {\bf id}_A) \TENS A) \TENS A) \TENS {\bf bool^k}$
into
the next configuration
$\langle \langle a_3, a_4 \rangle, bk' \rangle$ 
with 
type
$ (A \TENS A) \TENS {\bf bool^k}$.
Data $b$ with type ${\bf bool^4}$ and $bk$ with type ${\bf bool^k}$ in the input 
are used in order to choose one of select functions (which are defined later).

\begin{figure}[htbp]
\begin{center}
\includegraphics[scale=0.5]{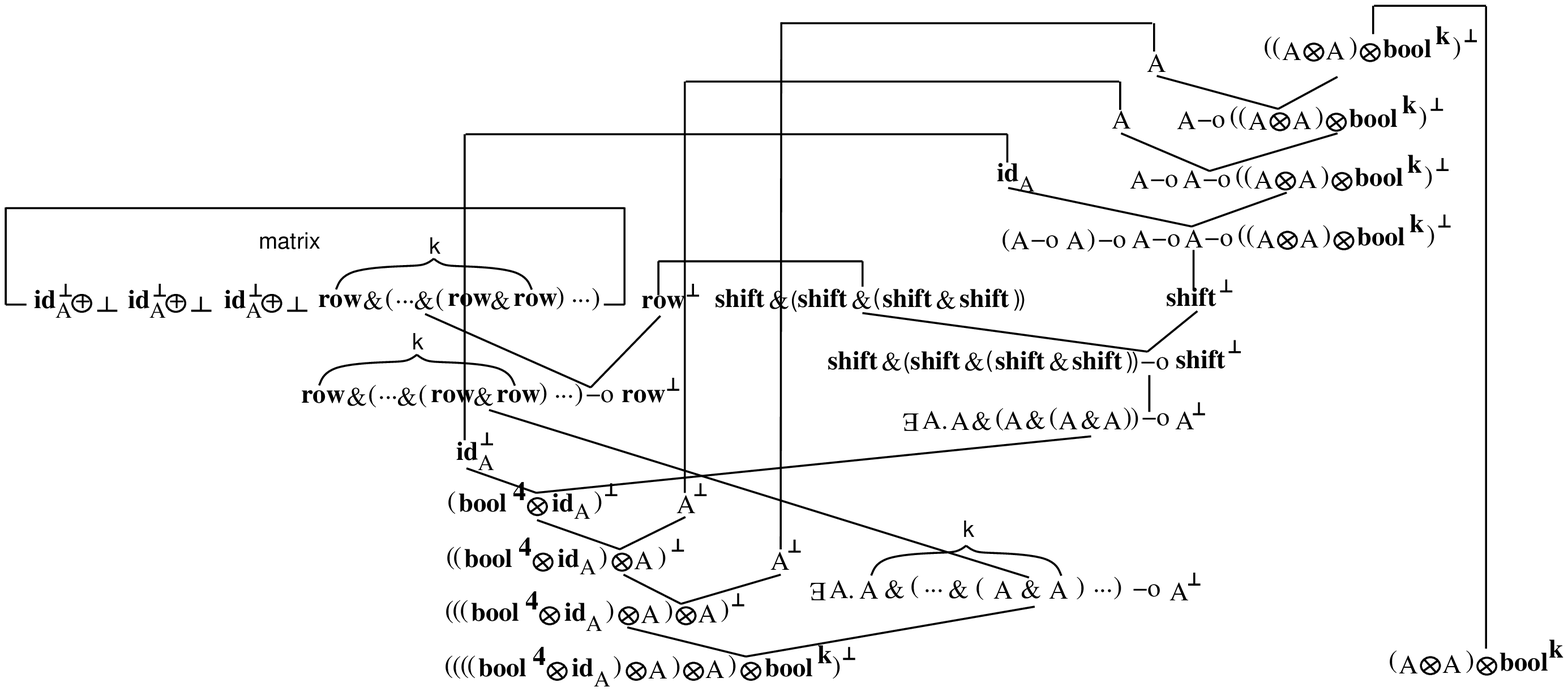}
\caption[{\tt comp}]{{\tt comp}}
\label{comp}
\end{center}
\end{figure}

Proof net {\tt matrix}
in Figure~\ref{comp} 
is shown in Figure~\ref{matrix}, 
where $r_1, \ldots, r_{k-1}, r_k$ are 
proof nets that have the form of Figure~\ref{row}.
The main purpose of {\tt matrix} is to retain $k$ proof nets of the form of Figure~\ref{row}.
Proof nets $s_1,s_2,s_3,$ and $s_4$ in Figure~\ref{row} have the form of 
Figure~\ref{shiftleft} or Figure~\ref{shiftright}.
We call such proof nets {\it shift functions}.
Proof nets that have the form of Figure~\ref{shiftleft} 
represent left moves of the head of $M$.
On the other hand 
proof nets that have the form of Figure~\ref{shiftright} 
represent right moves of the head of $M$.

\begin{figure}[htbp]
\begin{center}
\includegraphics[scale=0.5]{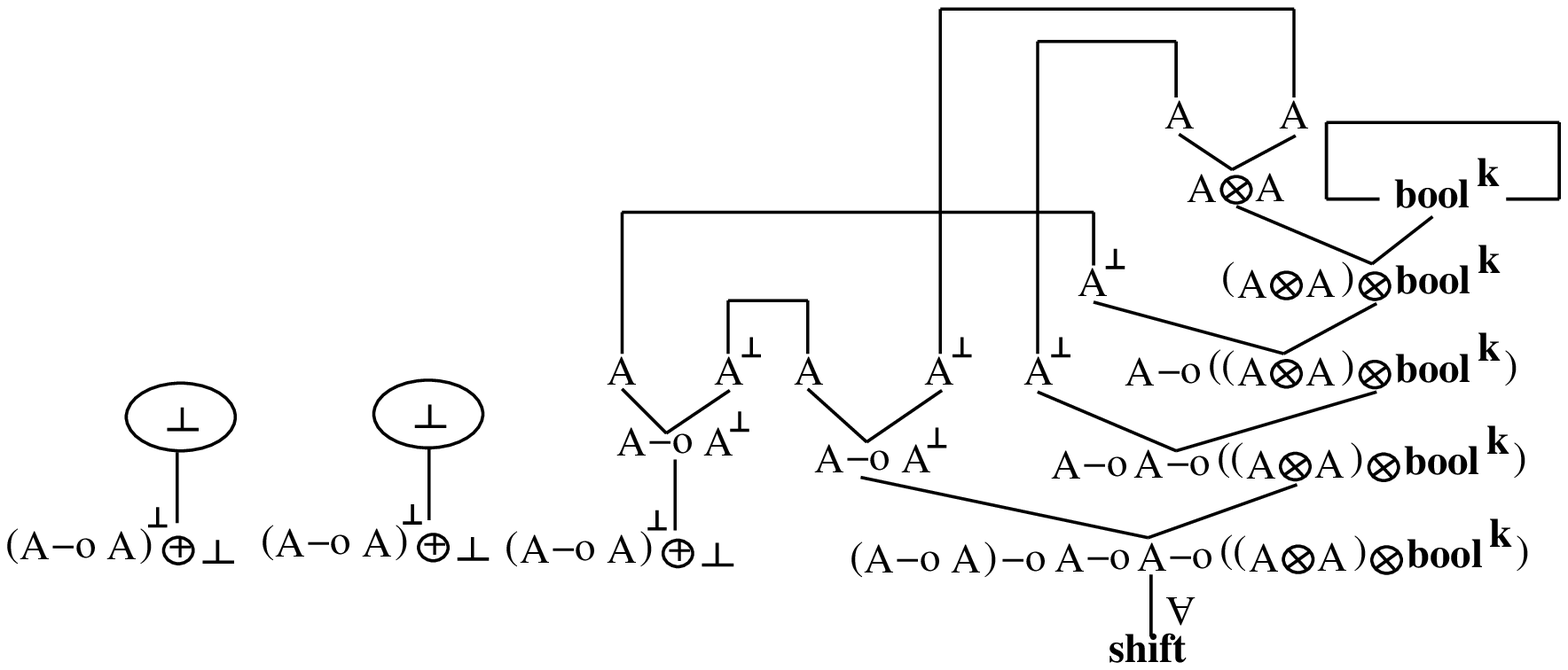}
\caption[an example of shft functions (left move)]{an example of ${\bf shft}$ proofs (left move)}
\label{shiftleft}
\end{center}
\end{figure}

\begin{figure}[htbp]
\begin{center}
\includegraphics[scale=0.5]{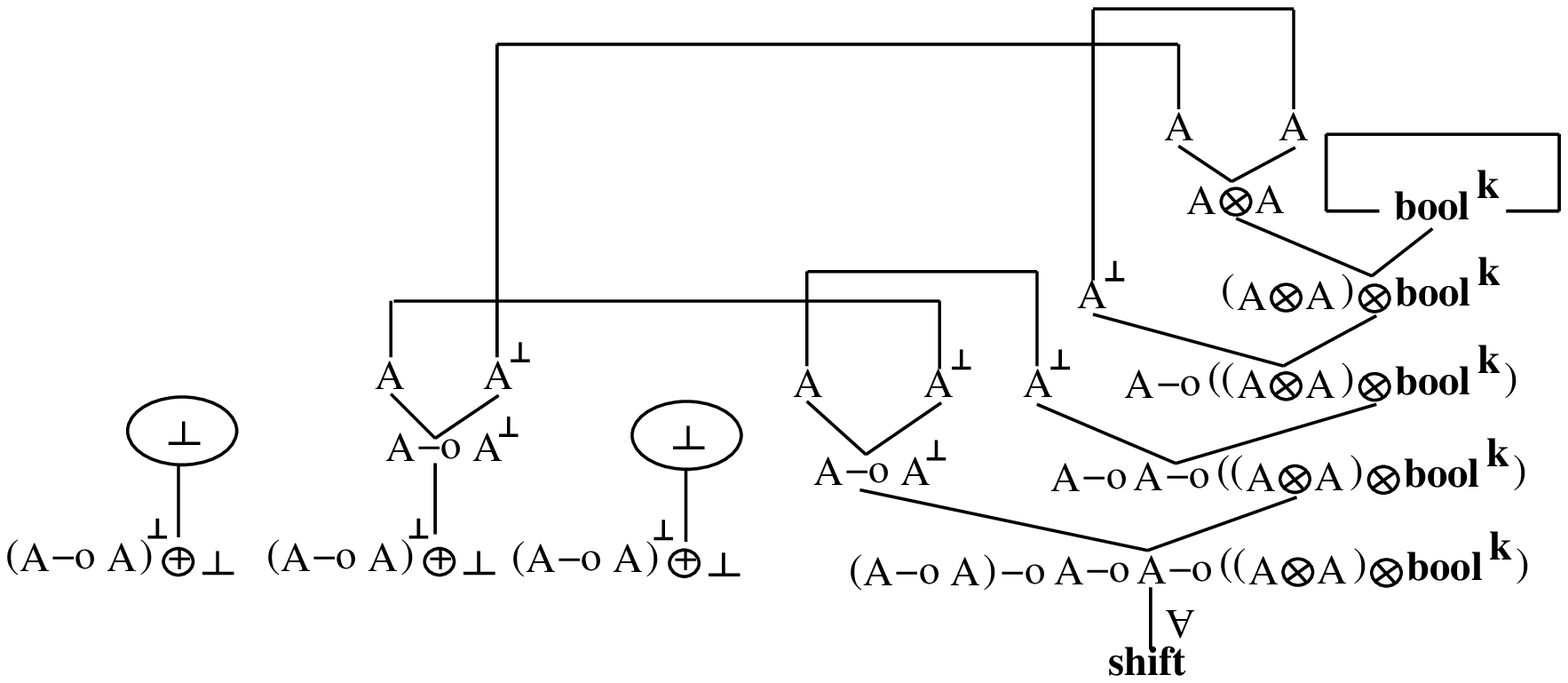}
\caption[an example of shft functions (right move)]{an example of ${\bf shft}$ proofs (right move)}
\label{shiftright}
\end{center}
\end{figure}


\begin{figure}[htbp]
\begin{center}
\includegraphics[scale=0.5]{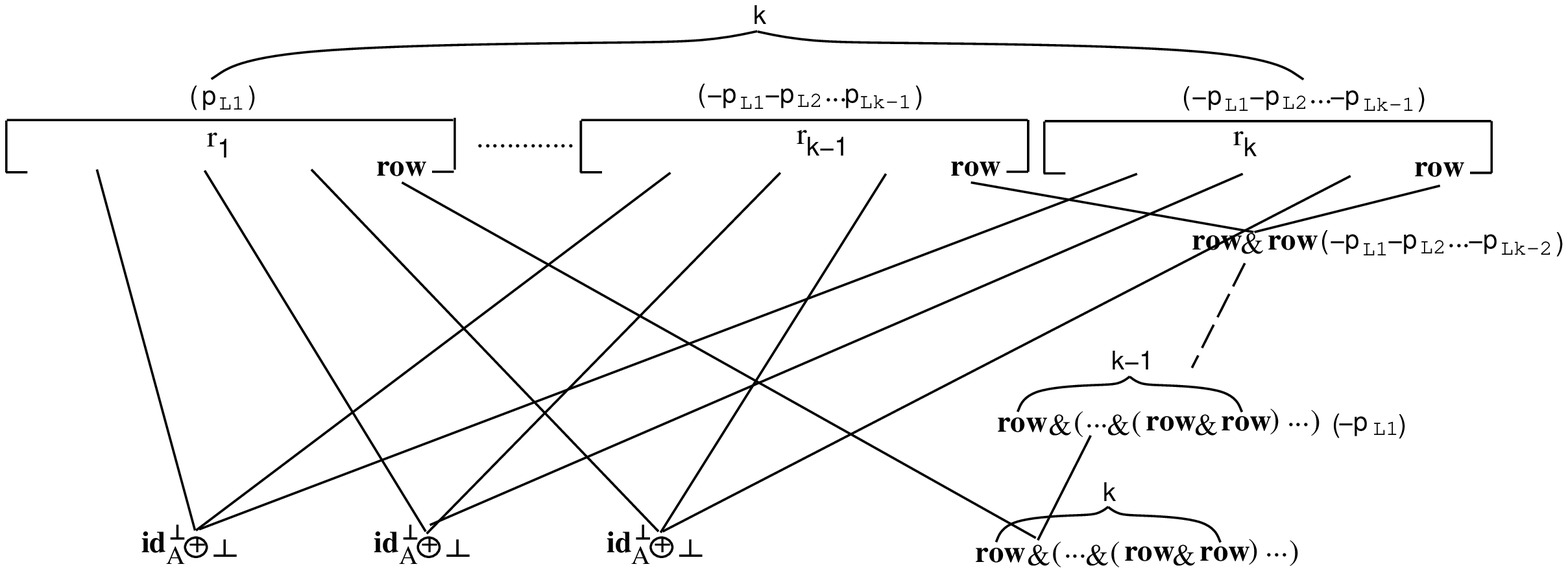}
\caption[{\tt matrix}]{{\tt matrix}}
\label{matrix}
\end{center}
\end{figure}


\begin{figure}[htbp]
\begin{center}
\includegraphics[scale=0.5]{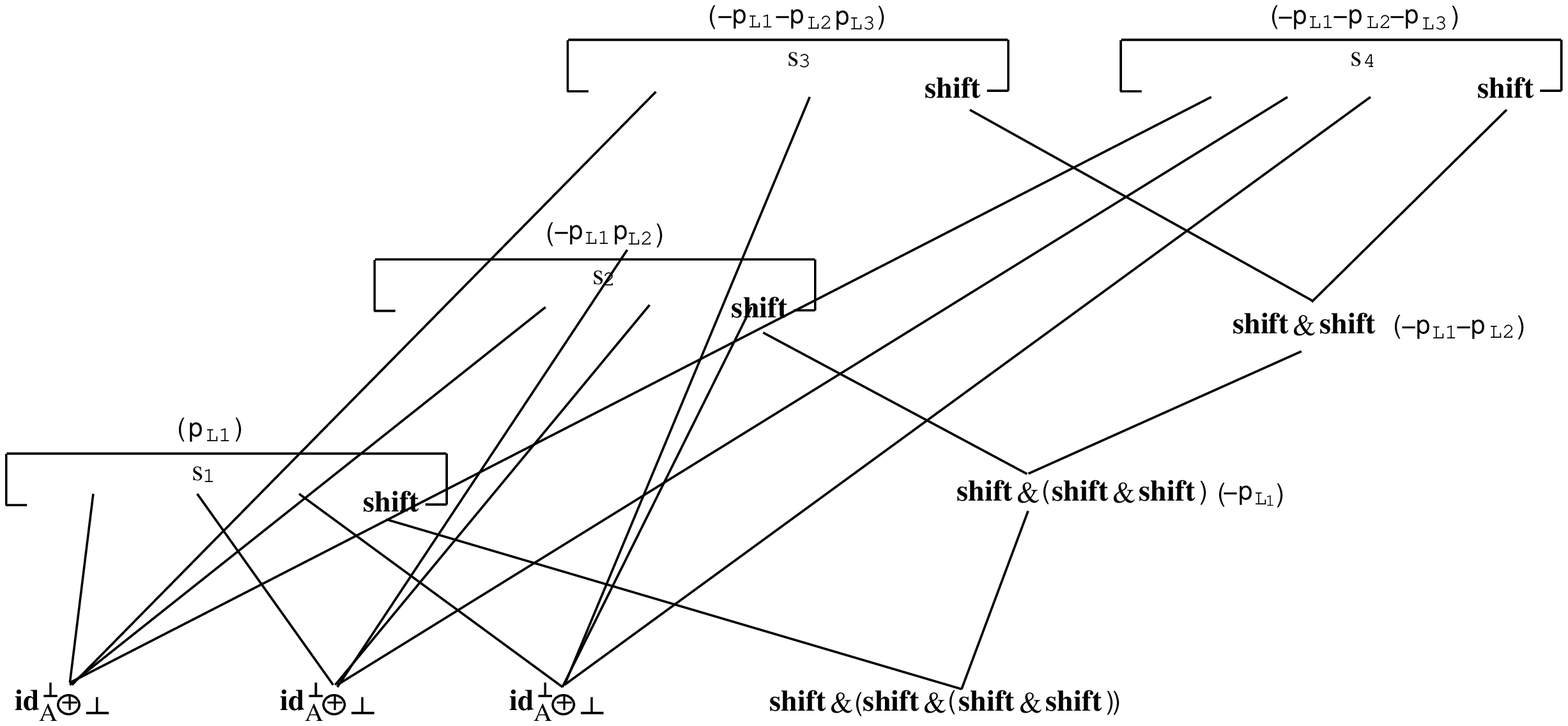}
\caption[{\tt row\_net}]{{\tt row\_net}}
\label{row}
\end{center}
\end{figure}

From what precedes it is obvious that we can encode the transition function of $M$
into a proof net with conclusions $? \bot, {\bf config}^\bot, {\bf config}$ of Light Linear Logic.
By using the proof net, as shown in Appendix~\ref{appendtm}, 
{\bf P}-time Turing machines can be encoded.
In other terms, we obtain the following theorem:
\begin{theorem}
\label{maintheorem}
Let ${\bf bint}$ be $\forall X. !(X \LIMP X) \LIMP !(X \LIMP X) \LIMP \$(X \LIMP X)$.
Let $M$ be a Turing machine with time bound of a polynomial with degree $k$.
In Light Linear Logic $M$ can be represented by a proof net with conclusions $\bot^{k+4}, {\bf bint}^\bot, \$^{k+3} {\bf config}$.
\end{theorem}

Furthermore, we can strengthen the above theorem as follows:
\begin{theorem}
\label{maintheorem2}
Let $f$ be a polynomial-time function with degree $k$.
In Light Linear Logic $f$ can be represented by a proof net with conclusions $\bot^{k+6}, {\bf bint}^\bot, \$^{k+5} {\bf bint}$.
\end{theorem}

In order to prove the theorem, we need a proof net 
that transforms ${\bf config}$ into ${\bf bint}$. 
In Appendix~\ref{appendcfgtobint}, we show a proof net that performs the translation.

\section{Our Nondeterministic Extension of the Light Linear Logic System}
In this section we consider a nondeterministic extension of the LLL system called the {\it NDLLL} system.
In this extended system we introduce a new self-dual additive connective ``nondeterministic with'' $\NDWITH$.
Then the formulas of NDLLL are constructed by adding the following clause to that of LLL:
\[ F = ... | F \NDWITH F \]
Since $\NDWITH$ is a self-dual connective, the negation of the formula $A \NDWITH B$ is defined as follows:
	\begin{itemize}
	\item $(A \NDWITH B)^\bot  \equiv_{\mathdef} A^\bot \NDWITH B^\bot$
	\end{itemize}
The link newly introduced in NDLLL is the form of Figure~\ref{nddef}. 
Proof nets for the NDLLL system are inductively defined from the rules of Figure~\ref{simple} 
and in the middle of Figure~\ref{nddef}. 
In a simple proof net for NDLLL a unique eigenweight is assigned to each 
$\NDWITH$-link occurrence in the same manner as that of $\WITH$-link. \\
Finally the rewrite rules of NDLLL are that of LLL plus the nondeterministic rewrite rule of Figure~\ref{nddef}.
In the rewrite rule for $\NDWITH$ any of the two contractums is nondeterministically selected.
If the left contractum (resp. the right contractum) of Figure~\ref{nddef} is selected, then 
all the occurrences of both eigenweights for $A \NDWITH B$ and $A^\bot \NDWITH B^\bot$ 
are assigned to 1 (resp. 0). 
In the next section we explain a usage for the $\NDWITH$. 
\begin{figure}[htbp]
\begin{center}
\includegraphics[scale=.5]{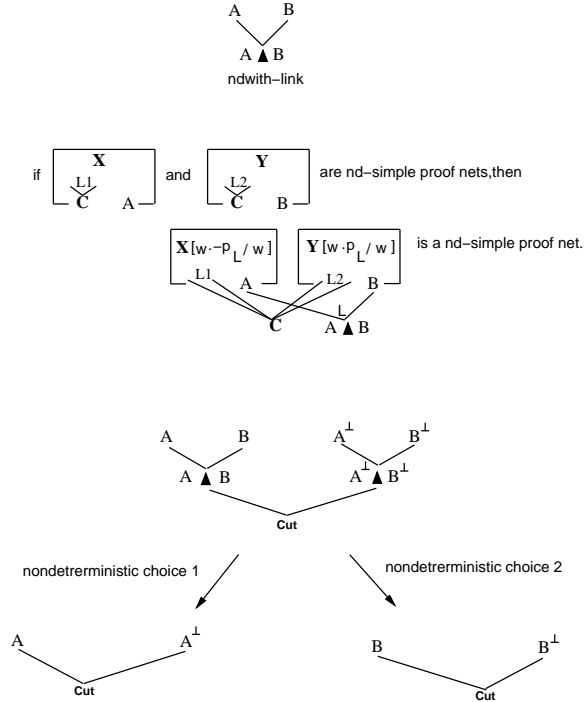}
\caption[Our nondeterministic extension of the LLL system]{Our nondeterministic extension of the LLL system}  
\label{nddef}
\end{center}
\end{figure}

\subsection{A Nondeterministic Turing Machine Encoding}
Our usage of $\NDWITH$-connective is to use $\NDWITH$ in proof nets on datatypes like 
${\bf bool} \equiv_{\mathdef} \forall X. X \WITH X \LIMP X$ which use the standard additive connectives 
$\WITH$ and $\PLUS$. 
As an example, we consider cut-elimination of Figure~\ref{ndwith-ex1}, where
note that the sub-proof net with conclusion ${\bf bool} \LIMP {\bf bool}$ of the right premise of Cut 
is constructed by using $\NDWITH$. 
From Figure~\ref{ndwith-ex1} to Figure~\ref{ndwith-ex5} the standard lazy cut elimination procedure
is performed. 
In Figure~\ref{ndwith-ex5} the nondeterministic cut elimination procedure defined in previous section 
is performed.
Figure~\ref{ndwith-ex6-a} is one choice and Figure~\ref{ndwith-ex6-b} the other choice.

\begin{figure}[htbp]
\begin{center}
\includegraphics[scale=.5]{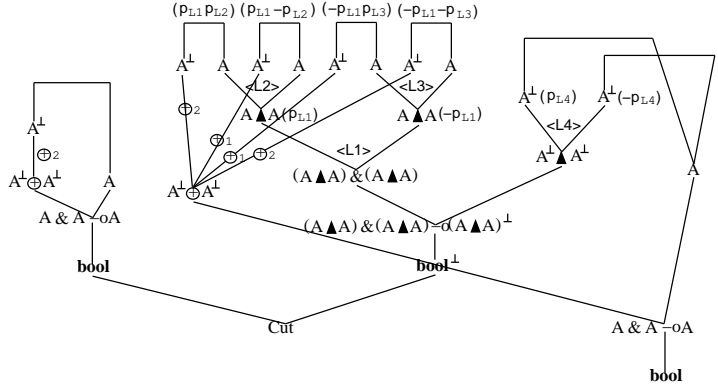}
\caption[An example: the starting point]{An example: the starting point}  
\label{ndwith-ex1}
\end{center}
\end{figure}

\begin{figure}[htbp]
\begin{center}
\includegraphics[scale=.5]{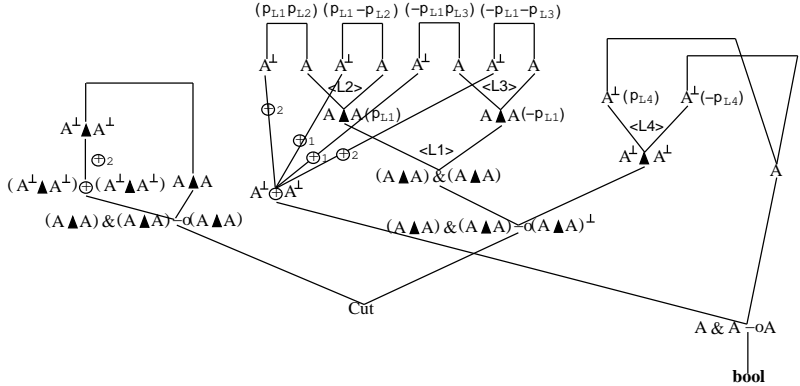}
\caption[An example: step 1]{An example: step 1}  
\label{ndwith-ex2}
\end{center}
\end{figure}

\begin{figure}[htbp]
\begin{center}
\includegraphics[scale=.5]{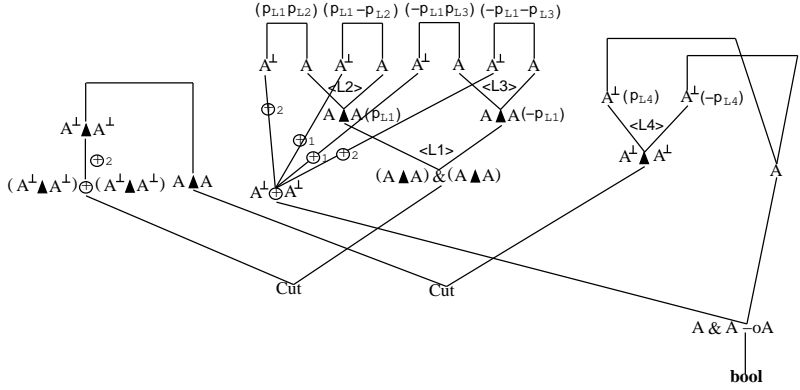}
\caption[An example: step 2]{An example: step 2}  
\label{ndwith-ex3}
\end{center}
\end{figure}

\begin{figure}[htbp]
\begin{center}
\includegraphics[scale=.5]{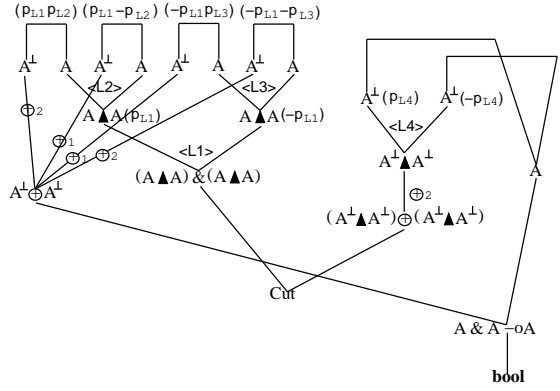}
\caption[An example: step 3]{An example: step 3}  
\label{ndwith-ex4}
\end{center}
\end{figure}

\begin{figure}[htbp]
\begin{center}
\includegraphics[scale=.5]{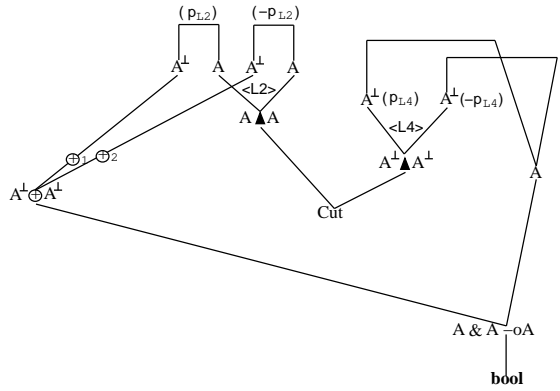}
\caption[An example: step 3]{An example: step 4}  
\label{ndwith-ex5}
\end{center}
\end{figure}

\begin{figure}[htbp]
\begin{center}
\includegraphics[scale=.5]{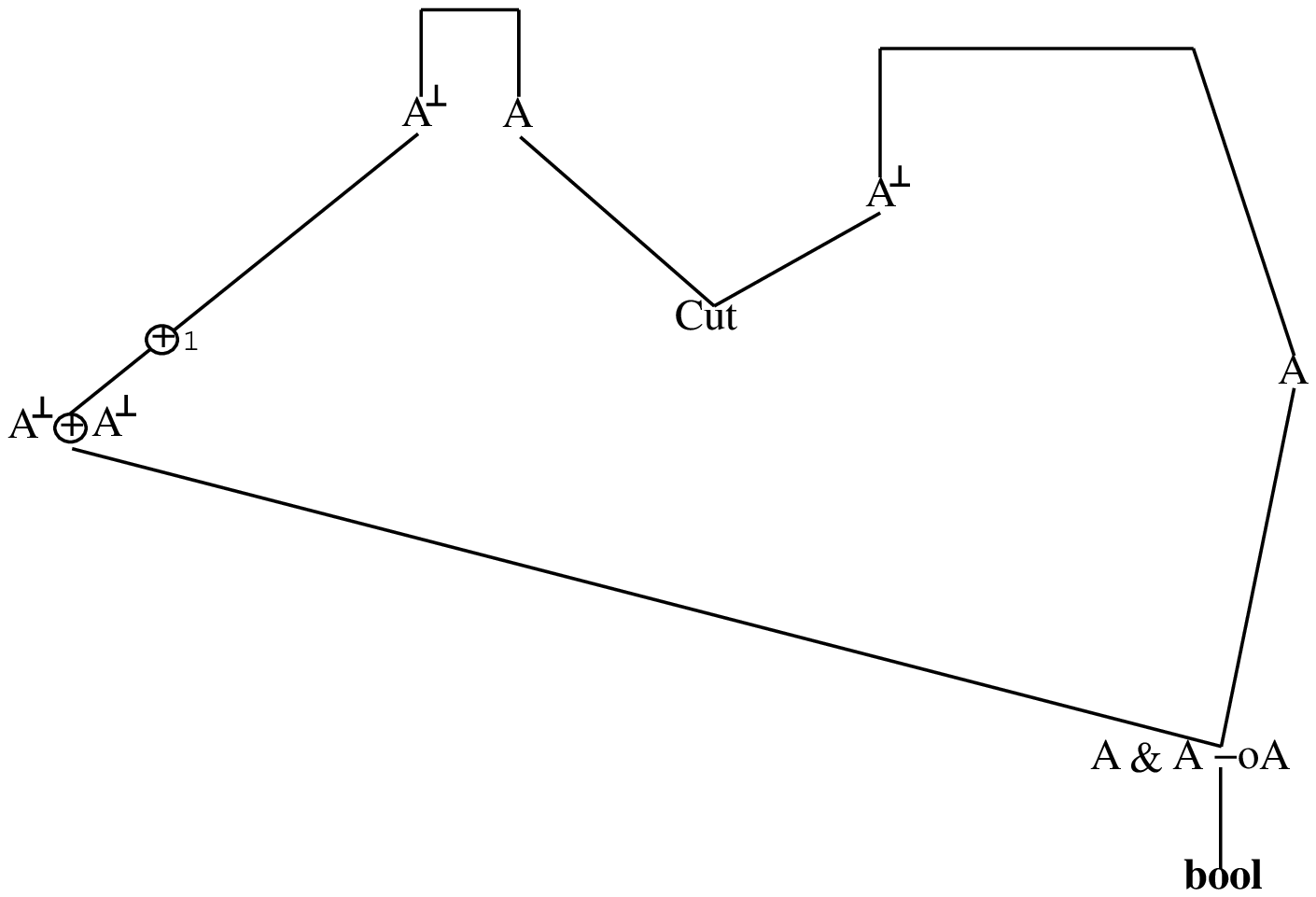}
\caption[An example: nondeterministic choice 1]{An example: nondeterministic choice 1}  
\label{ndwith-ex6-a}
\end{center}
\end{figure}

\begin{figure}[htbp]
\begin{center}
\includegraphics[scale=.5]{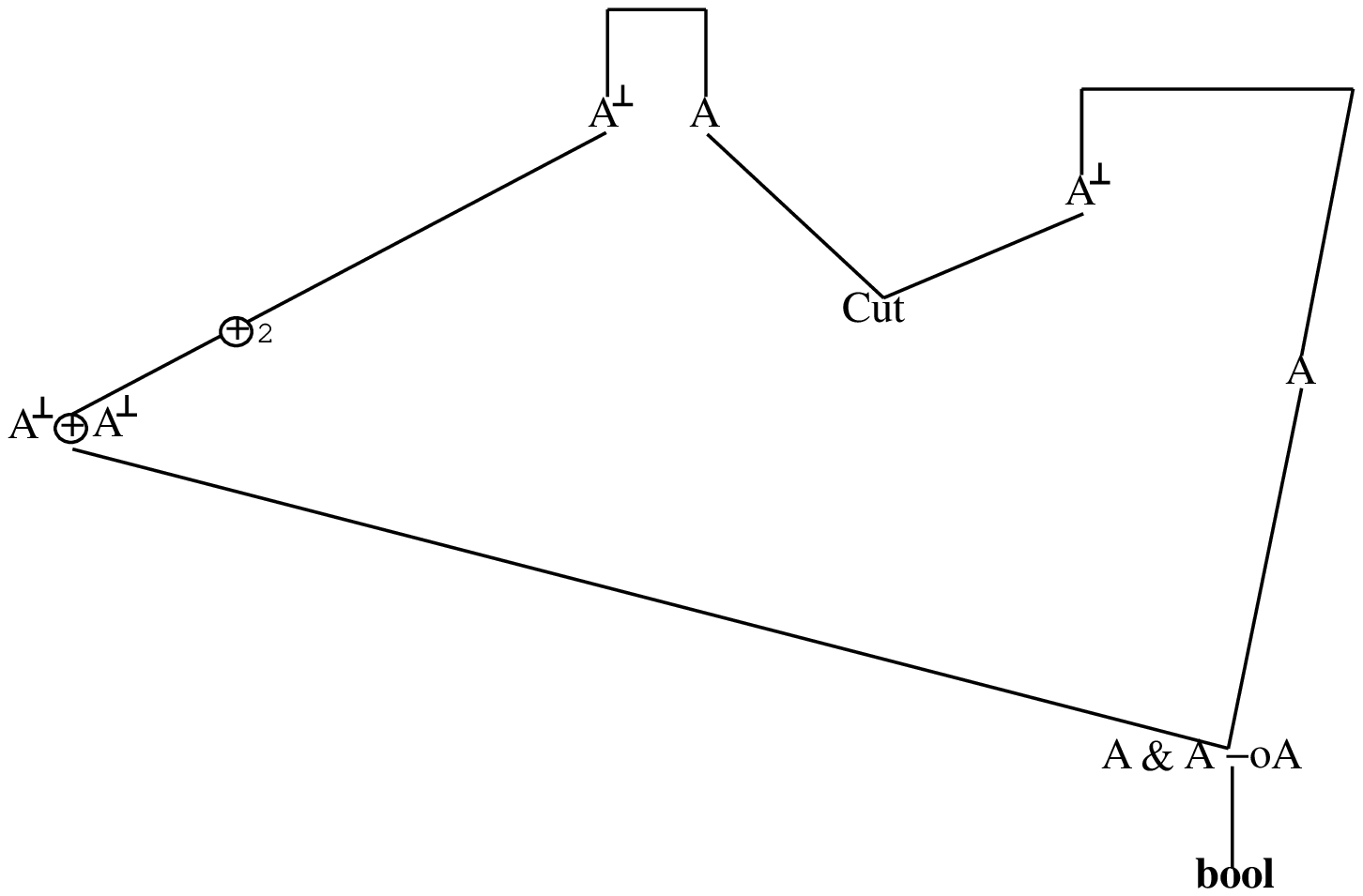}
\caption[An example: nondeterministic choice 2]{An example: nondeterministic choice 2}  
\label{ndwith-ex6-b}
\end{center}
\end{figure}

An encoding of a nondeterministic Turing machine into NDLLL uses the same idea. 
The encoding is the same as that of a deterministic Turing machine into LLL except for
{\tt comp} of Figure~\ref{comp}. 
The {\tt comp} proof net is replaced by the {\tt nd-comp} of Figure~\ref{comp-ndlll}, 
where 
\[{\bf ndrow} \equiv_{\mathdef}
({\bf shift} \NDWITH {\bf shift}) \WITH (({\bf shift} \NDWITH {\bf shift}) 
\WITH (({\bf shift} \NDWITH {\bf shift}) \WITH ({\bf shift} \NDWITH {\bf shift}))). \]
The idea is completely the same as that of the above example.
The information about nondeterministic transitions of a nondeterministic Turing machine 
is stored in a proof of
\[ \overbrace{{\bf ndrow} \WITH (\cdots \WITH ({\bf ndrow} \WITH {\bf ndrow} }^{k} ) \cdots ). \]

\begin{figure}[htbp]
\begin{center}
\includegraphics[scale=.5]{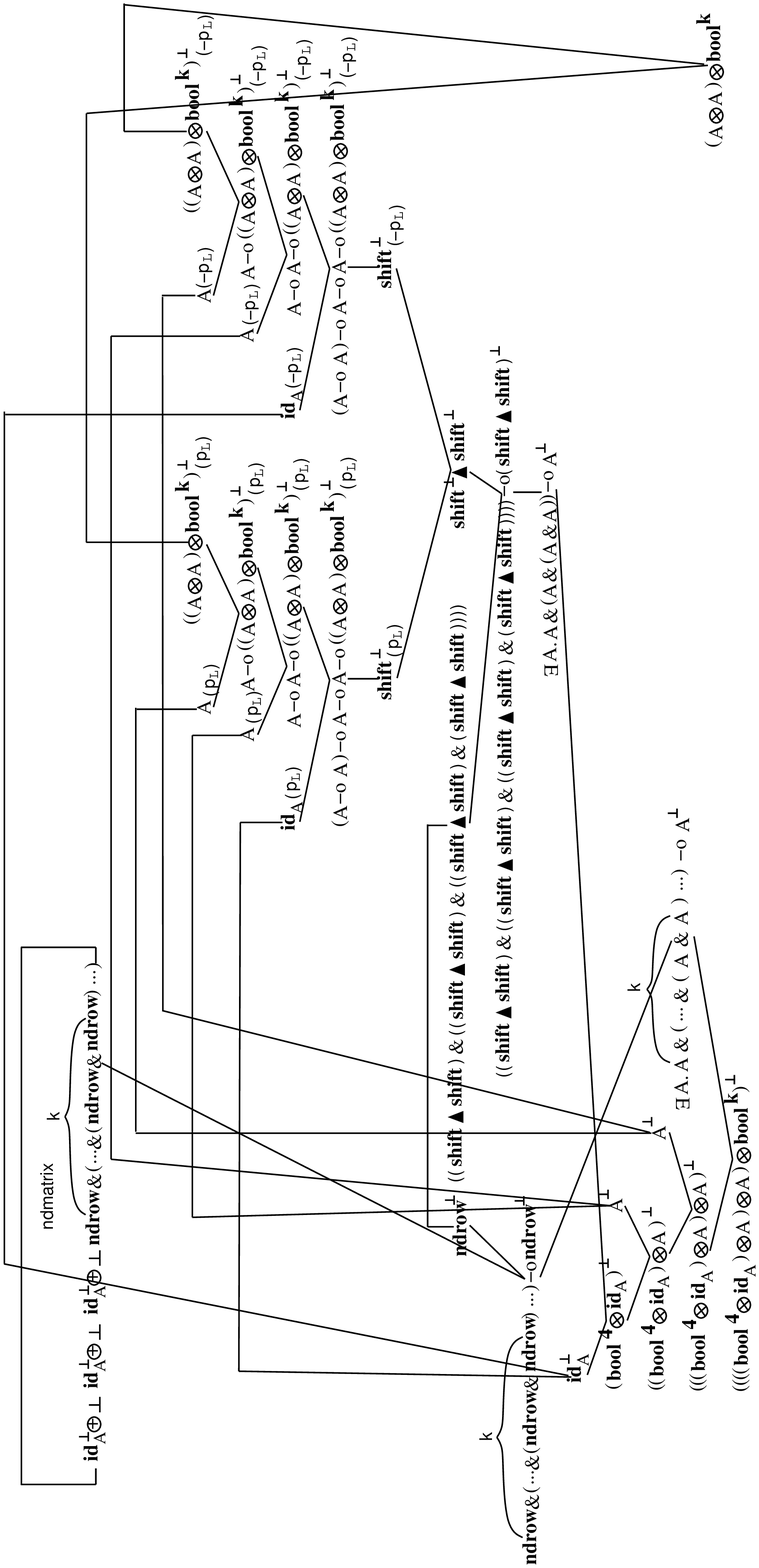}
\caption[{\tt nd-comp}]{{\tt nd-comp}}  
\label{comp-ndlll}
\end{center}
\end{figure}

By the completely same manner as Theorem~\ref{maintheorem} except for {\tt nd-comp} 
the following theorem holds.

\begin{theorem}
\label{maintheoremndlll}
Let $M$ be a nondeterministic Turing machine with time bound of a polynomial with degree $k$.
In Nondeterministic Light Linear Logic $M$ can be represented by a proof net with conclusions $\bot^{k+4}, {\bf bint}^\bot, \$^{k+3} {\bf config}$.
\end{theorem}

Usually a nondeterministic Turing machine characterizes a language accepted by the machine. 
Without loss of generality, we can assume that nondeterministic Turing machine has 
two special state symbols ${\tt yes}$ and ${\tt no}$ which judge whether a word is accepted by the machine.
Moreover we can prove the following theorem.
\begin{theorem}
\label{maintheorem2ndlll}
Let $L \subseteq \{ 0, 1 \}^\ast$ be a language whose is accepted by a nondeterministic polynomial-time 
Turing machine $M$ with degree $k$. 
In Nondeterministic Light Linear Logic the characterization function of $L$ from $\{ 0 , 1 \}$ 
to  $\{ 0, 1 \}$ can be represented by a proof net with conclusions $\bot^{k+5}, {\bf bint}^\bot, \$^{k+4} {\bf bool}$.
\end{theorem}

In order to obtain such a proof net from the proof net with 
conclusions $\bot^{k+4}, {\bf bint}^\bot, \$^{k+3} {\bf config}$ constructed from 
Theorem~\ref{maintheoremndlll}, 
at first we construct a proof net which extracts a ${\bf bool^k}$ proof from a ${\bf config}$ proof.
Figure~\ref{config2boolk} shows the proof net.
Next we construct a proof net which maps a ${\bf bool^k}$ proof to a ${\bf bool}$ proof.
The specification of proof net is that 
\begin{enumerate}
	\item if a given ${\bf bool^k}$ proof net represents ${\tt yes}$, then 
	the return value is a ${\bf bool}$ proof net that represents ${\tt yes}$;
	\item otherwise, 
	the return value is a ${\bf bool}$ proof net that represents ${\tt no}$.
\end{enumerate}
We can easily construct such a proof net.

\begin{figure}[htbp]
\begin{center}
\includegraphics[scale=.5]{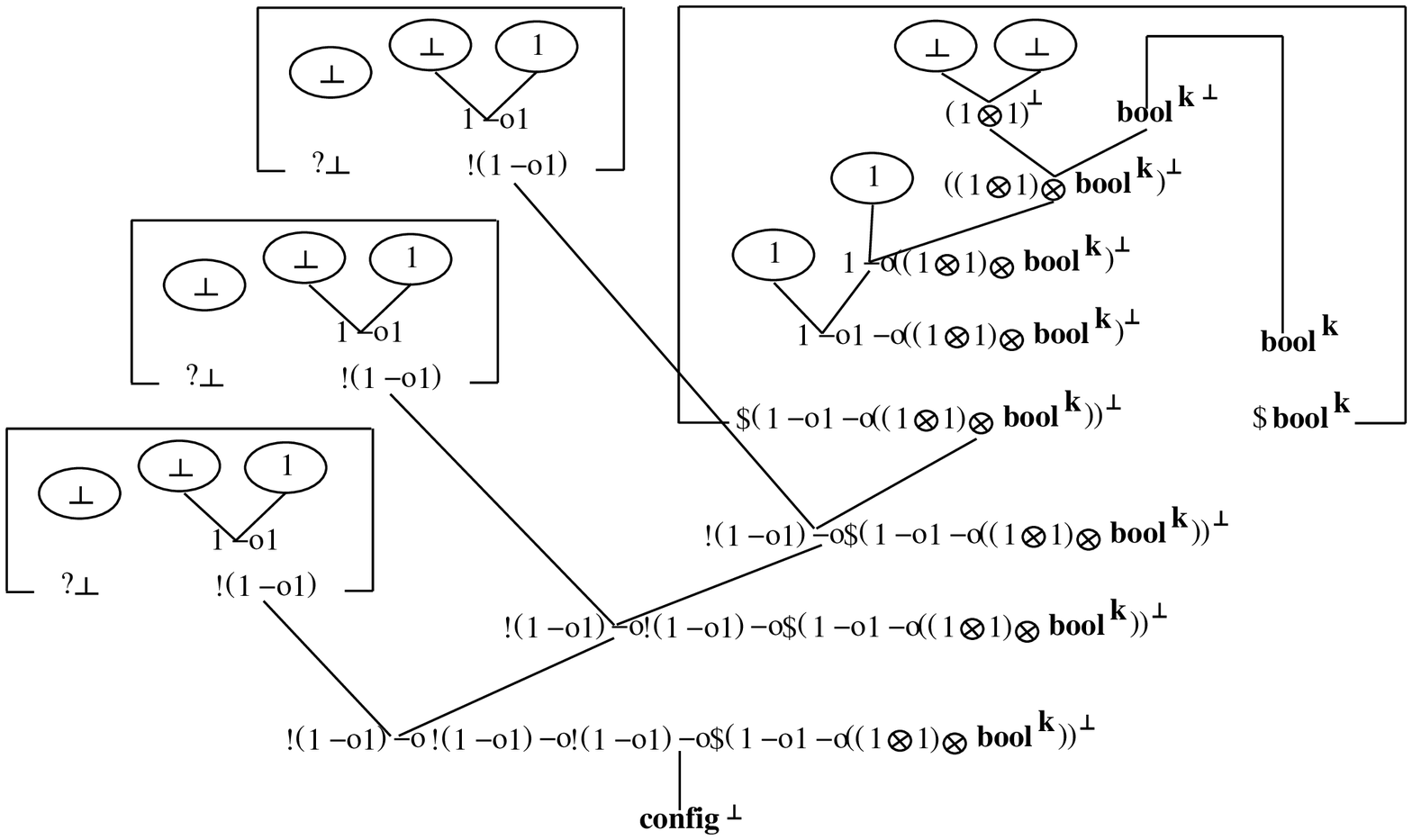}
\caption[{\tt config2boolk}]{{\tt config2boolk}}  
\label{config2boolk}
\end{center}
\end{figure}


\subsection{Time Bound of Nondeterministic Light Linear Logic}
Next we discuss the {\bf P}-time bound of lazy cut elimination.
We define the {\it size} of a link  to be the number of the conclusions of the link.
Moreover we define the {\it size} of a nd-simple proof net $\Theta$ (denoted by $\size(\Theta)$ to be
the sum of the sizes of the links in $\Theta$.
The {\it depth} of $\Theta$ (denoted by $\depth(\Theta)$) is defined to be the maxmal nesting number of the boxes ($!$-boxes or $\$$-boxes) 
in $\Theta$. 
As discussed in \cite{Gir98}, $\size(\Theta)$  is quadratic w.r.t the encoded data generated from $\Theta$.
When $\Theta_1 \to_{\lazy} \Theta_2$ by nondeterministic choice, 
it is obvious that the size of $\Theta_2$ is strictly less than that of $\Theta_1$.
 We suppose that  $\Theta \to_{\lazy}^\ast \Theta'$,
where $\to_{\lazy}^{\ast}$ is the reflexive transitive closure of $\to_{\lazy}$.
From the above observation and the discussion in \cite{Gir98} on the LLL system, 
the following proposition is obvious.
\begin{proposition}
In the NDLLL system, if $\Theta \to_{\lazy}^\ast \Theta'$, then
the size of $\Theta'$ is bounded by $\size(\Theta)^{2^{\depth(\Theta)}}$.
\end{proposition}
It is easy to see that on the above proposition we can lazily reduce $\Theta$ to $\Theta'$ 
in a polynomial time w.r.t $\size(\Theta)^{2^{\depth(\Theta)}}$, 
because $\size(\Theta)$ is quadratic w.r.t the encoded data generated from $\Theta$.
\begin{proposition}
\label{ndptime}
In the NDLLL system, if $\Theta \to_{\lazy}^\ast \Theta'$, then
$\Theta$ is reduced to $\Theta'$ in a polynomial time w.r.t $\size(\Theta)^{2^{\depth(\Theta)}}$.
\end{proposition}
Let $M$ be 
a nondeterministic polynomial-time 
Turing machine with degree $k$.
Then by Theorem~\ref{maintheoremndlll} we can construct 
a nd-simple proof net $\Theta_1$ with conclusions $\bot^{k+4}, {\bf bint}^\bot, \$^{k+3} {\bf config}$.
Then let $\Theta_2$ be a simple proof net with the conclusion ${\bf bint}$ 
representing a binary integer with length $n$. 
Then from Proposition~\ref{ndptime} we can see the proof net constructed by connecting $\Theta_1$ and 
$\Theta_2$ via Cut-link is lazily and nondeterministically reduced to a normal form 
in a polynomial time w.r.t $n^{2^{k}}$. 
%

\section{Concluding Remarks}
It seems possible that a {\bf P}-time Turing machine encoding in Light Affine Logic 
is mechanically translated into that in Light Linear Logic. 
A given proof of the {\bf P}-time Turing machine encoding in Light Affine Logic,
we replace all the formula occurrences $A$ in the proof  by $A \WITH 1$
and then apply an extract function like Figure~\ref{extract} to the resulting proof.
But we did not adopt the method, since the simple transition makes 
a too redundant proof in Light Linear Logic.
So we made some optimizations. For example,
In \cite{Rov99} $\ALL X. X \TENS X \LIMP X$ was used as the boolean type.
The above mentioned translation makes 
$\ALL X. ((((X \WITH 1) \TENS (X \WITH 1)) \WITH 1) \LIMP (X \WITH 1)) \WITH 1$.
The study to find optimal translations seems interesting.
\begin{ack}
The author thanks Luca Roversi for discussions at his visit to University of Torino.
\end{ack}

\clearpage

\appendix

\section{Our encoding of Turing machines(continued)}
\label{appendtm}
For shorthand, we use $!^k A$ to represent 
$\overbrace{! (\cdots ! (! }^{k} A) \cdots )$.
We also use $?^k A \equiv_{\mathdef} \overbrace{? (\cdots ? (? }^{k} A) \cdots )$ and 
$\$^k A \equiv_{\mathdef} \overbrace{\$ (\cdots \$ (\$ }^{k} A) \cdots )$
In addition, we implicitly assume coercion for $\bot$.
In other terms, we assume that 
by using a proof net with conclusions $?^{k_1} \bot,\ldots,?^{k_p} \bot, \Gamma$
we can construct a proof net with conclusions $?^{k} \bot, \Gamma$ 
provided $k \ge k_1,\ldots,k_p$.
This is done by using $p$ proof nets that have the forms of Figure~\ref{botcoercion}, Cut-links, 
and one contraction-link. \\
\begin{figure}[htbp]
\begin{center}
\includegraphics[scale=0.5]{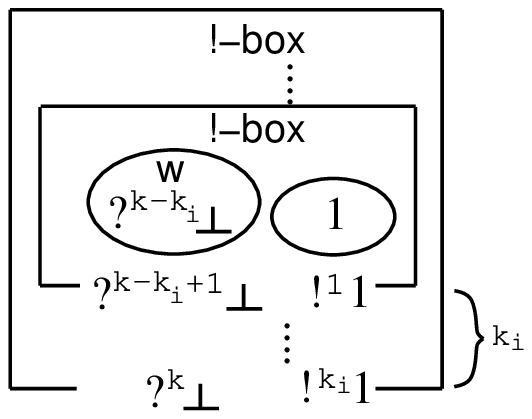}
\caption[coercion of $\bot$]{coercion of $\bot$}
\label{botcoercion}
\end{center}
\end{figure}
Unlike \cite{Gir98}, we do not use ${\bf int} \equiv_{\mathdef} \forall X. !(X \LIMP X) \LIMP \$(X \LIMP X)$
for our Turing machine encoding: 
we only use ${\bf bint} \equiv_{\mathdef} \forall X. !(X \LIMP X) \LIMP !(X \LIMP X) \LIMP \$(X \LIMP X)$ 
instead of ${\bf int}$ since we would like to reduce the number of proof nets
appearing in this paper. It is possible to construct a Turing machine encoding 
from our transition function encoding by using ${\bf int}$.
In the following we show three basic functions for ${\bf bint}$: 
two successor function and addition.
Figure~\ref{suc0} and Figure~\ref{suc1} show two successor functions for ${\bf bint}$:
we call these {\tt suc0} and {\tt suc1} respectively.
Unlike ${\bf int}$, ${\bf bint}$ has two successor functions.\\
\begin{figure}[htbp]
\begin{center}
\includegraphics[scale=0.5]{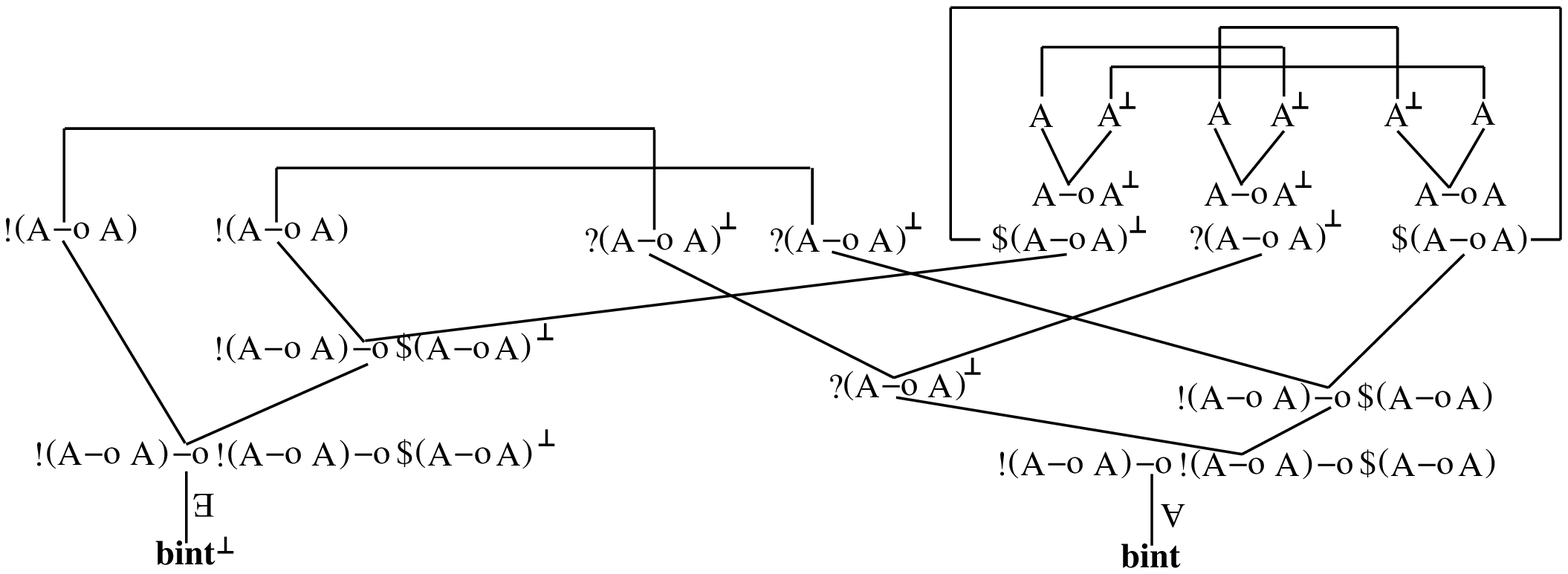}
\caption[{\tt suc0}]{{\tt suc0}}
\label{suc0}
\end{center}
\end{figure}
\begin{figure}[htbp]
\begin{center}
\includegraphics[scale=0.5]{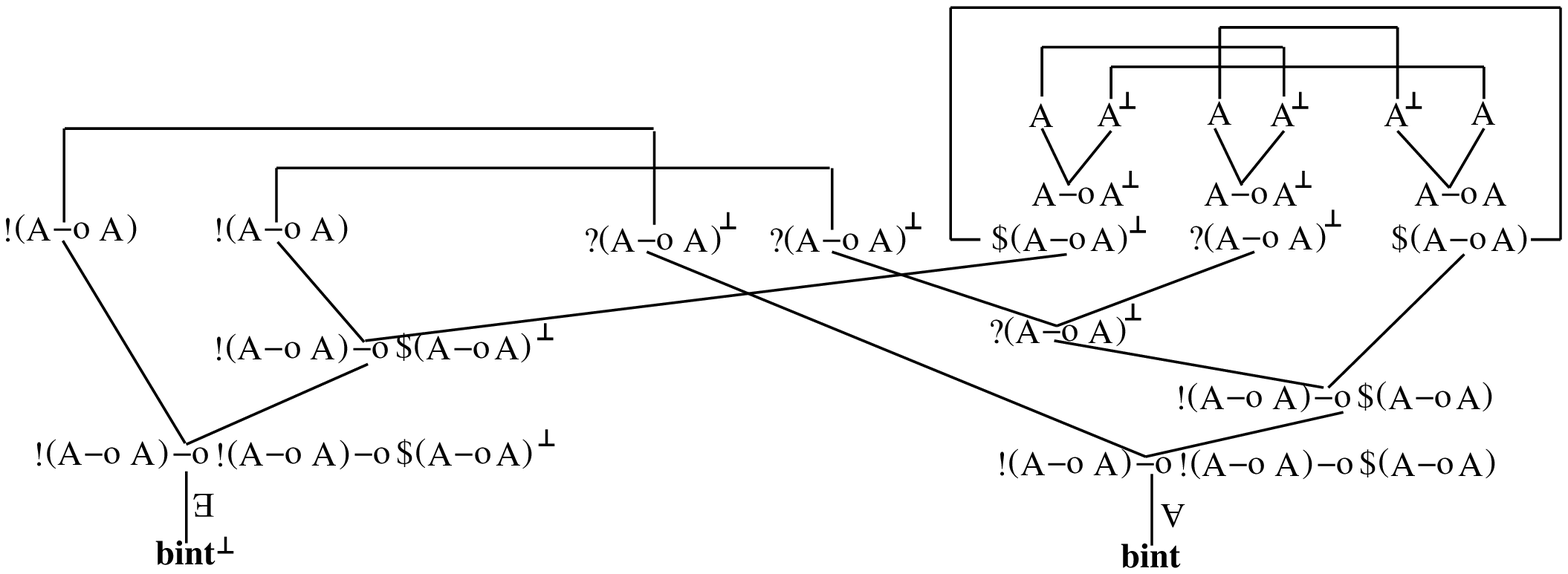}
\caption[{\tt suc1}]{{\tt suc1}}
\label{suc1}
\end{center}
\end{figure}
Figure~\ref{badd} shows the analogue in ${\bf bint}$ to the addition in ${\bf int}$: we call the proof net {\tt badd}. 
If we regard two inputs proofs of ${\bf bint}$ of the proof net 
as two lists which only have $0$ and $1$, then 
we can regard {\tt badd} as a concatenation function of two inputs.\\
\begin{figure}[htbp]
\begin{center}
\includegraphics[scale=0.5]{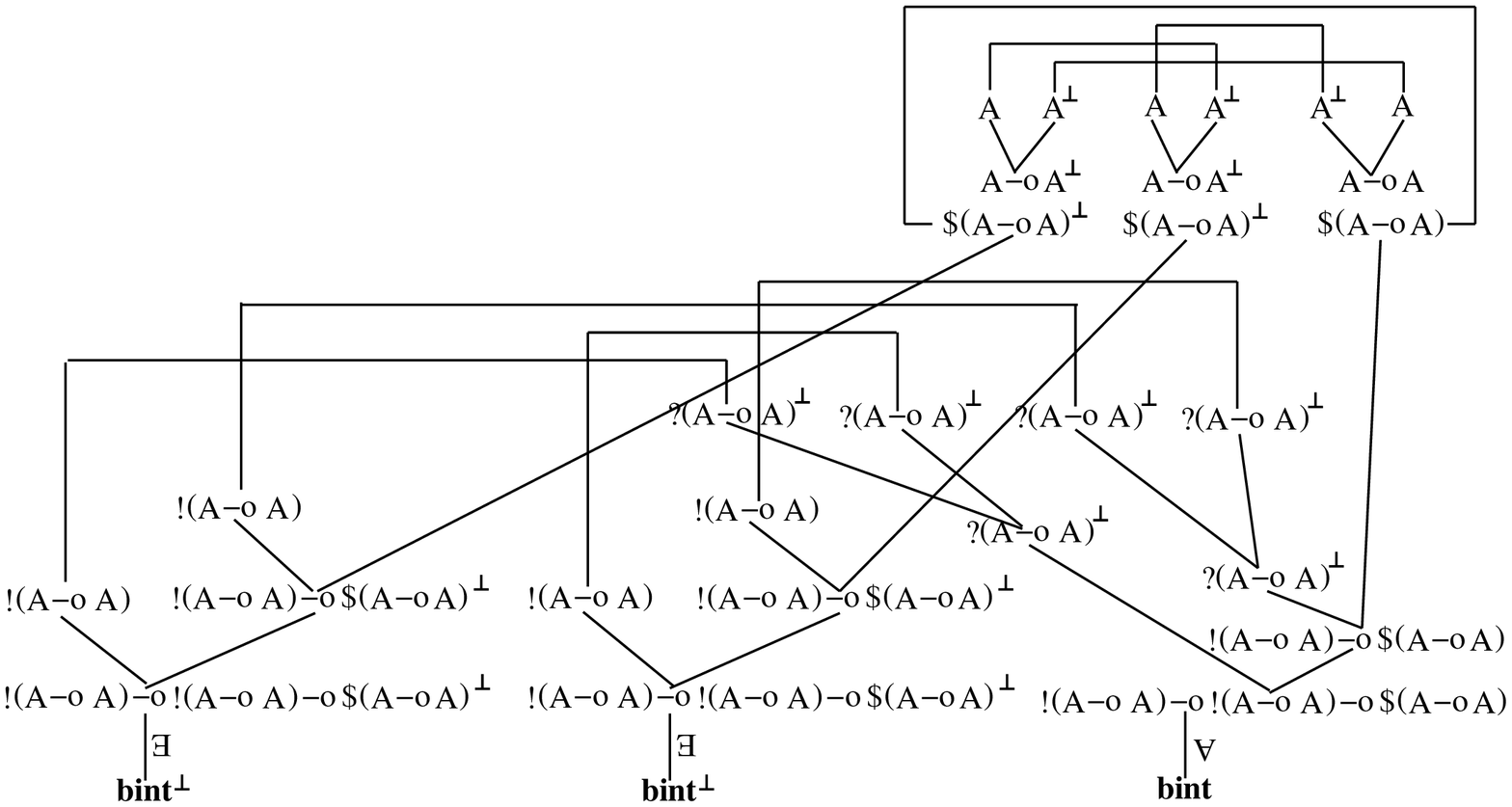}
\caption[{\tt badd}]{{\tt badd}}
\label{badd}
\end{center}
\end{figure}
Figure~\ref{bmul} shows a proof net called {\tt bmul}. 
In the figure, {\tt empty} is a proof net of ${\bf bint}$ 
that does not have exponential-links except for two weakening-links with $?(A \LIMP A)^\bot$. 
Let $\Theta_1$ be a ${\bf bint}$ proof that is supplied to ${\bf bint}^\bot$ port of {\tt bmul}
and $\Theta_2$ be a proof net with $!{\bf bint}$ as one of conclusions
that is supplied to $?{\bf bint}^\bot$ port of {\tt bmul}.
Let $\ell$ be the length of $\Theta_1$.
The evaluated result of {\tt bmul} provided inputs $\Theta_1$ and $\Theta_2$ are given, 
is $\ell$ copies of $\Theta_2$. Let $m$ be the length of $\Theta_2$. 
The length of the result is $\ell \times m$.
The proof net {\tt bmul} is analogous to multiplication of ${\bf int}$.\\
\begin{figure}[htbp]
\begin{center}
\includegraphics[scale=0.5]{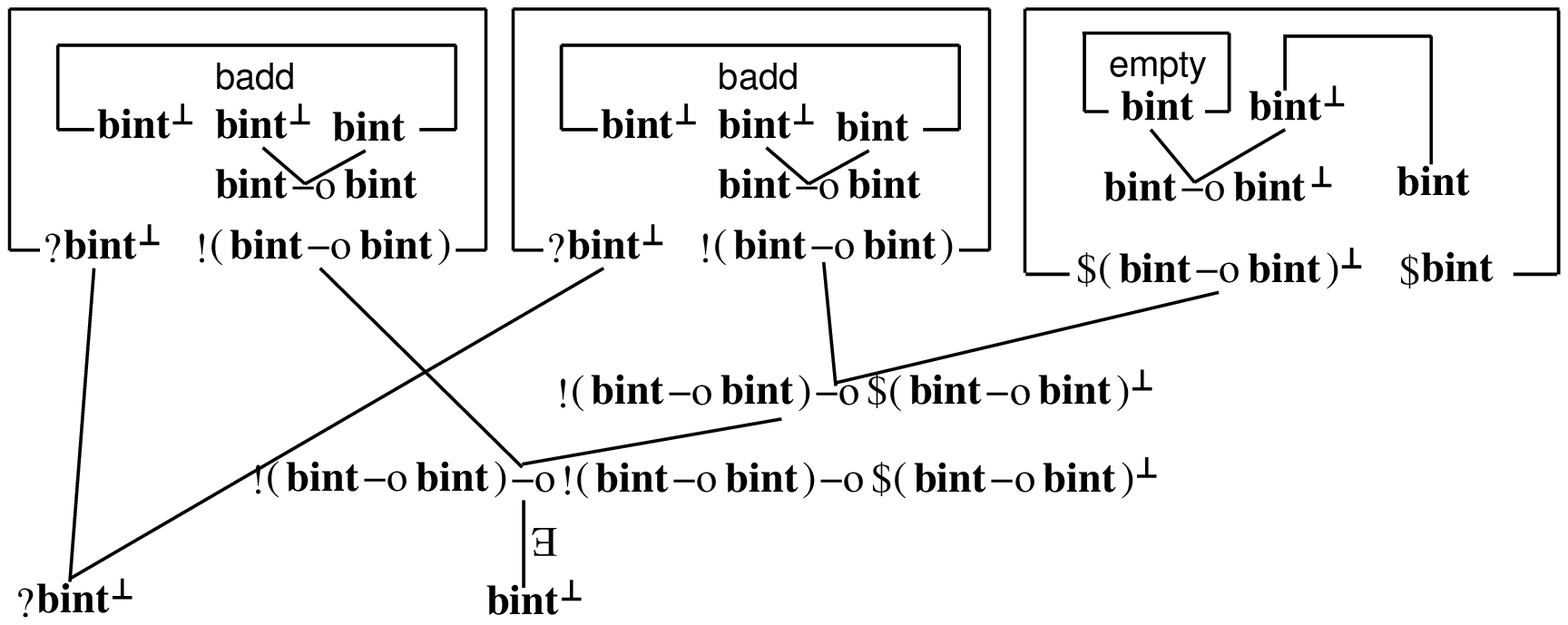}
\caption[{\tt bmul}]{{\tt bmul}}
\label{bmul}
\end{center}
\end{figure}
The proof net shown in Figure~\ref{bint2config} transform a ${\bf bint}$ proof into 
a ${\bf config}$ proof that is a initial configuration of Turing machines. 
We call the proof net {\tt bint2config}. 
By using {\tt bint2config} and {\tt transition}, our encoding of the transition function of $M$, 
we can construct the engine part of Turing machines shown in Figure~\ref{tmengine}.
But it is not sufficient for a proof of Theorem~\ref{maintheorem}: 
besides we need constructions for polynomial time bound.
To do this, we prepare several proof nets.\\
\begin{figure}[htbp]
\begin{center}
\includegraphics[scale=0.5]{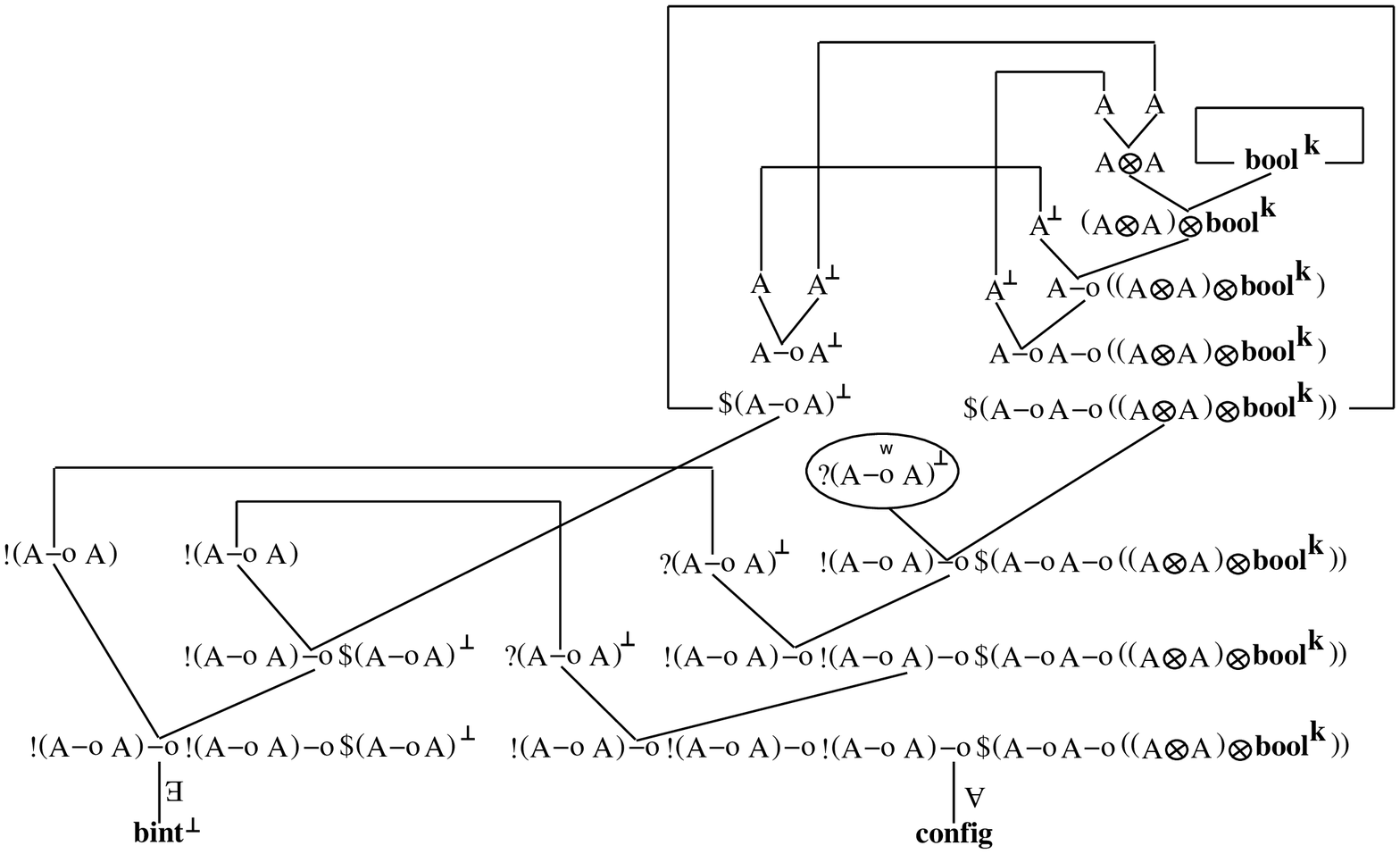}
\caption[{\tt bint2config}]{{\tt bint2config}}
\label{bint2config}
\end{center}
\end{figure}
\begin{figure}[htbp]
\begin{center}
\includegraphics[scale=0.5]{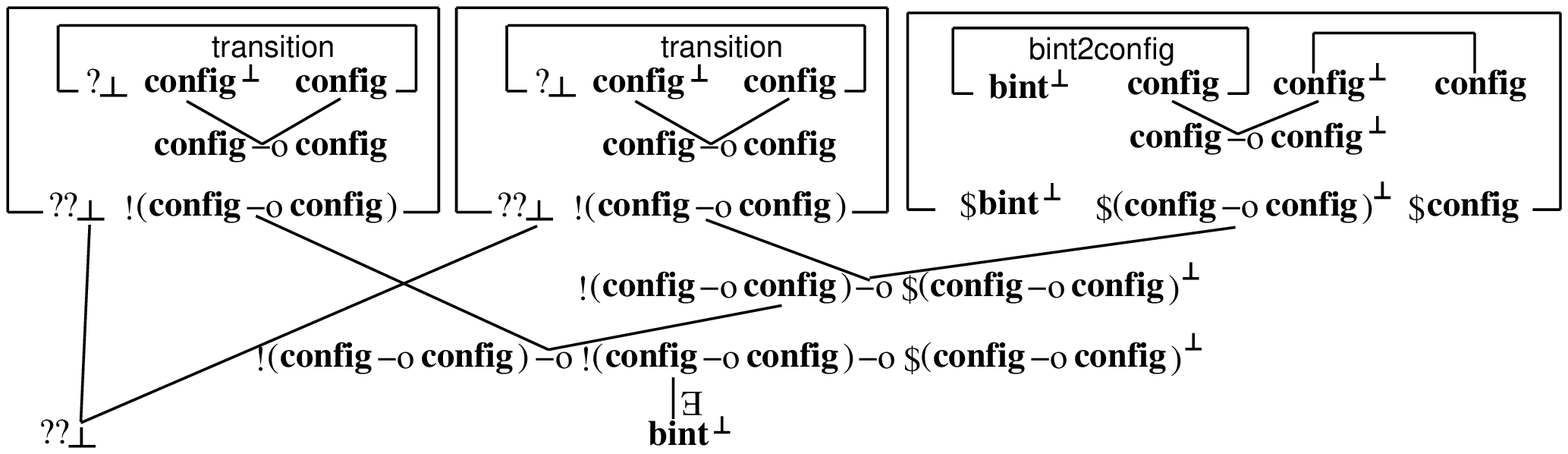}
\caption[{\tt tm\_engine}]{{\tt tm\_engine}}
\label{tmengine}
\end{center}
\end{figure}
The proof net ${\tt coer}^{p,q}$ of Figure~\ref{coercion} is the ${\bf bint}$ version 
of ${\bf int}$ coercion of \cite{Gir98}. Then $p$ must be greater than $0$.
This proof net is used in Figure~\ref{kcomposition} and Figure~\ref{TM}.
\begin{figure}[htbp]
\begin{center}
\includegraphics[scale=0.5]{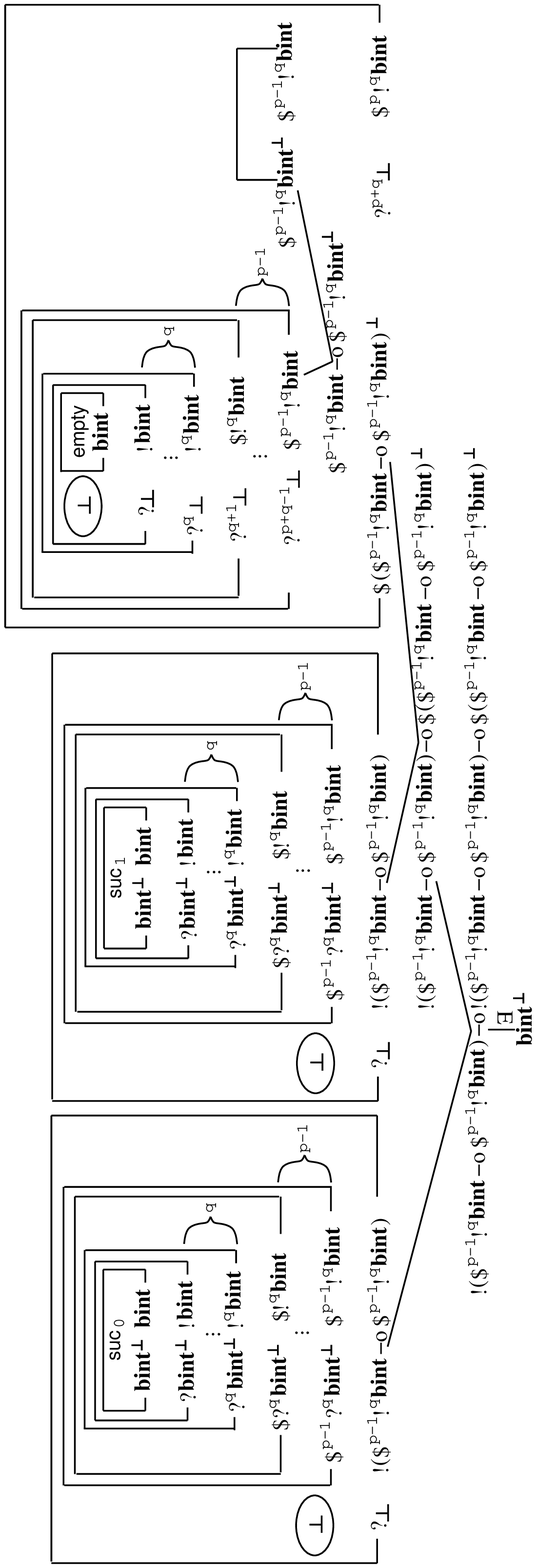}
\caption[${\tt coer}^{p,q}$]{${\tt coer}^{p,q}$}
\label{coercion}
\end{center}
\end{figure}
The proof net {\tt k-contraction} of Figure~\ref{kcontraction} is also the ${\bf bint}$ version 
of ${\bf int}$ contraction of \cite{Gir98}. 
This proof net is used in Figure~\ref{kcomposition} and Figure~\ref{kpolynomial}.\\
\begin{figure}[htbp]
\begin{center}
\includegraphics[scale=0.48]{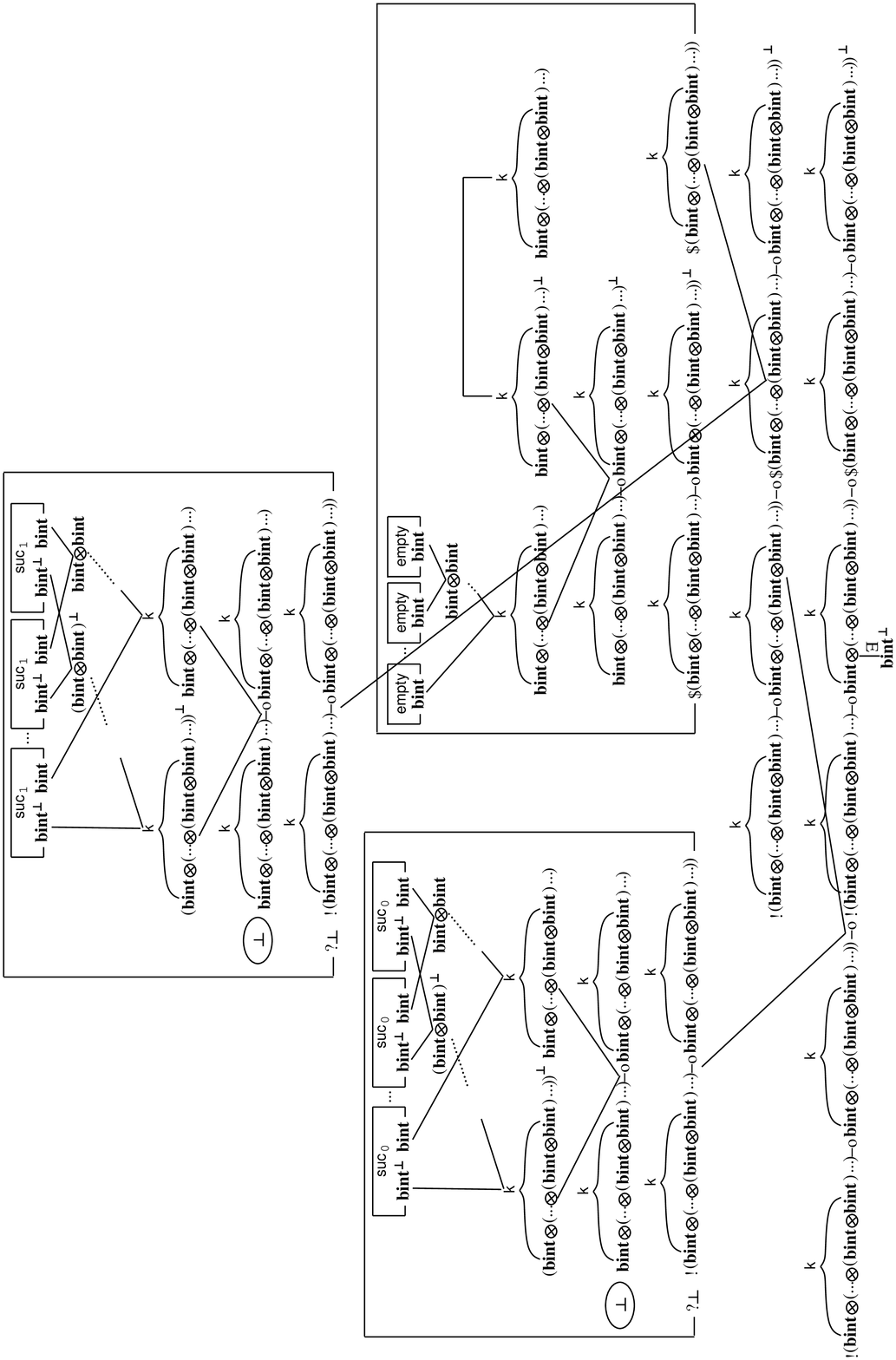}
\caption[{\tt k-contraction}]{{\tt k-contraction}}
\label{kcontraction}
\end{center}
\end{figure}
The proof net in Figure~\ref{kcomposition} is used in Figure~\ref{kpolynomial}.
This is basically $k$ compositions of {\tt bmul}.
The ${\bf bint}$ proof $c_{\cconst}$ in Figure~\ref{kcomposition} is 
a constant that does not depend on the lengths of inputs of Turing machine $M$.\\
\begin{figure}[htbp]
\begin{center}
\includegraphics[scale=0.5]{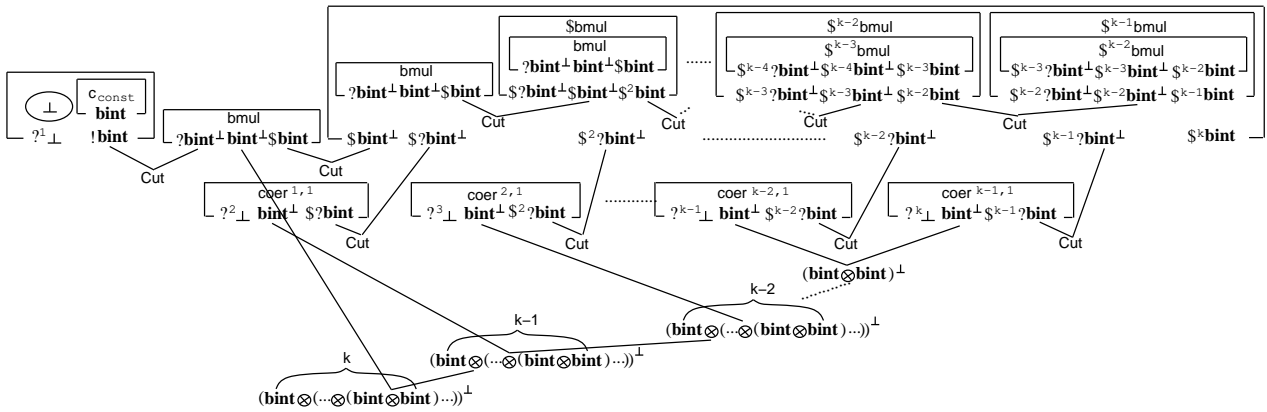}
\caption[{\tt k-composition} of multiplications]{{\tt k-composition} of multiplications}
\label{kcomposition}
\end{center}
\end{figure}
The proof net {\tt kpolynomial} of Figure~\ref{kpolynomial} is our polynomial construction with degree $k$.
Let $\Theta$ be a proof net of ${\bf bint}$ and $\ell$ be the length of $\Theta$.
The evaluated result of {\tt kpolynomial} provided an input $\Theta$ is given,  
is a nest of $\$$-boxes which has an inside proof net 
of ${\bf bint}$ with the length $c_{\cconst} \times \ell^k$.\\
\begin{figure}[htbp]
\begin{center}
\includegraphics[scale=0.5]{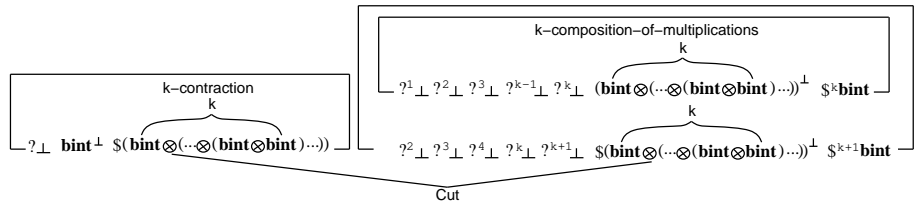}
\caption[{\tt kpolynomial}]{{\tt kpolynomial}}
\label{kpolynomial}
\end{center}
\end{figure}
Finally we obtain our encoding of a Turing machine with polynomial time bound of Figure~\ref{TM}.
This completes our proof of Theorem~\ref{maintheorem}.
\begin{figure}[htbp]
\begin{center}
\includegraphics[scale=0.5]{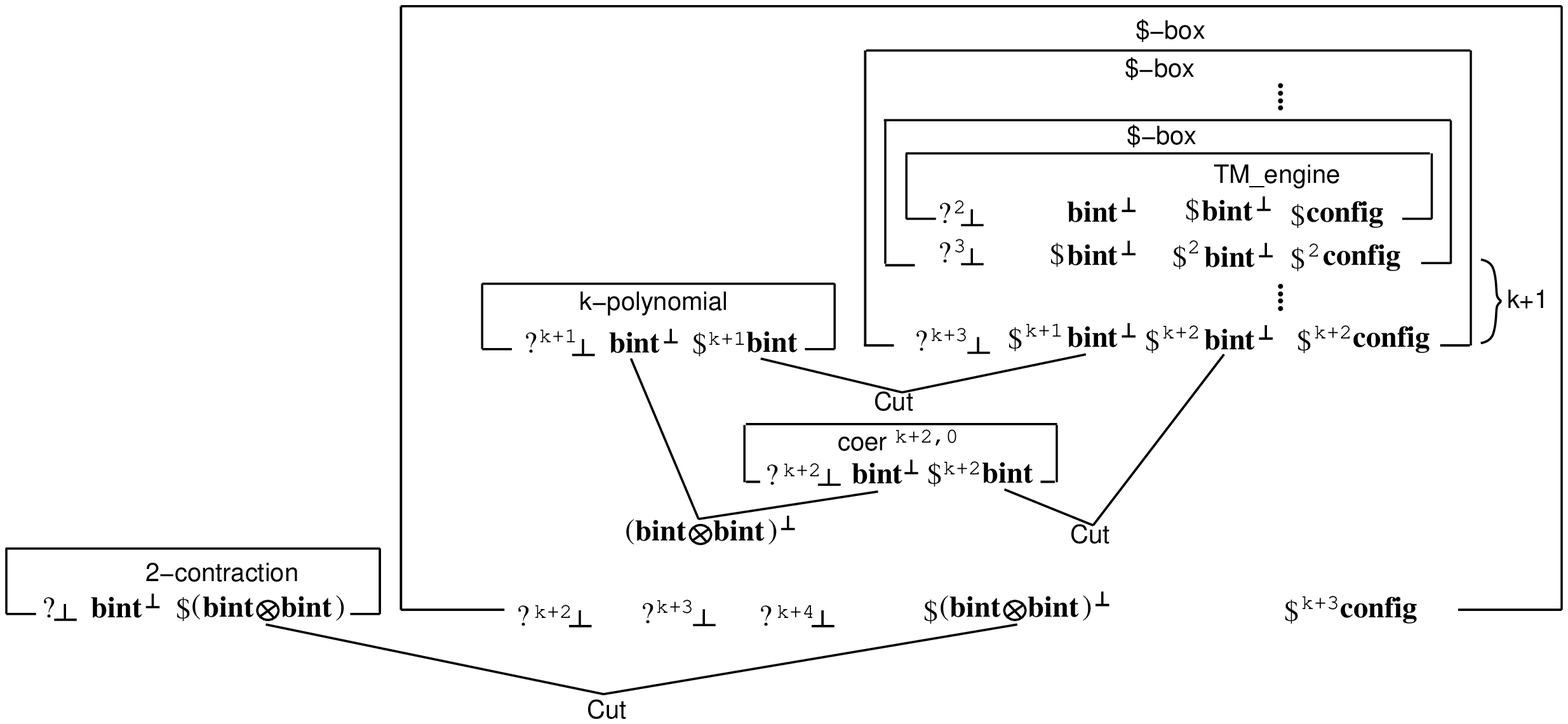}
\caption[{\tt TM}]{{\tt TM}}
\label{TM}
\end{center}
\end{figure}

\clearpage

\section{A transformation from {\bf config} proofs into {\bf bint} proofs}
\label{appendcfgtobint}

At first we remark that
when a given proof net with conclusion $\Gamma$, we can construct a proof net
with ${\bf bool^2}^\bot, \Gamma$ as shown in Figure~\ref{bool-weakening}. 
It is easy to extend the remark to the general ${\bf bool^k}$ case for $k \le 2$. \\
Then based on the above remark, as a derived rule, we introduce ${\bf bool^k}$-axiom 
as shown in Figure~\ref{boolk-axiom} in order to keep figures as simple as possible.

\begin{figure}[htbp]
\begin{center}
\includegraphics[scale=0.5]{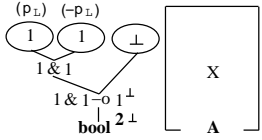}
\caption[${\bf bool^2}$-weakening]{${\bf bool^2}$-weakening}
\label{bool-weakening}
\end{center}
\end{figure}

\begin{figure}[htbp]
\begin{center}
\includegraphics[scale=0.5]{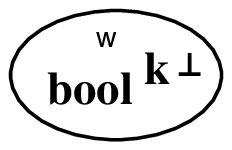}
\caption[${\bf bool^k}$-axiom]{${\bf bool^k}$-axiom}
\label{boolk-axiom}
\end{center}
\end{figure}

In order to translate {\bf config} proofs into {\bf bint} proofs, 
we introduce an immediate type 
${\bf tint} \equiv_{\mathdef} \forall X. !(X \LIMP X) \LIMP !(X \LIMP X) \LIMP !(X \LIMP X) \LIMP \$(X \LIMP X)$. \\
Figure~\ref{config2bint} shows our translator from {\bf config} proofs into {\bf bint} proofs.
When a given {\bf config} proof, at first we duplicate the proof by using
{\tt 2-contraction-config} proof net.
A construction of {\tt 2-contraction-config} proof net is not so easy as 
that of {\tt 2-contraction} for {\bf bint}. 
Appendix~\ref{constractionofcontractionconfig} is devoted to the construction.

\begin{figure}[htbp]
\begin{center}
\includegraphics[scale=0.5]{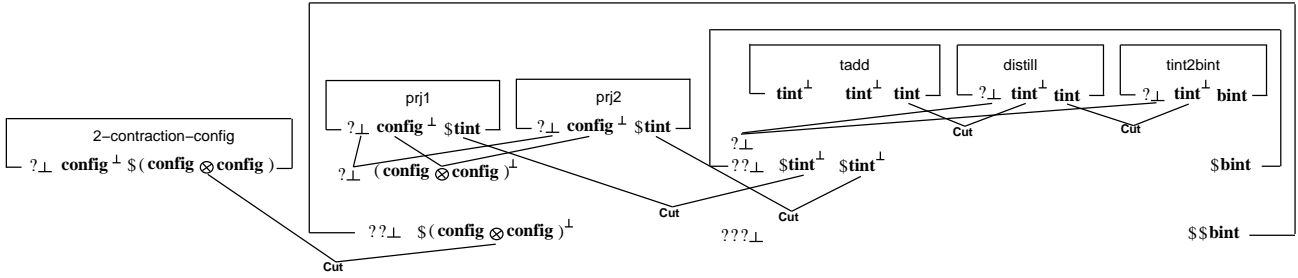}
\caption[{\tt config2bint}]{{\tt config2bint}}
\label{config2bint}
\end{center}
\end{figure}

After that, each duplicated {\bf config} proof net is projected
into a {\tt tint} proof by using {\tt prj1} or {\tt prj2} shown in Figure~\ref{Prj1OrPrj2-90}.
The purpose of {\tt prj1} is to extract the left parts of configurations of Turing machines and 
similarly that of {\tt prj2} is to extract the right parts. \\
Proof net {\tt prj1} has proof net {\tt prj1sub} shown in Figure~\ref{prj1sub} 
as a sub-proof net and
{\tt prj2} has {\tt prj2sub} shown in Figure~\ref{prj2sub}.
Proof net {\tt prj1} also has proof nets ${\tt tsuc0^r}$, ${\tt tsuc1^r}$, and ${\tt tsuc*^r}$
as sub-proof nets and 
{\tt prj2} has {\tt tsuc0}, {\tt tsuc1}, and {\tt tsuc*}.
Figure~\ref{tsuc0} and Figure~\ref{tsuc0r} show proof net {\tt tsuc0} and ${\tt tsuc0^r}$ 
respectively. 
We omit {\tt tsuc1}, {\tt tsuc*}, ${\tt tsuc1^r}$, and ${\tt tsuc*^r}$, 
since the constructions of these proof nets are easy exercise.
Note that in order to recover tapes correctly we need to reverse the left parts of tapes.
Next we concatenate obtained two {\bf tint} proofs by {\tt tadd} of Figure~\ref{tadd}.

\begin{figure}[htbp]
\begin{center}
\includegraphics[scale=0.5]{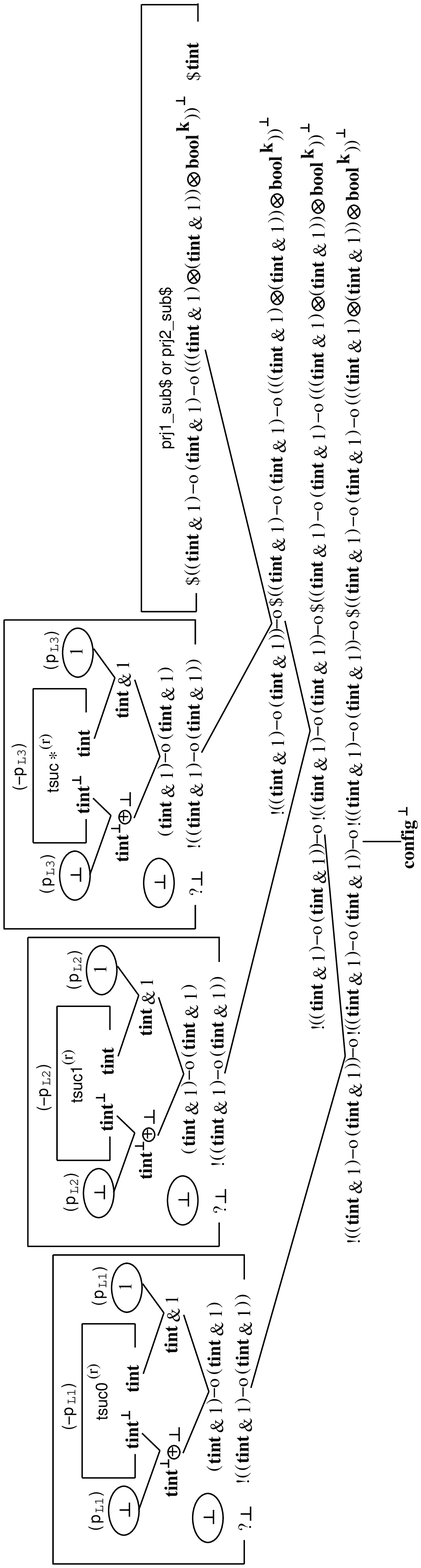}
\caption[{\tt prj1} or {\tt Prj2}]{{\tt prj1} or {\tt prj2}}
\label{Prj1OrPrj2-90}
\end{center}
\end{figure}

\clearpage 

\begin{figure}[htbp]
\begin{center}
\includegraphics[scale=0.5]{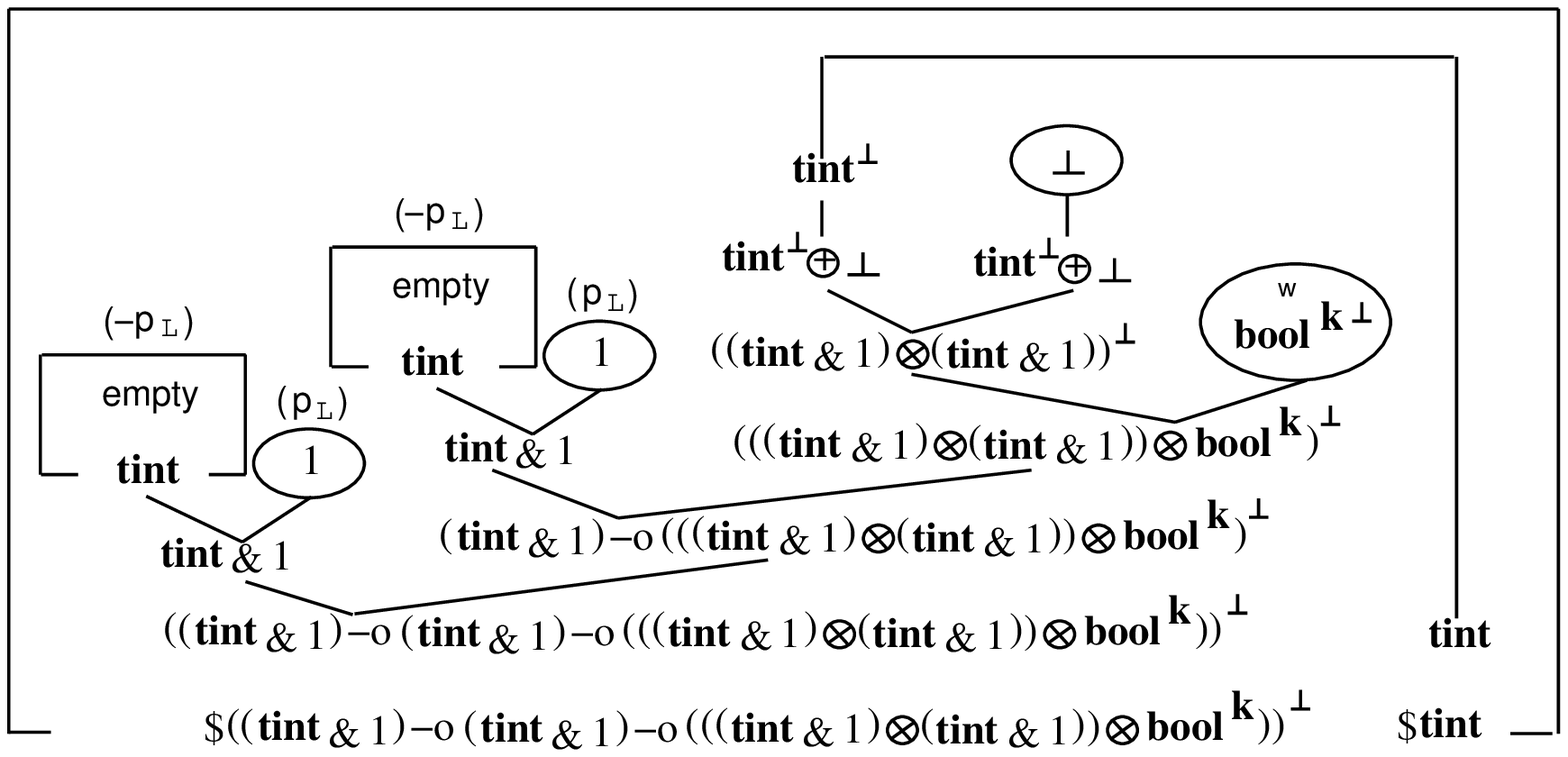}
\caption[{\tt prj1sub}]{{\tt prj1sub}}
\label{prj1sub}
\end{center}
\end{figure}

\begin{figure}[htbp]
\begin{center}
\includegraphics[scale=0.5]{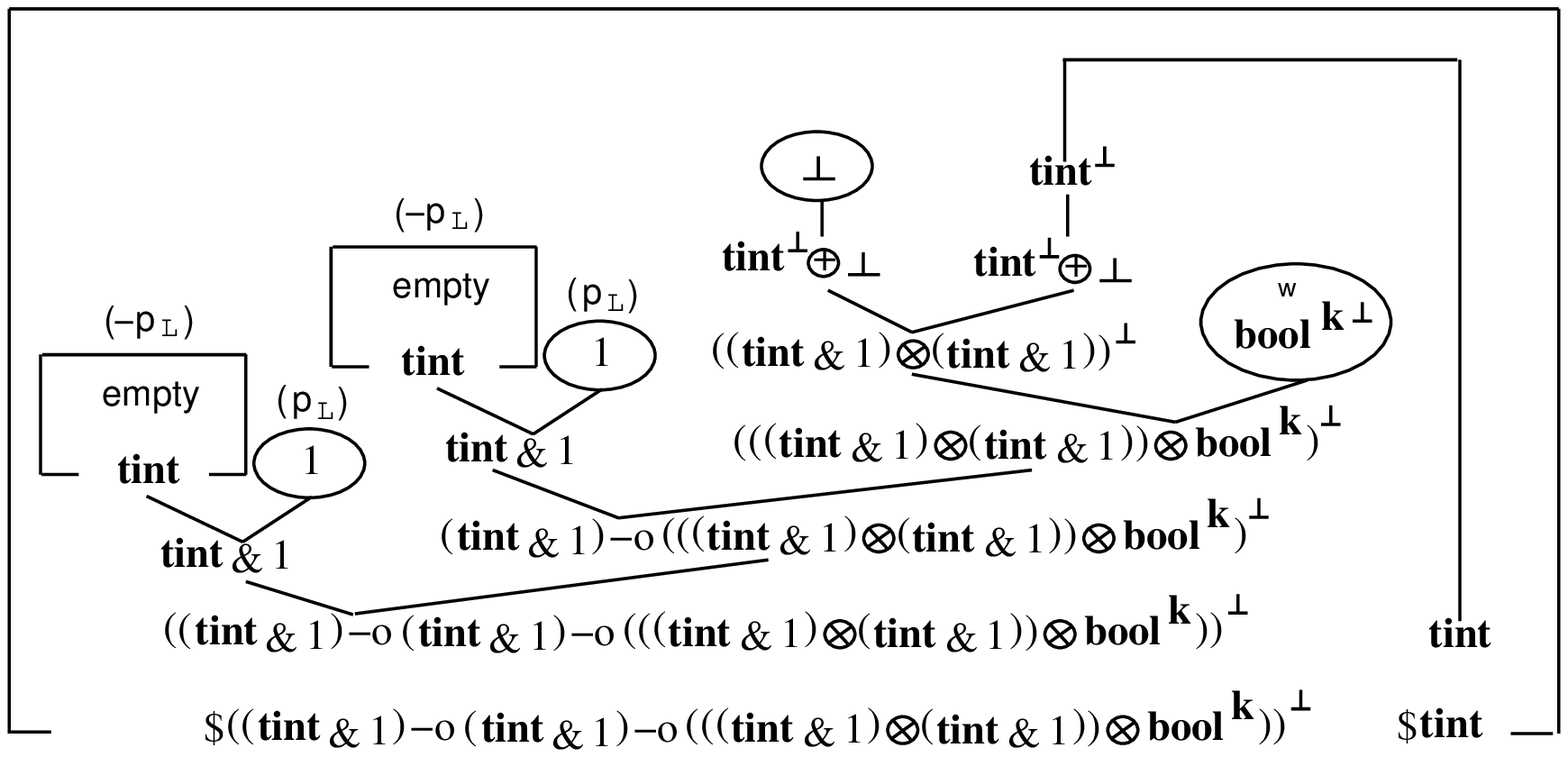}
\caption[{\tt prj2sub}]{{\tt prj2sub}}
\label{prj2sub}
\end{center}
\end{figure}

\begin{figure}[htbp]
\begin{center}
\includegraphics[scale=0.5]{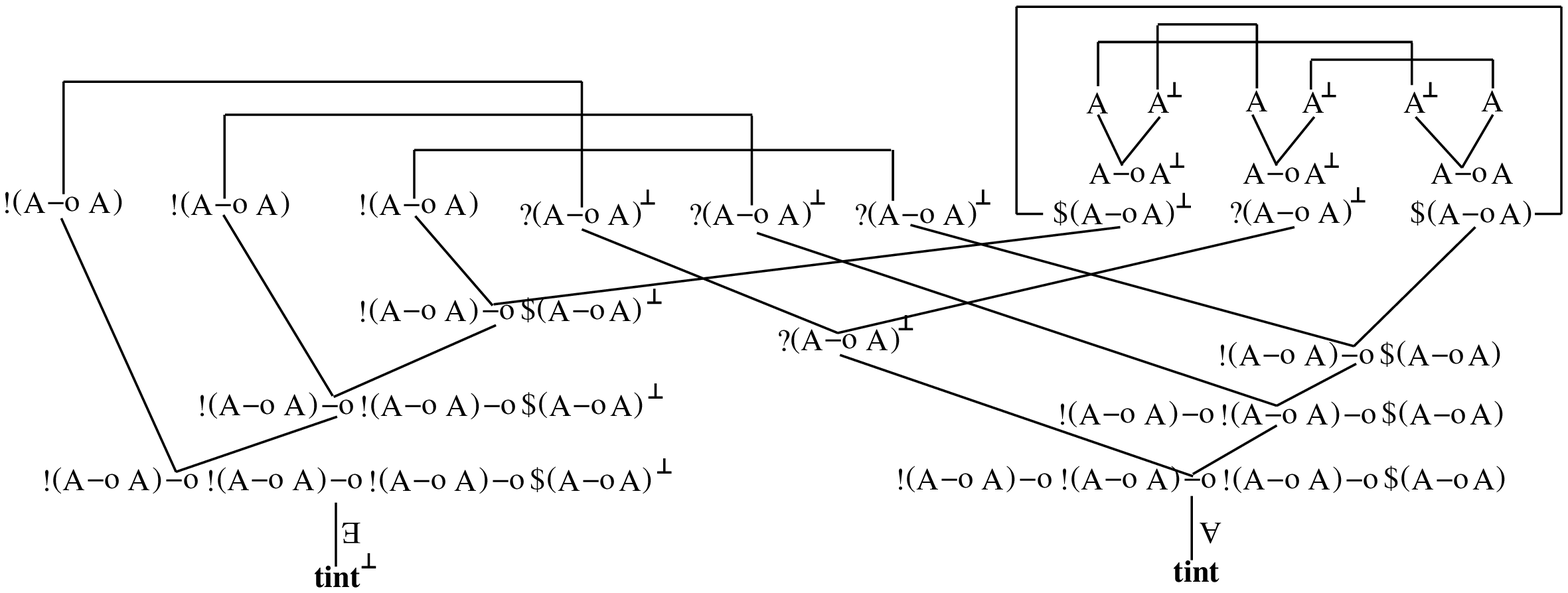}
\caption[{\tt tsuc0}]{{\tt tsuc0}}
\label{tsuc0}
\end{center}
\end{figure}

\begin{figure}[htbp]
\begin{center}
\includegraphics[scale=0.5]{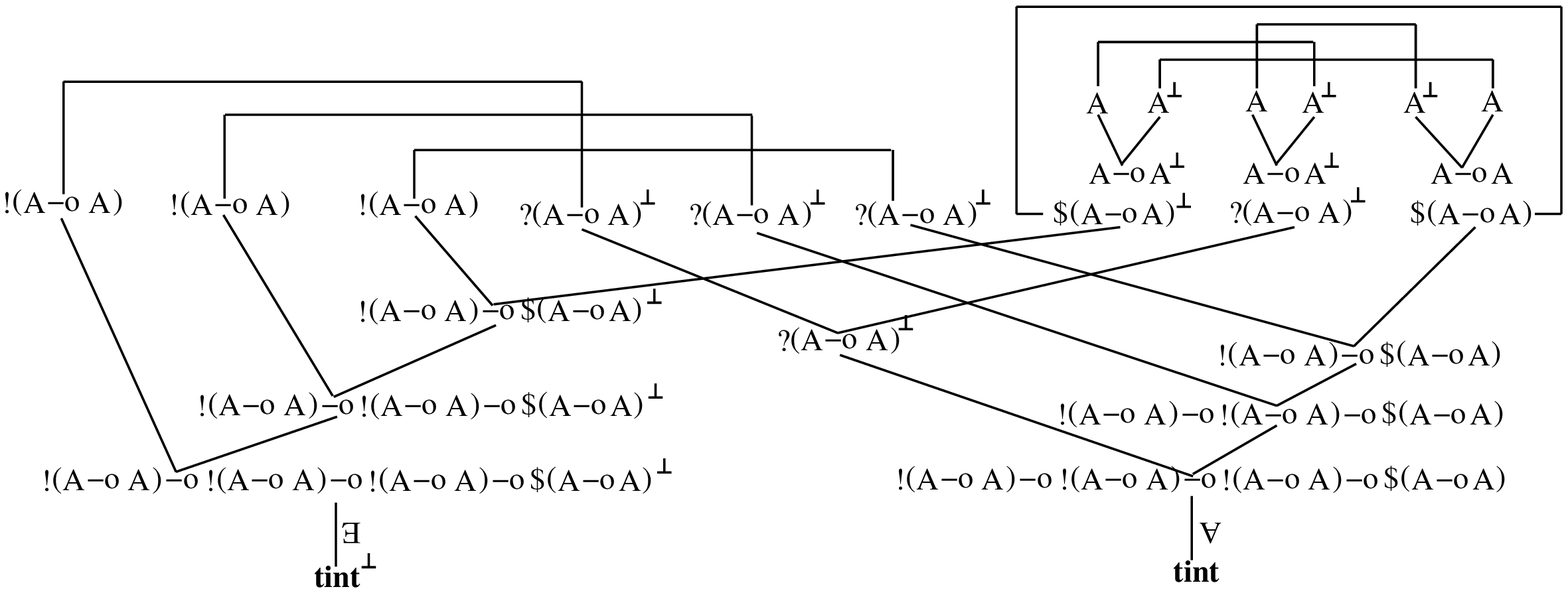}
\caption[${\tt tsuc0^r}$]{${\tt tsuc0^r}$}
\label{tsuc0r}
\end{center}
\end{figure}

\begin{figure}[htbp]
\begin{center}
\includegraphics[scale=0.5]{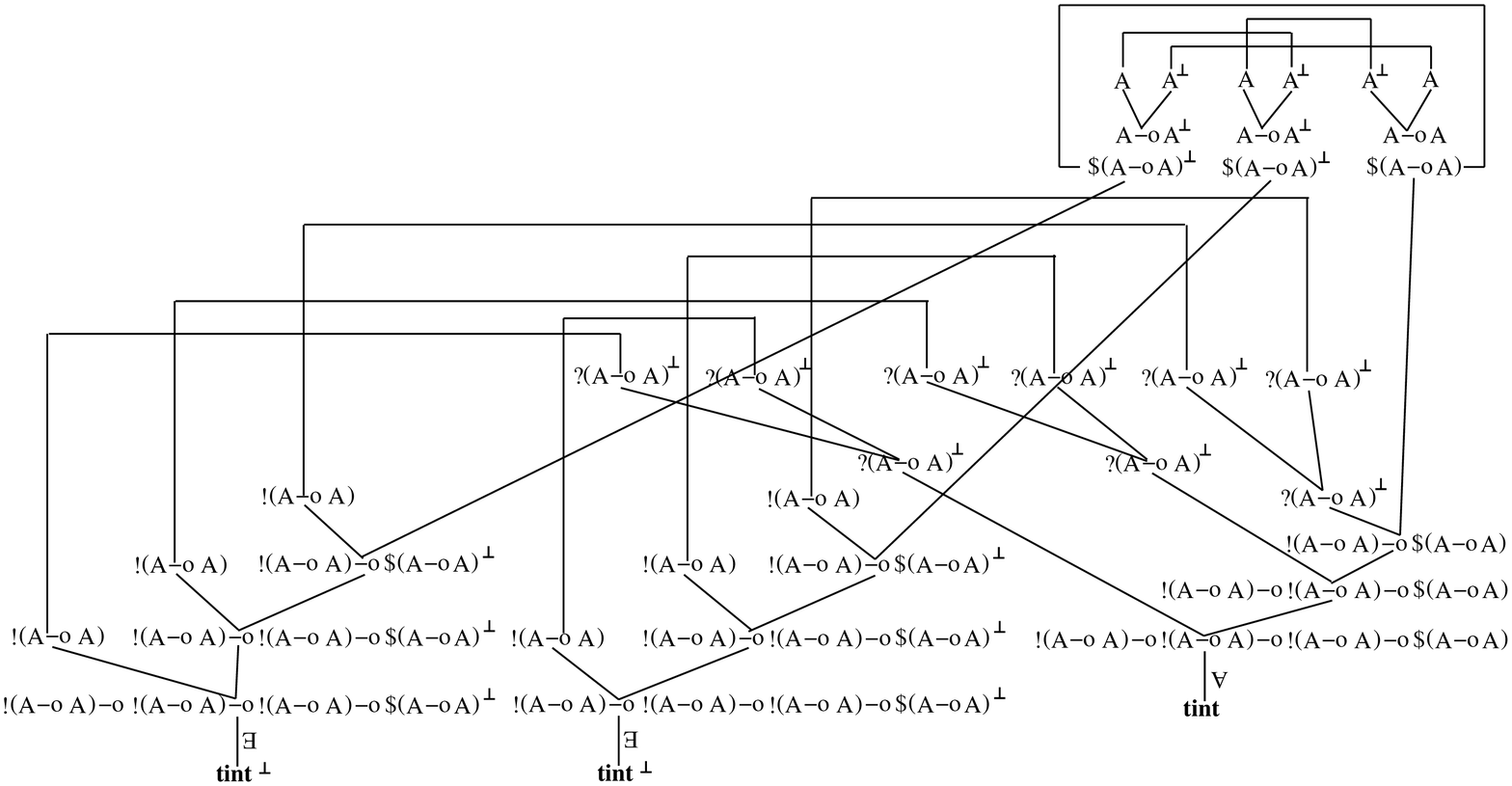}
\caption[{\tt tadd}]{{\tt tadd}}
\label{tadd}
\end{center}
\end{figure}

We distinguish the two normal proofs of ${\bf bool^2}$. 
One is called ${\tt Pi_{01}}$. and the other ${\tt Pi_\ast}$ (see Figure~\ref{pi01andpistar}). 
Next we apply {\tt distill} of Figure~\ref{distill} to the obtained {\bf tint} proof.
The construction of the proof net {\tt distill} is inspired by 
that of {\tt strip} term in \cite{MO00}.
The intention of the {\tt distill} proof net is to keep 
occurrences of $0$ and $1$  
until the first $\ast$ occurrence is reached.
After that, the rest are discarded.
Figure~\ref{distillstepX} shows  three sub-proof nets {\tt distill\_step\_X} ({\tt X}=1,2, and $\ast$) 
of the {\tt distill} proof net.
Moreover, two sub-proof nets ${\tt distill\_step\_sub\_X_L}$ and ${\tt distill\_step\_sub\_X_R}$ 
of ${\tt distill\_step\_X}$
have the forms of Figure~\ref{distillstepsubjoin} or Figure~\ref{distillstepsubdiscard}.
Table~\ref{distillstepsubXD} shows the correspondence.\\
Figure~\ref{tint2bint} shows {\tt tint2bint} proof.
The intention is to remove $\ast$-entry.

\begin{figure}[htbp]
\begin{center}
\includegraphics[scale=0.5]{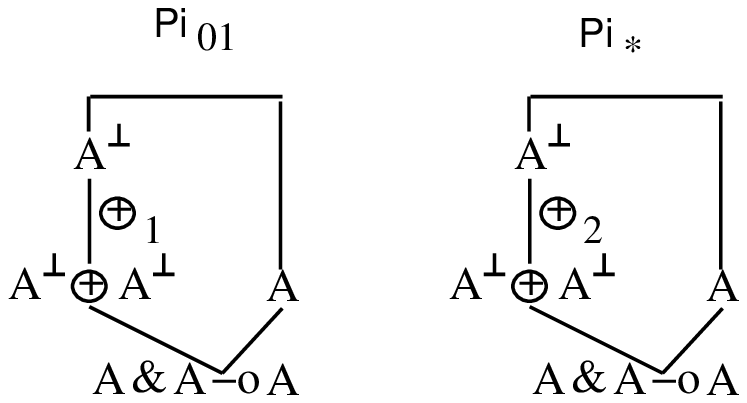}
\caption[${\tt Pi_{01}}$ and ${\tt Pi_\ast}$]{${\tt Pi_{01}}$ and ${\tt Pi_\ast}$}
\label{pi01andpistar}
\end{center}
\end{figure}

\begin{figure}[htbp]
\begin{center}
\includegraphics[scale=0.5]{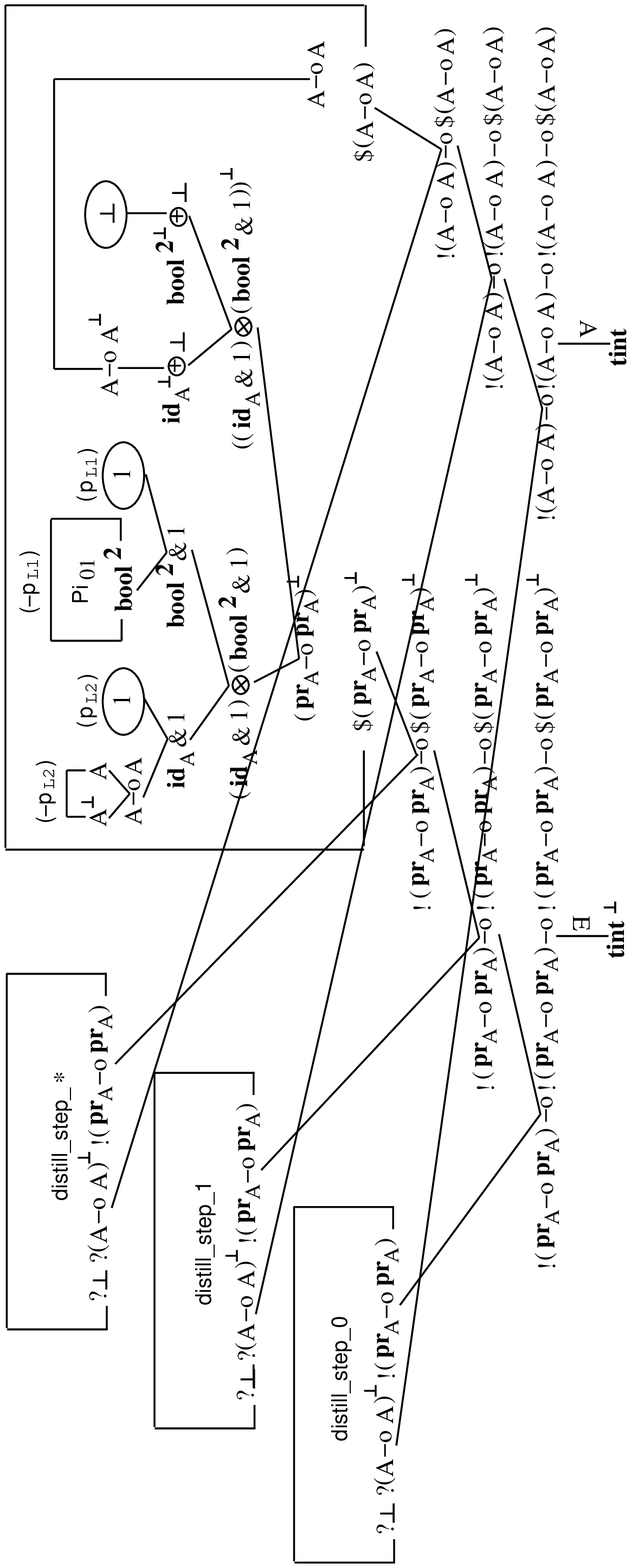}
\caption[{\tt distill}]{{\tt distill}}
\label{distill}
\end{center}
\end{figure}

\begin{figure}[htbp]
\begin{center}
\includegraphics[scale=0.5]{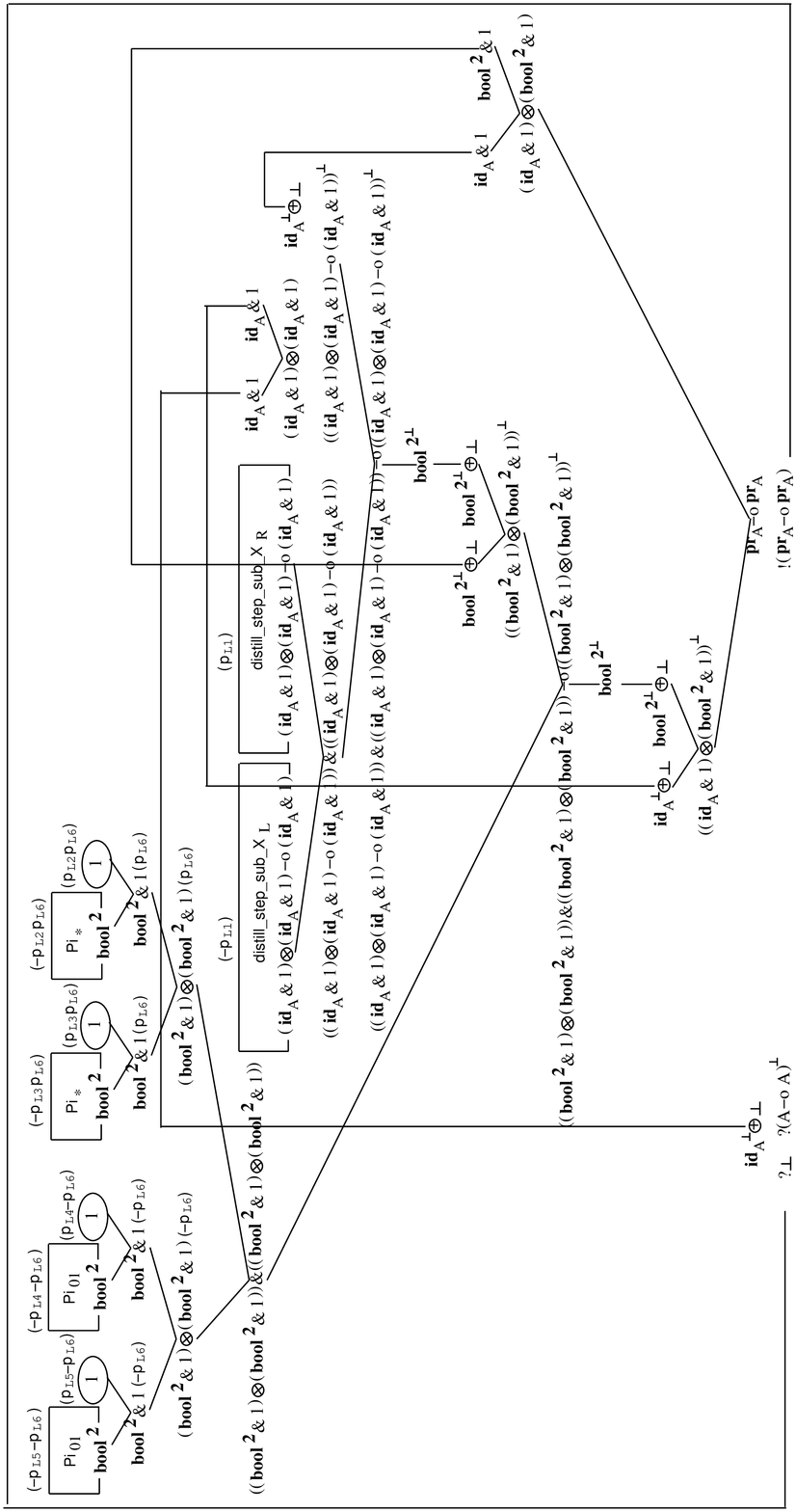}
\caption[{\tt distill\_step\_X}]{{\tt distill\_step\_X}}
\label{distillstepX}
\end{center}
\end{figure}

\begin{figure}[htbp]
\begin{center}
\includegraphics[scale=0.5]{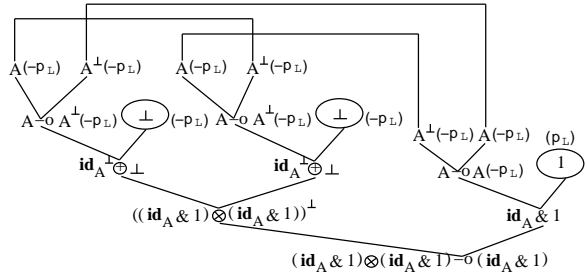}
\caption[{\tt distill\_step\_sub\_join}]{{\tt distill\_step\_sub\_join}}
\label{distillstepsubjoin}
\end{center}
\end{figure}

\begin{figure}[htbp]
\begin{center}
\includegraphics[scale=0.5]{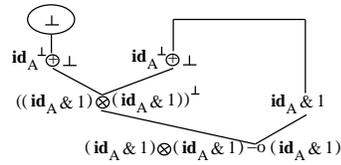}
\caption[{\tt distill\_step\_sub\_discard}]{{\tt distill\_step\_sub\_discard}}
\label{distillstepsubdiscard}
\end{center}
\end{figure}

\begin{table}[htbp]
\begin{center}
\begin{tabular}{|c|c|c|}
\hline
& D=L & D=R \\ 
\hline
X=0 & {\tt distill\_step\_sub\_join} & {\tt distill\_step\_sub\_discard} \\
\hline
X=1 & {\tt distill\_step\_sub\_join} & {\tt distill\_step\_sub\_discard} \\
\hline
X=$\ast$ & {\tt distill\_step\_sub\_discard} & {\tt distill\_step\_sub\_discard} \\
\hline
\end{tabular}
\caption[${\tt distill\_step\_sub\_X_D}$]{${\tt distill\_step\_sub\_X_D}$}
\label{distillstepsubXD}
\end{center}
\end{table}

\begin{figure}[htbp]
\begin{center}
\includegraphics[scale=0.5]{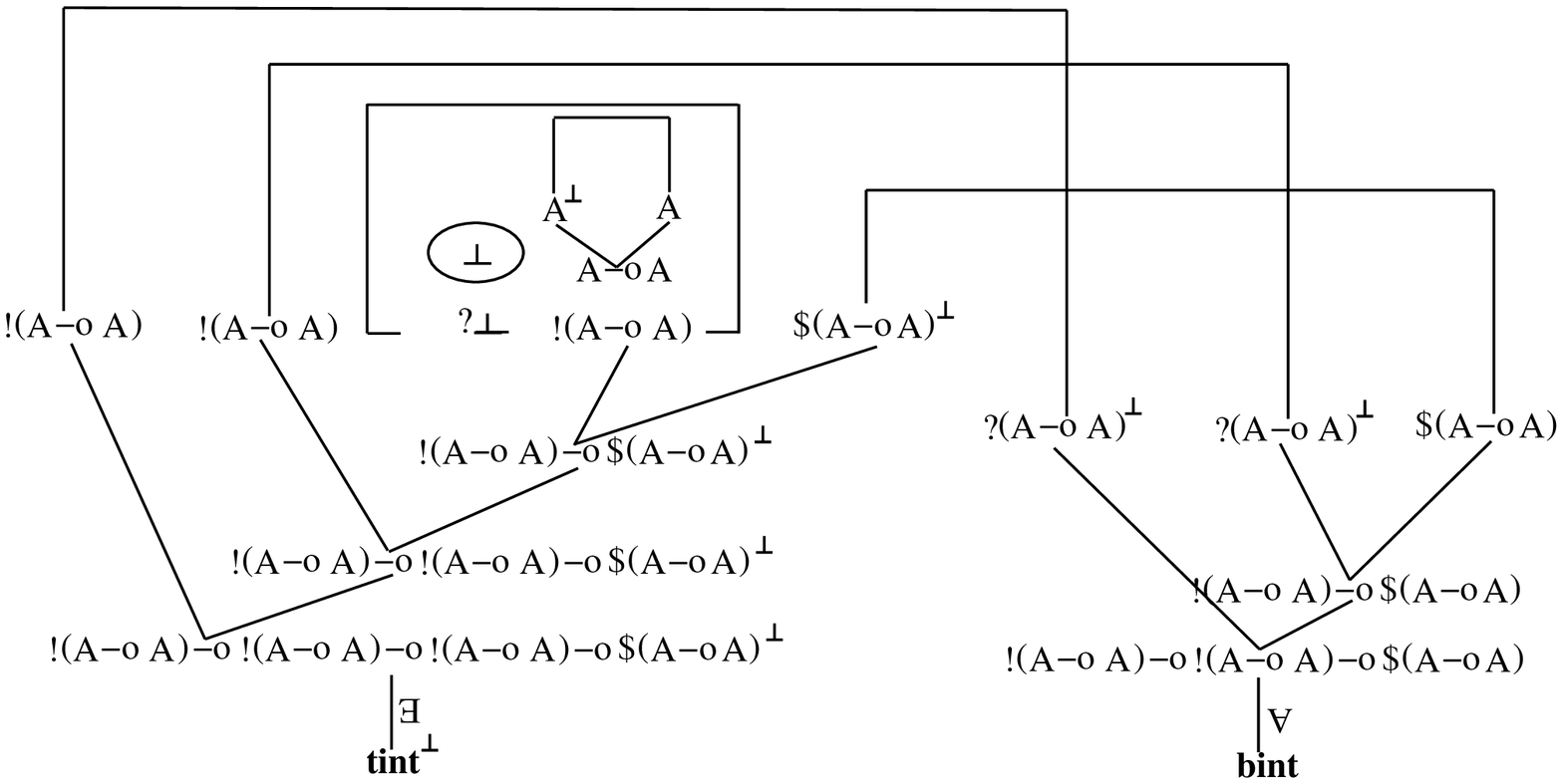}
\caption[{\tt tint2bint}]{{\tt tint2bint}}
\label{tint2bint}
\end{center}
\end{figure}

\section{Contraction on {\bf config}}
\label{constractionofcontractionconfig}
In this section we give how to construct {\tt 2-contraction-config} proof net. 
In {\tt config2bint} of Appendix~\ref{appendcfgtobint} we do not need the ${\bf book^k}$-part 
of a given {\bf config} proof. 
Hence in the construction of {\tt 2-contraction-config} proof net 
we could discard ${\bf book^k}$-parts. 
But we give a general construction that duplicates ${\bf book^k}$-parts here. 
Figure~\ref{config2contraction} shows {\tt 2-contraction-config} proof net.
In this proof net, 
\begin{enumerate}
\item when given a {\bf config} proof net, 
  {\tt pre\_config\_dup} of Figure~\ref{preconfigdup} outputs a quartet of {\bf config} proof nets, 
  where two {\bf config} proofs are the same and only keep the left part and ${\bf book^k}$ of the input, and
  the rest, which are two {\bf config} proofs, are also the same and only keep the right part and ${\bf book^k}$ of the input; 
\item each {\tt configadd} of Figure~\ref{configadd} concatenate 
  two {\bf config} proof nets in the quartet.
\end{enumerate}

Type ${\bf dconfig_2^k}$ of proof net {\tt pre\_config\_dup} is defined as follows:
\[
{\bf config_2^k} \equiv_{\mathdef}  
\overbrace{({\bf config} \TENS {\bf config}) \WITH (\cdots \WITH (({\bf config} \TENS {\bf config}) \WITH ({\bf config} \TENS {\bf config})) \cdots ) }^{k} 
\]
\[
{\bf dconfig_2^k} \equiv_{\mathdef}  {\bf config_2^k} \PLUS {\bf config_2^k}
\]

In {\tt pre\_config\_dup}, at first, we make $4k$ {\bf config} proofs. 
Then according to the ${\bf book^k}$-value of the input ${\bf config}$ proof, 
we choose 4 {\bf config} proofs.
That is why we use $k$-ary tuples by $\WITH$-connectives in ${\bf config_2^k}$.
In addition we need to distinguish the left part and the right part of the input ${\bf config}$ proof.
That is why we use one $\PLUS$-connective in ${\bf dconfig_2^k}$. \\
Figure~\ref{preconfigdupmain} shows sub-proof net {\tt pre\_config\_dup\_main} of 
{\tt pre\_config\_dup}.
Note that as shown in Figure~\ref{bdup}, we can duplicate ${\bf book^2}$ proof without using $\$$
(of course we can easily extend this construction to the ${\bf bool^k}$ case). \\
Proof nets ${\tt dconfig\_sucX}$ (where $X=1,2,\, $ and $\, \ast$) shown in Figure~\ref{dconfigk2} 
occur in \\
{\tt pre\_config\_dup\_main} as sub-proof nets.
Figure~\ref{suc0L} and Figure~\ref{suc0R} show proof nets {\tt suc0L} and {\tt suc0R}.
We omit {\tt suc1L}, {\tt suc1R}, {\tt suc$\ast$L} and {\tt suc$\ast$R} 
since the constructions of these proof nets are easy exercise.

\begin{figure}[htbp]
\begin{center}
\includegraphics[scale=0.5]{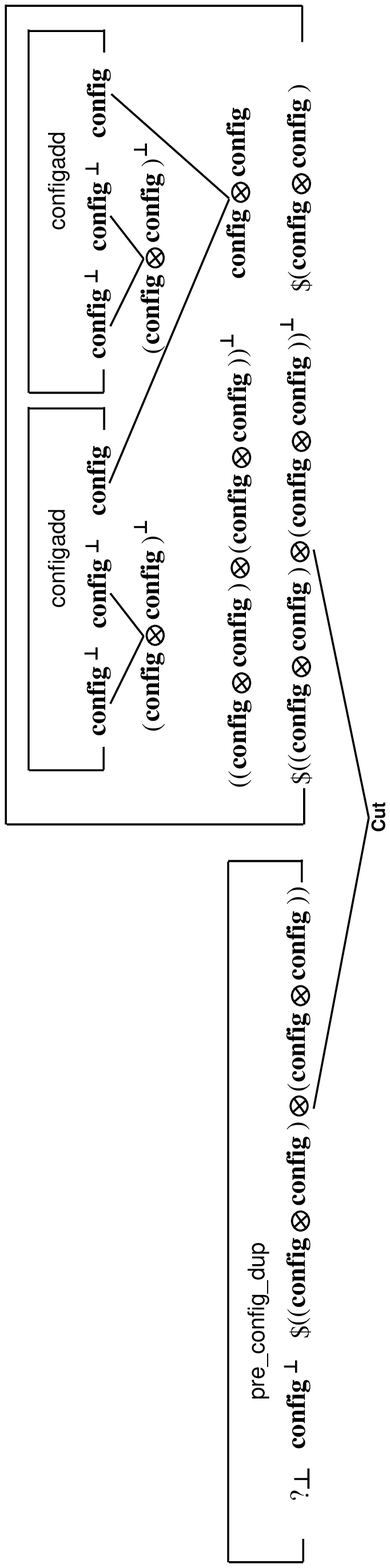}
\caption[{\tt 2-contraction-config}]{{\tt 2-contraction-config}}
\label{config2contraction}
\end{center}
\end{figure}

\begin{figure}[htbp]
\begin{center}
\includegraphics[scale=0.5]{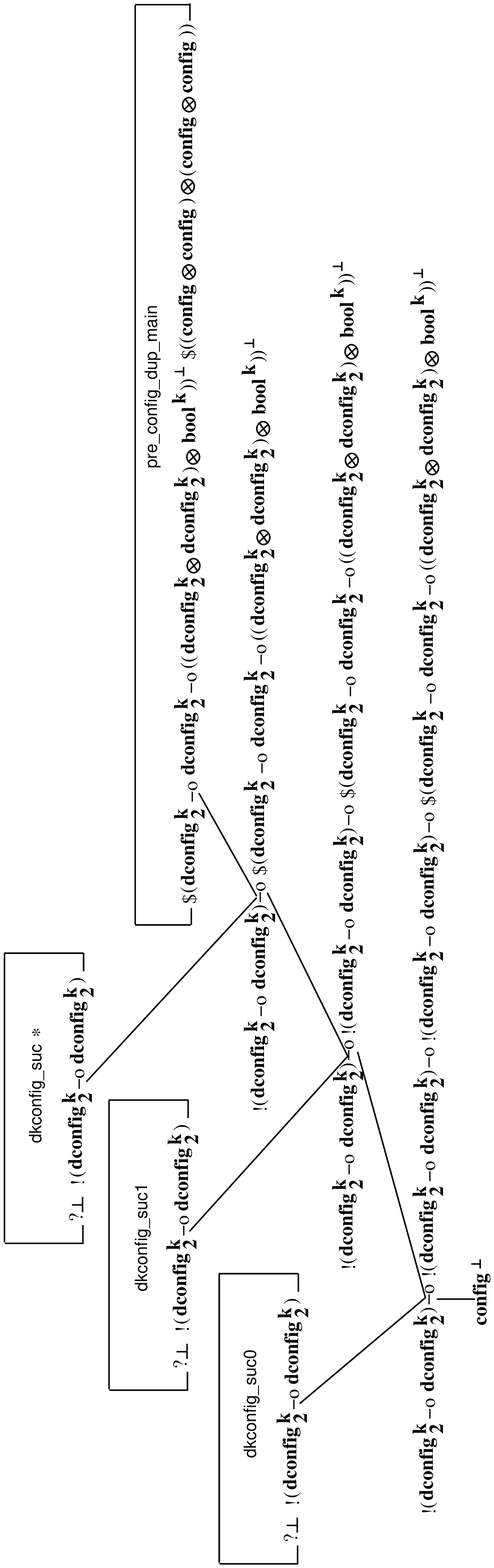}
\caption[{\tt pre\_config\_dup}]{{\tt pre\_config\_dup}}
\label{preconfigdup}
\end{center}
\end{figure}

\begin{figure}[htbp]
\begin{center}
\includegraphics[scale=0.5]{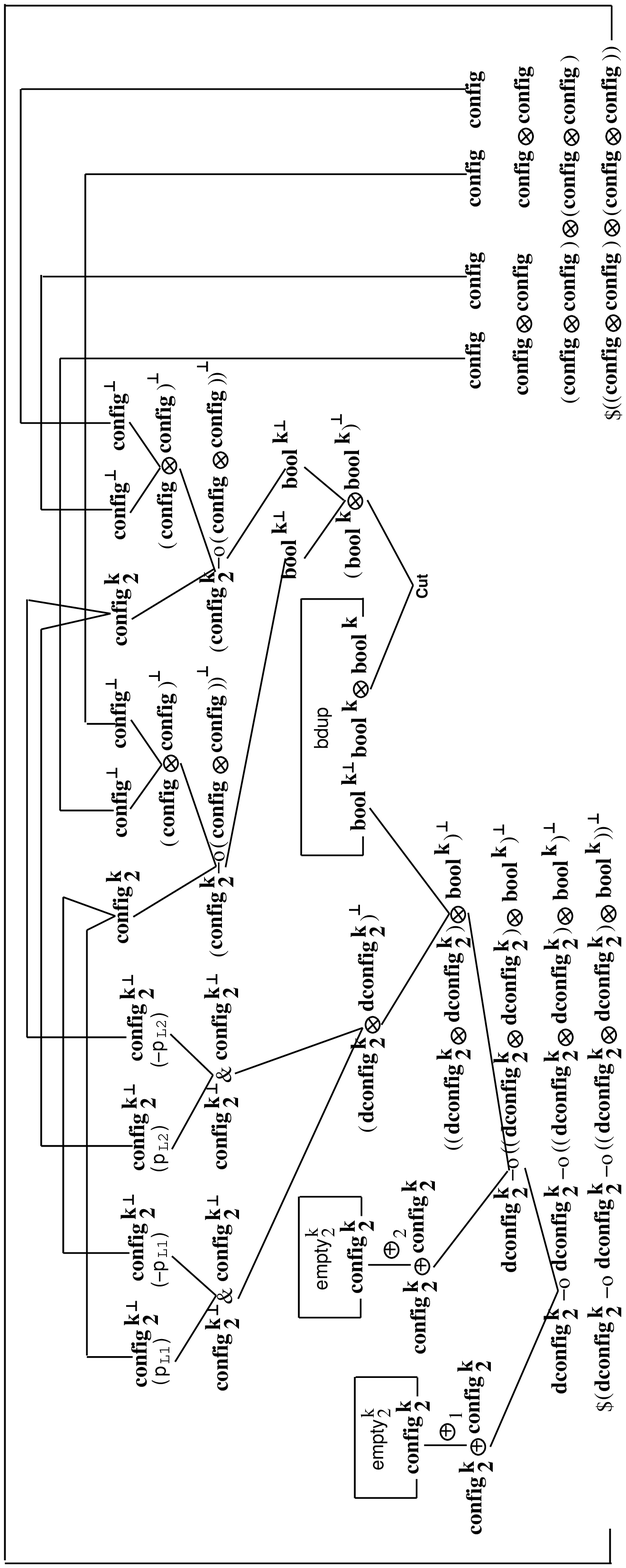}
\caption[{\tt pre\_config\_dup\_main}]{{\tt pre\_config\_dup\_main}}
\label{preconfigdupmain}
\end{center}
\end{figure}

\begin{figure}[htbp]
\begin{center}
\includegraphics[scale=0.5]{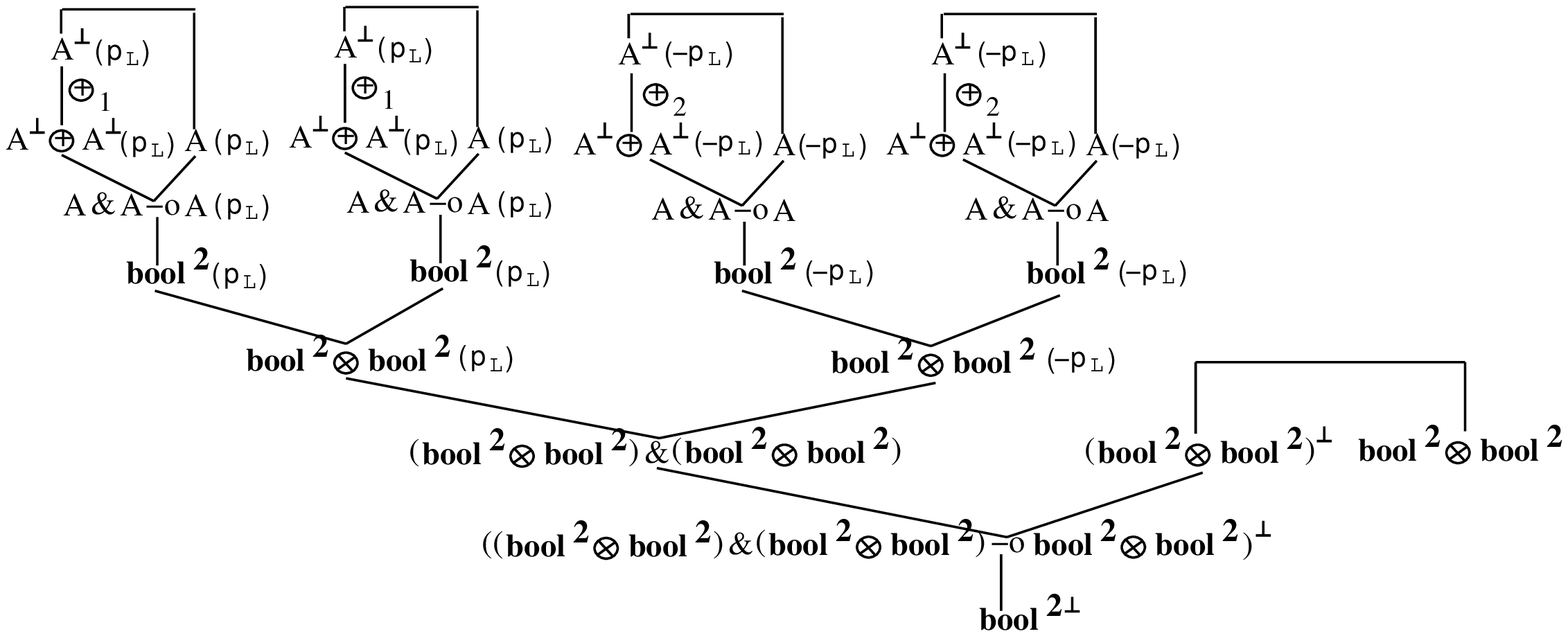}
\caption[{\tt bdup}]{{\tt bdup}}
\label{bdup}
\end{center}
\end{figure}

\begin{figure}[htbp]
\begin{center}
\includegraphics[scale=0.5]{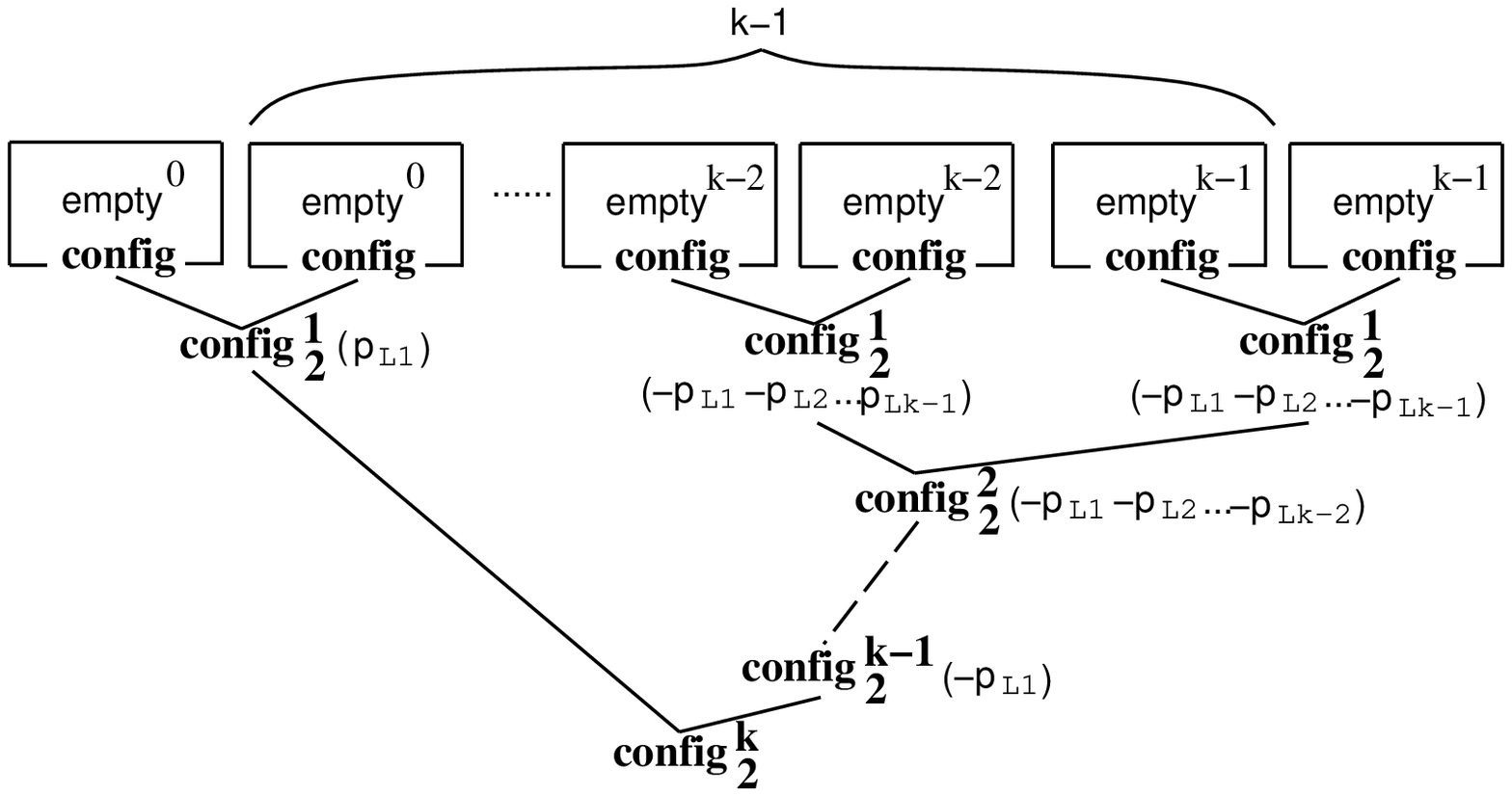}
\caption[${\tt empty}^k_2$]{${\tt empty}^k_2$}
\label{emptyk2}
\end{center}
\end{figure}

\begin{figure}[htbp]
\begin{center}
\includegraphics[scale=0.5]{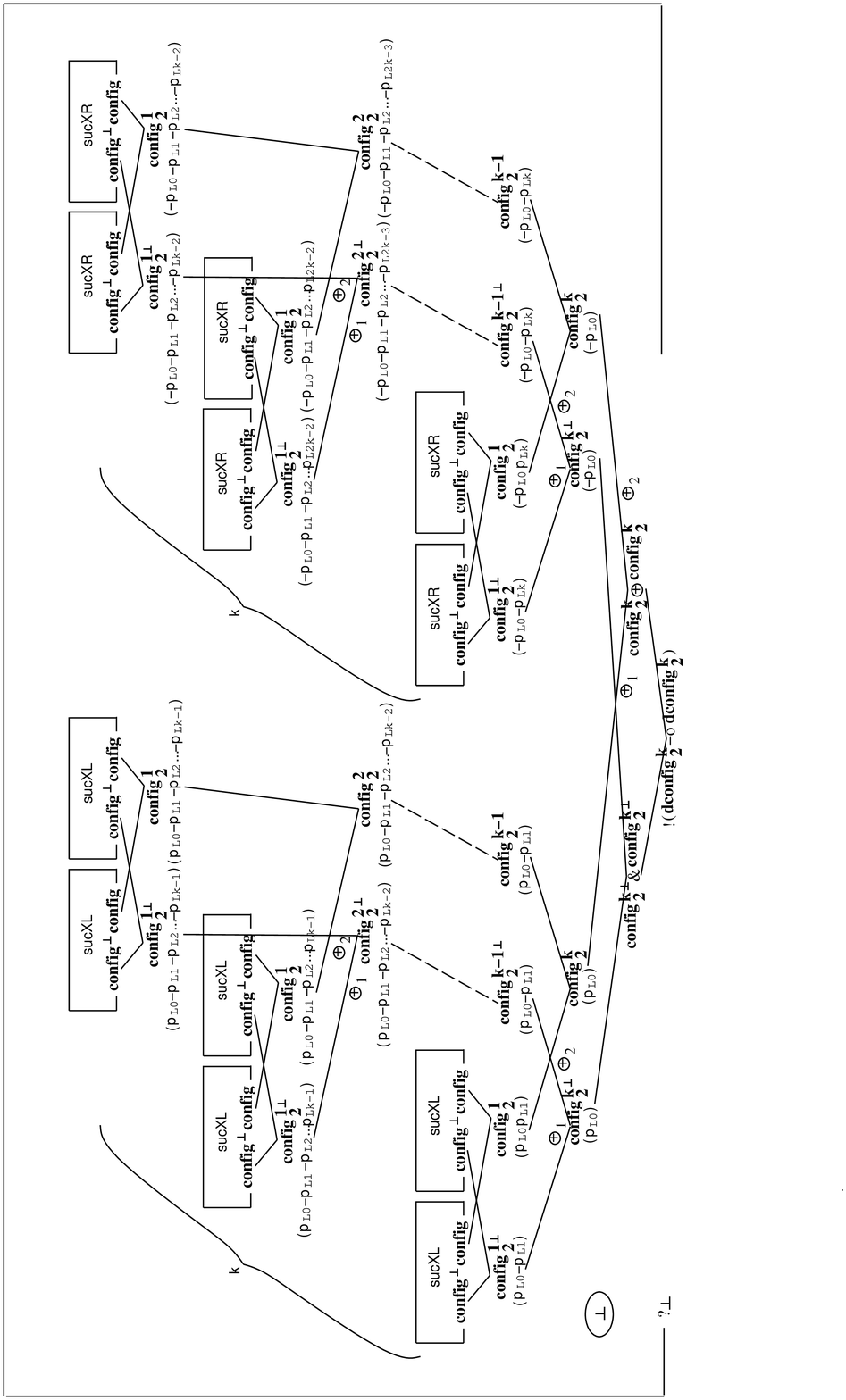}
\caption[${\tt dconfig\_sucX}$]{${\tt dconfig\_sucX}$}
\label{dconfigk2}
\end{center}
\end{figure}

\begin{figure}[htbp]
\begin{center}
\includegraphics[scale=0.5]{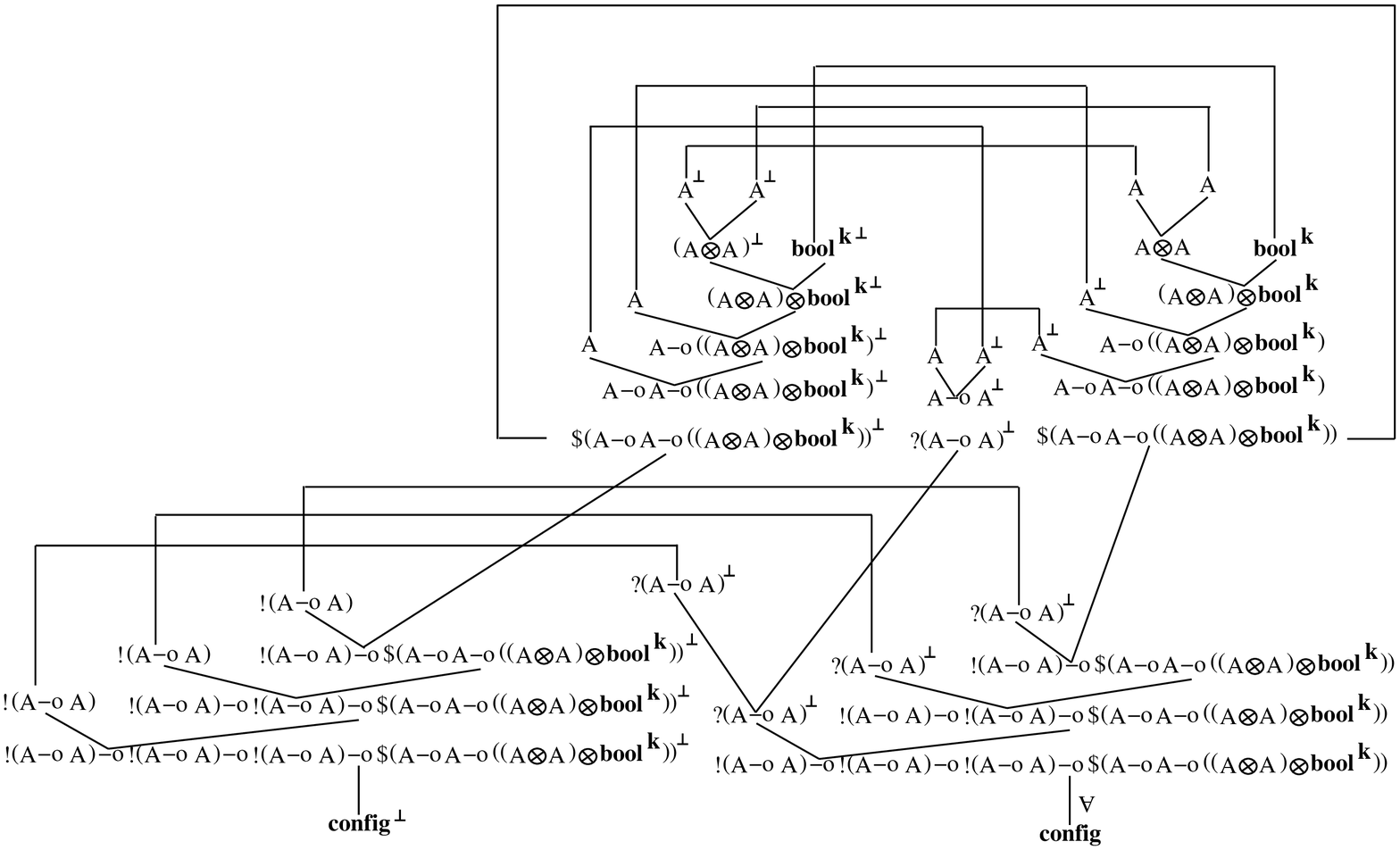}
\caption[{\tt suc0L}]{{\tt suc0L}}
\label{suc0L}
\end{center}
\end{figure}

\begin{figure}[htbp]
\begin{center}
\includegraphics[scale=0.5]{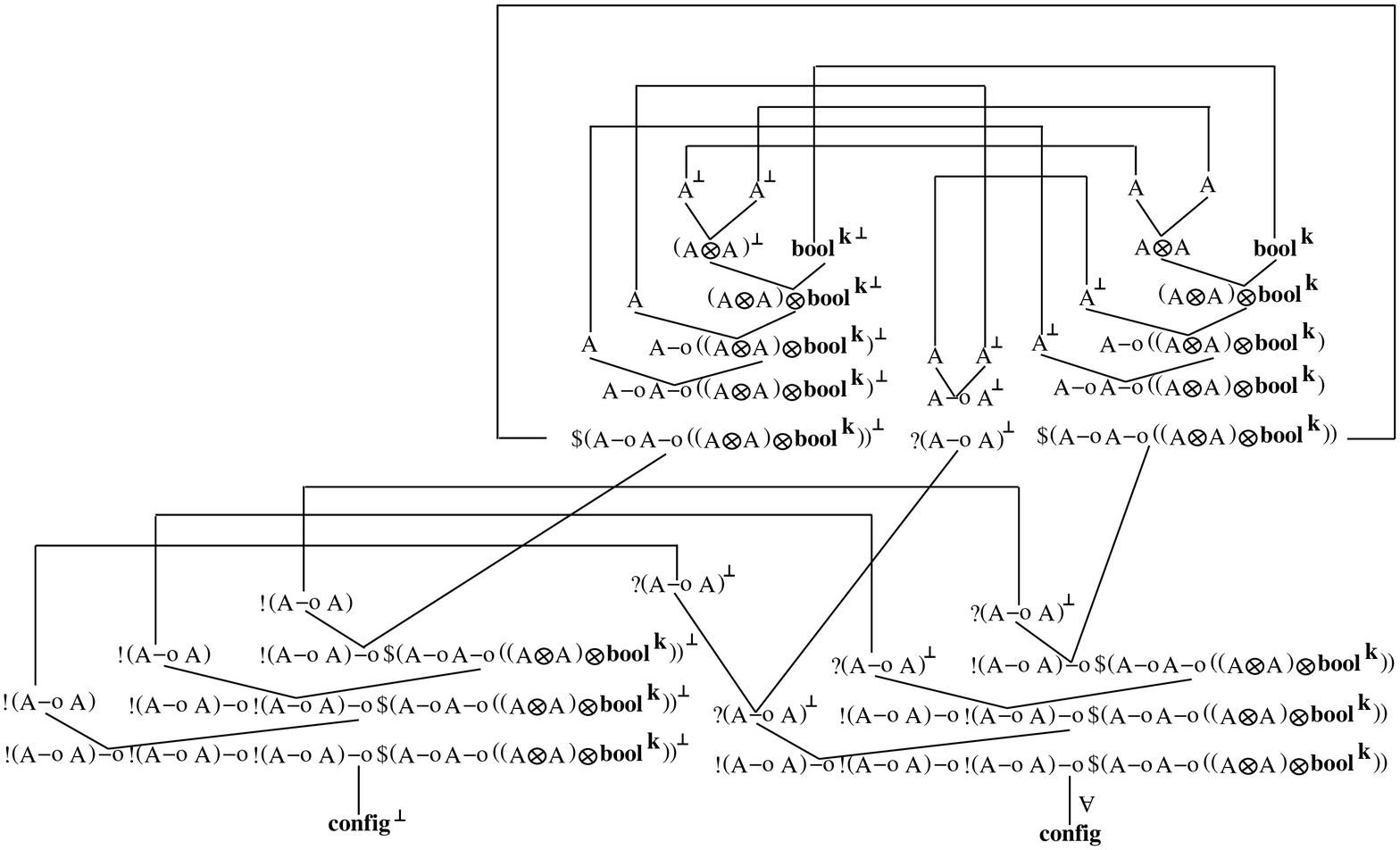}
\caption[{\tt suc0R}]{{\tt suc0R}}
\label{suc0R}
\end{center}
\end{figure}

\begin{figure}[htbp]
\begin{center}
\includegraphics[scale=0.5]{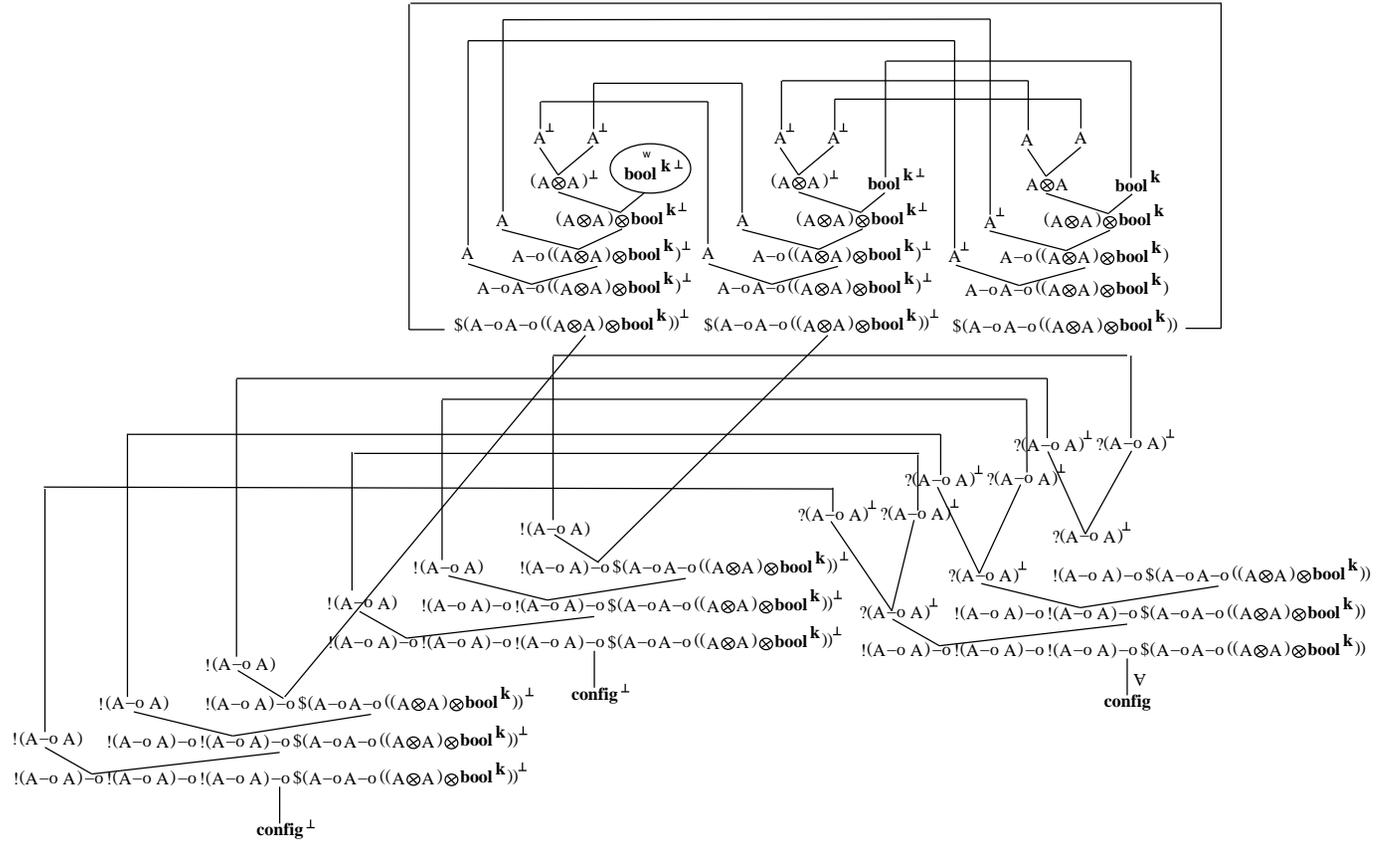}
\caption[{\tt configadd}]{{\tt configadd}}
\label{configadd}
\end{center}
\end{figure}

\end{paper}
\end{document}